\documentclass[twocolumn]{aastex63}

\usepackage{epsfig}
\usepackage{graphicx}
\usepackage{amsthm}
\usepackage{amssymb}
\usepackage{rotating}
\usepackage{multirow}
\usepackage{longtable}
\usepackage{savesym}
\savesymbol{tablenum}
\restoresymbol{SIX}{tablenum}
\usepackage{booktabs}

\defcitealias{2019ApJ...873...51L}{Paper~I}
\defcitealias{2020ApJ...888...98L}{Paper~II}

\submitjournal{ApJ}

\shorttitle{Surveying G\ion{H}{2} Regions: III. W49A}
\shortauthors{De\,Buizer et al.}

\begin{document}

\title{Surveying the Giant \ion{H}{2} Regions of the Milky Way with \textit{SOFIA}: III. W49A}

\email{jdebuizer@sofia.usra.edu}

\author[0000-0001-7378-4430]{James M. De Buizer}
\affil{\textit{SOFIA}-USRA, NASA Ames Research Center, MS 232-12, Moffett Field, CA 94035, USA}

\author[0000-0003-4243-6809]{Wanggi Lim}
\affil{\textit{SOFIA}-USRA, NASA Ames Research Center, MS 232-12, Moffett Field, CA 94035, USA}

\author[0000-0001-6159-2394]{Mengyao Liu}
\affil{Dept. of Astronomy, University of Virginia, Charlottesville, Virginia 22904, USA}

\author[0000-0003-3682-854X]{Nicole Karnath}
\affil{\textit{SOFIA}-USRA, NASA Ames Research Center, MS 232-12, Moffett Field, CA 94035, USA}

\author{James T. Radomski}
\affil{\textit{SOFIA}-USRA, NASA Ames Research Center, MS 232-12, Moffett Field, CA 94035, USA}

\begin{abstract}
We present our third set of results from our mid-infrared imaging survey of Milky Way Giant \ion{H}{2} regions with our detailed analysis of W49A, one of the most distant, yet most luminous, G\ion{H}{2} regions in the Galaxy. We used the FORCAST instrument on the Stratospheric Observatory For Infrared Astronomy ({\it SOFIA}) to obtain 20 and 37\,$\mu$m images of the entire $\sim$5.0$\arcmin\times 3.5\arcmin$ infrared-emitting area of W49A at a spatial resolution of $\sim$3$\arcsec$. Utilizing these {\it SOFIA} data in conjunction with previous multi-wavelength observations from the near-infrared to radio, including  {\it Spitzer}-IRAC and {\it Herschel}-PACS archival data, we investigate the physical nature of individual infrared sources and sub-components within W49A. For individual compact sources we used the multi-wavelength photometry data to construct spectral energy distributions (SEDs) and fit them with massive young stellar object (MYSO) SED models, and find 22 sources that are likely to be MYSOs. Ten new sources are identified for the first time in this work. Even at 37\,$\mu$m we are unable to detect infrared emission from the sources on the western side of the extremely extinguished ring of compact radio emission sources known as the Welch Ring. Utilizing multi-wavelength data, we derived luminosity-to-mass ratio and virial parameters of the extended radio sub-regions of W49A to estimate their relative ages and find that overall the sub-components of W49A have a very small spread in evolutionary state compared to our previously studied G\ion{H}{2} regions.

\end{abstract}


\section{Introduction} 

It is believed that the vast majority of all stars in a galaxy form within OB clusters \citep{2003ARA&A..41...57L}. Galactic giant \ion{H}{2} (G\ion{H}{2}) regions are hosts to the largest and most massive star-forming clusters in the Milky Way, and thus have been used as laboratories for the study of the extreme environments of the earliest stages of clustered massive star formation \citep[e.g., ][]{1978A&A....66...65S, 2004MNRAS.355..899C, 2011MNRAS.411..705M}. An overview of the nature of G\ion{H}{2} regions and why they are important to study is highlighted in detail in the introduction of Lim \& De Buizer (2019; hereafter ``\citetalias{2019ApJ...873...51L}''),  which focused on the infrared properties of the G\ion{H}{2} region W51A. That was followed up by a paper dedicated to the study of G\ion{H}{2} region M17 (Lim et al. 2020; hereafter ``\citetalias{2020ApJ...888...98L}''). Building off those studies, this is the third paper in our continuing series concentrating on the study of the infrared properties of Galactic G\ion{H}{2} regions utilizing new data obtained from SOFIA, this time focusing on the well-known source W49A.

Even though W49A is located on the far side of our Galaxy at a distance of 11.1~kpc \citep{2013ApJ...775...79Z}, it has an infrared emitting region subtending more than 4 arcminutes ($\sim$13\,pc) on a side. This makes W49A one of the largest and most luminous \citep[$M\sim10^6\,M_\sun$, $L > 10^7\,L_\sun$:][]{1973ApL....13..147B,1990MNRAS.244..458W} G\ion{H}{2} regions in the Galaxy. Given its nature as a massive star forming region, observations towards the massive young stellar objects (MYSOs) it contains are subject to a large amount of local extinction. Additionally, due to its location in the plane of the Galaxy and its large distance, it is observed through the obscuring gas and dust of the Milky Way's Sagittarius spiral arm, which crosses the line of sight to W49A twice \citep{2004ApJ...605..247P}. The combined level of local and galactic extinction \citep[estimates are as high as A\textsubscript{V} $\sim$ 300 mag;][] {1976ApJ...209...94W} means that even at near- and mid-infrared wavelengths there is considerable obscuration. Given its significant brightness at longer wavelengths, however, W49A has been the subject of numerous studies from the infrared to radio wavelengths. A consequence of the large distance to W49A is that, unlike the previously studied G\ion{H}{2} regions from this project, the physical scales we are probing will be much larger ($\theta\textsubscript{resolution}$ $\sim$ 3$\farcs$1 $\sim$ 35000~au). While more than 50 O-star candidates have been identified in the central region ($r \sim45\arcsec$) of W49A alone \citep{2003ApJ...589L..45A}, it is believed to harbor one of the largest concentrations of compact \ion{H}{2} (C\ion{H}{2}) and ultracompact \ion{H}{2} (UC\ion{H}{2}) regions \citep{1987Sci...238.1550W}, signifying that it may be a relatively young G\ion{H}{2} region.

In Section \ref{sec:obs}, we will discuss the new \textit{SOFIA} observations and give information on the data obtained. In Section \ref{sec:W49Asources}, we will give more background on W49A as we compare our new data to previous observations and discuss individual sources and regions in-depth. In Section \ref{sec:data}, we will discuss our data analysis, modeling, and derivation of physical parameters of sources and regions. Our results and   conclusions are summarized in Section \ref{sec:sum}.

\section{Observations and Data Reduction} \label{sec:obs}

The observational techniques and reduction processes employed on the data were, for the most part, identical to those described in \citetalias{2019ApJ...873...51L} for W51A. Below we will highlight some of observation and reduction details specific to these new observations. For a more in-depth discussion of these details and techniques, refer to \citetalias{2019ApJ...873...51L}. 

All observations were made with the airborne astronomical observatory, \textit{SOFIA} \citep{2012ApJ...749L..17Y}, utilizing the FORCAST instrument \citep{2013PASP..125.1393H}. All data were taken of W49A on the night of 2015 September 16 (Flight 239). All observations were taken at an altitude of 41000\,ft, which typically yields precipitable water vapor overburdens of less than 8\,$\mu$m. FORCAST is a facility imager and spectrograph that employs a Si:As 256$\times$256 blocked-impurity band (BIB) detector array to cover a wavelength range of 5 to 25\,$\mu$m and a Si:Sb 256$\times$256 BIB array to cover the range from 25 to 40\,$\mu$m. Imaging data were obtained in the 20\,$\mu$m ($\lambda_{eff}$ = 19.7\,$\mu$m; $\Delta\lambda$ = 5.5\,$\mu$m) and 37\,$\mu$m ($\lambda_{eff}$ = 37.1\,$\mu$m; $\Delta\lambda$ = 3.3\,$\mu$m) filters simultaneously using an internal dichroic. In imaging mode the arrays cover a 3$\farcm$40$\times$3$\farcm$20 instantaneous field-of-view with a pixel scale of 0$\farcs$768 pixel$^{-1}$ after distortion correction. 

\begin{figure*}[htb!]
\epsscale{1.1}
\plotone{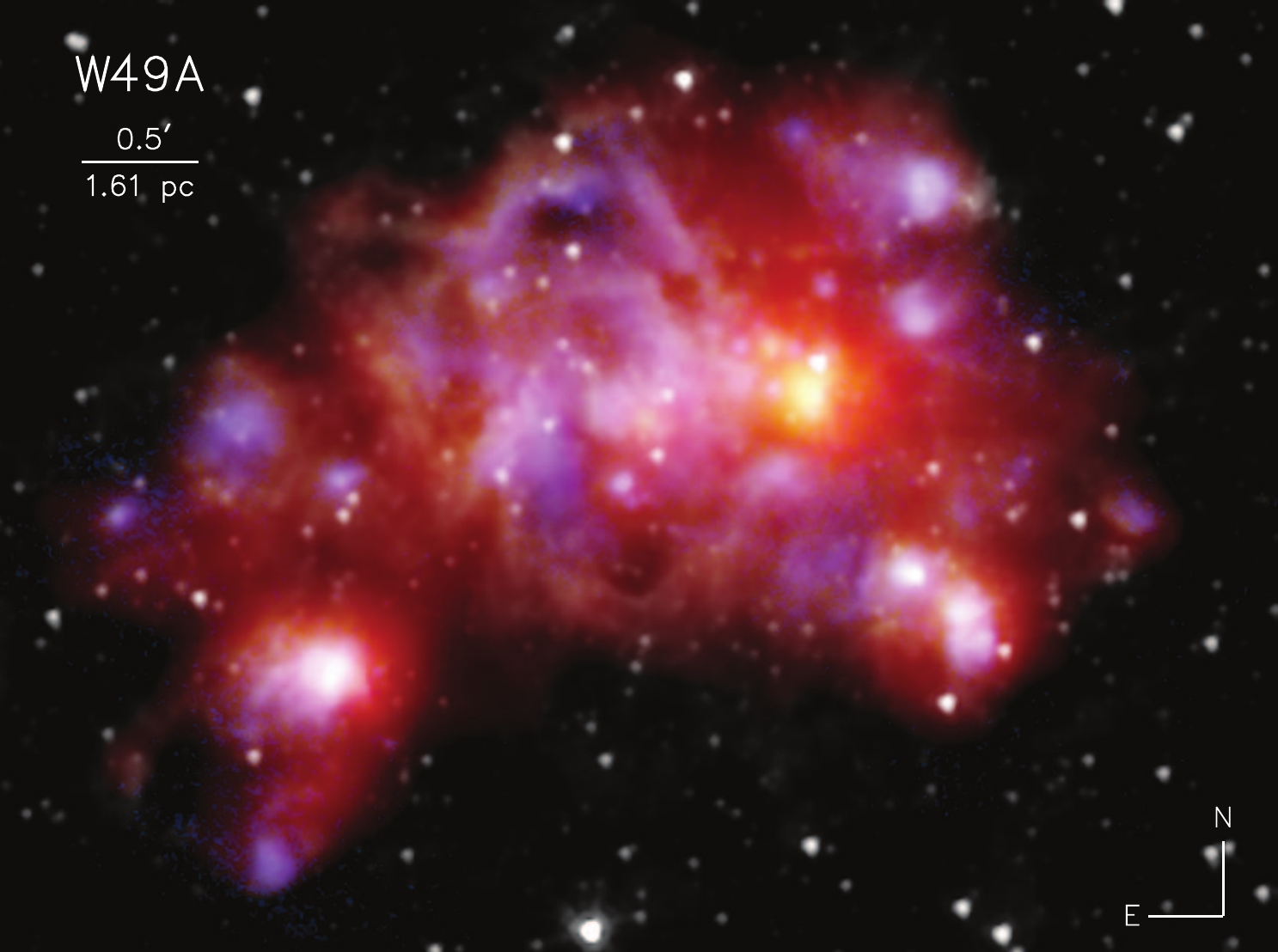}
\caption{A 3-color image of a $\sim$5$\arcmin\times 4\arcmin$ field centered on W49A. Blue is the \textit{SOFIA}-FORCAST 20\,$\mu$m data, green is the \textit{SOFIA}-FORCAST 37\,$\mu$m data, and red is the \textit{Herschel}-PACS 70\,$\mu$m data. Overlaid in white is the \textit{Spitzer} -IRAC 3.6\,$\mu$m data, which traces the revealed stars within W49A, field stars, and hot dust. \label{fig:fig1}}
\end{figure*} 

All images were obtained by employing the standard chop-nod observing technique used in ground-based thermal infrared observing, with chop throws of 4.2$\arcmin$ and nod throws of 15$\arcmin$ which were sufficiently large enough to sample clear off-source sky regions uncontaminated by the extended emission of W49A. The mid-infrared emitting area of W49A is much larger than the FORCAST field of view, and thus had to be mapped using multiple pointings. We created a mosaic from 3 individual pointings with each pointing having an average on-source exposure time of about 180s at both 20\,$\mu$m and 37\,$\mu$m. Final mosaicked images made from the individual pointing images were stitched together using the \textit{SOFIA} Data Pipeline software REDUX \citep{2015ASPC..495..355C}.

\textit{SOFIA} data were calibrated by the \textit{SOFIA} pipeline with a system of stellar calibrators taken across all flights in the flight series and applied to all targets within that flight series \citep[see also the FORCAST calibration paper by][]{2013PASP..125.1393H}. Corrections are made for airmass of the science data as well. 

In order to try to resolve sources in crowded regions, the \textit{SOFIA} images at 20 and 37\,$\mu$m were deconvolved using the maximum likelihood method (Richardson 1972; Lucy 1974). Like all deconvolution methods, knowledge of the point-spread function (PSF) of an unresolved source is needed at each wavelength, and with high signal-to-noise. The PSF of \textit{SOFIA} is slightly variable from flight to flight and observations to observation, so using a bright, isolated point-source in the W49A mosaic would be most accurate. However, no such point-source is present in our data. Therefore, using standard stars observed throughout several flights, an average full width at half maximum (FWHM) for each wavelength was determined. Then artificially generated PSFs (an Airy pattern calculated from the wavelength, telescope diameter, and central obscuration diameter) were constructed and convolved with a Gaussian to achieve PSFs with FWHMs that equaled the measured average FWHMs of the standard stars. These idealized PSFs were then used in the deconvolution procedure. The deconvolution routine was stopped at 100 iterations for both the 20 and 37\,$\mu$m images. Deconvolutions with the average PSF from the standard star observations were also tried, and yielded nearly identical results. The deconvolved images using the idealized PSF also  compared favorably to simple unsharp masking of the original images, and thus the substructures revealed in the deconvolved images are believed to be reliable.  Reliable flux conservation for deconvolved data requires flat and/or zero-mean backgrounds. However, the data for W49A are pervaded with diffuse and extensive nebular dust emission, and this makes the derived fluxes of the embedded compact sources from the deconvolved data less reliable than from the natural resolution data. Therefore, the deconvolved data are only used in this study for morphological comparisons to data at other wavelengths with higher spatial resolution. 

In addition to the \textit{SOFIA} data, we also utilize science-ready imaging data from the \textit{Spitzer Space Telescope} and \textit{Herschel Space Telescope} archives. In addition to having access to the \textit{Very Large Array} 3.6\,cm data with $\sim$0.8$\arcsec$ spatial resolution used in \citet{1997ApJ...482..307D}, we additionally obtained \textit{VLA} archival data at 3.6\,cm  with a spatial resolution of 9.8$\arcsec$ and a field of view of 2$\farcm$5 centered on W49A.  

\section{Comparing \textit{SOFIA} Images of W49A to Previous Imaging Observations} \label{sec:W49Asources}

W49 was first discovered as a radio continuum region by \citet{1958BAN....14..215W} at 22\,cm, and consists of the high-mass star-forming region W49A and the nearby ($\sim$12$\arcmin$ away) supernova remnant W49B. The giant \ion{H}{2} region of W49A displays extensive extended radio continuum emission ($d \gtrsim2\arcmin$), while at the same it is believed to harbor the highest concentration of compact and ultra compact \ion{H}{2} regions in the Galaxy \citep[18;][]{2013MNRAS.435..400U}, with a central group of UC\ion{H}{2} regions distributed in a ring-like structure \citep{1987Sci...238.1550W}. In addition to these UC\ion{H}{2} regions, the entire region around W49A ($r \sim8$\,pc) has more than 250 massive O-type stars \citep{2005A&A...430..481H}.

\begin{figure*}[htb!]
\epsscale{0.95}
\plotone{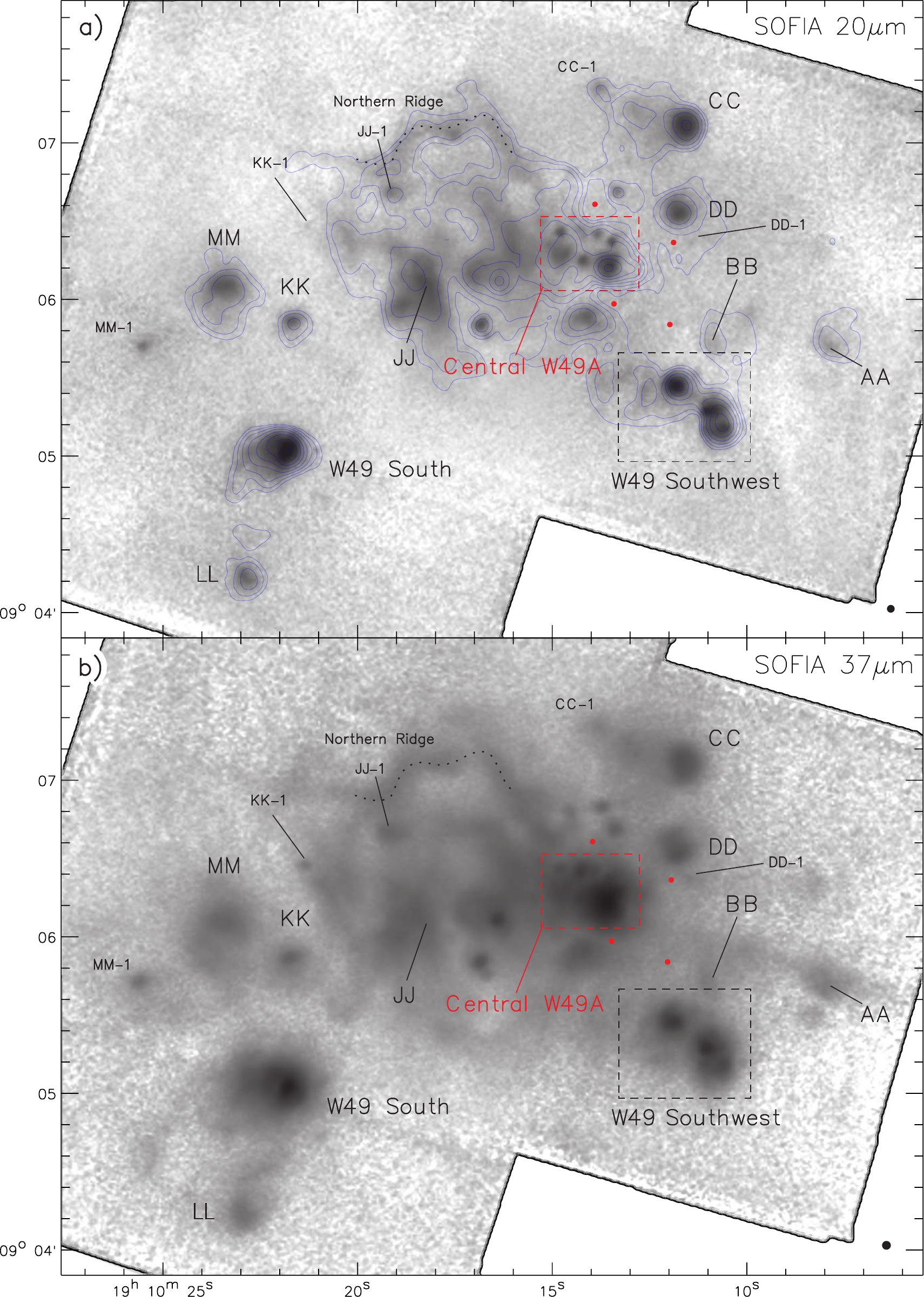}
\caption{W49A image mosaic taken at a) 20\,$\mu$m and b) 37\,$\mu$m by \textit{SOFIA} shown in inverse gray scale (i.e. brighter features are  darker in color). The light blue contours in the top panel are the 3.6\,cm radio continuum emission from \citet{1997ApJ...482..307D}. All sources are labeled except in the higher source-density areas, where only the region name is given (dashed boxes). For the W49\,Southwest region see Figure \ref{fig:W49Southwest2} and for the Central W49A region see Figure \ref{fig:W49A_center} for more details and to see the individual sources labeled within those areas. The curvy dotted line represents the location of the Northern Ridge. The red dots are the locations of the sources identified in \citet{2000ApJ...540..316S} as (going north to south) EE East, DD South, HH West, and BB East. The black dot in the lower right of each panel indicates the resolution of the image at each wavelength. \label{fig:SOFIAdata}}
\end{figure*}

The infrared observations from \textit{SOFIA} show the region to have structured but extended dust emission spread over an approximately 4$\arcmin\times 3\arcmin$ area (Figure \ref{fig:fig1}), which corresponds generally to the extent of the cm radio continuum emission seen by \citet{1997ApJ...482..307D}, as can be seen in Figure \ref{fig:SOFIAdata}a. While most of the infrared features are seen at both \textit{SOFIA} wavelengths, the dust is more pronounced and extended at 37\,$\mu$m compared to 20\,$\mu$m (Figure \ref{fig:SOFIAdata}), indicative of widespread cool dust.  

\subsection{Discussion of Individual Sources in W49A}

\begin{figure*}[htb!]
\epsscale{1.1}
\plotone{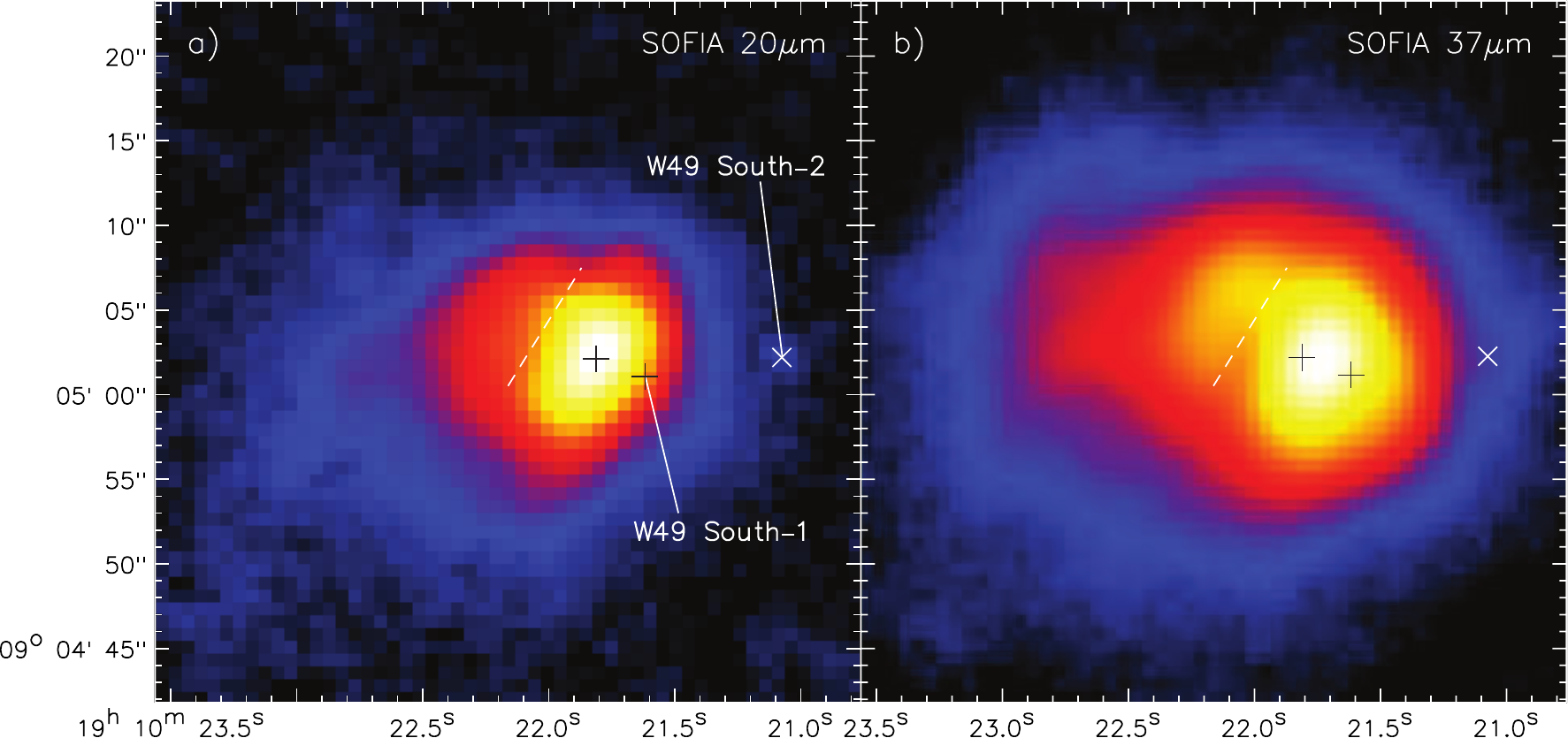}
\caption{W49\,South with false-color images showing emission at a) \textit{SOFIA} 20\,$\mu$m and b) \textit{SOFIA} 37\,$\mu$m. The black crosses show the locations of the cm radio continuum peaks for W49\,South (east cross) and W49\,South-1 (west cross). Newly identified infrared source W49\,South-2 is marked with an x. The location of the dark lane discussed in the text is indicated by the dashed white line.  \label{fig:W49South1}}
\end{figure*}

From the first observations at arcminute resolution \citep{1967ApJ...150..807M}, the radio continuum structure of W49A was seen to display two intensity peaks separated by $\sim$2$\farcm$5. Multiple monikers were used in naming these sources in early observations, however they are now most often denoted as W49\,North and W49\,South, or W49N and W49S \citep[e.g.,][]{1984ApJ...283..632D}. \citet{1990ApJ...351..189D} observed the radio continuum emission at 2$\arcsec$ resolution with the \textit{VLA} and labeled the peaks in emission they found as A through S by increasing right ascension. W49A was revisited with the \textit{VLA} by \citet{1997ApJ...482..307D}, and even more radio continuum emission peaks were found. Those newly detected peaks that were resolved from or close to already known peaks were indexed with a number (e.g., G$_1$, G$_2$, etc.), however entirely new peaks independent of already known sources were labeled with a double letter (e.g., AA, BB, etc.), in order of increasing right ascension. Even higher spatial resolution observations by \citet{2000ApJ...540..308D} and \citet{2020AJ....160..234D} at 3.6\,cm ($\theta_{beam} \sim$0$\farcs$15) and at several mm wavelengths (0$\farcs$04 $<\theta_{beam}<$ 0$\farcs$35), further resolved details and even more sub-sources in the crowded south-western side of the Welch Ring (i.e. sources A through H). 

The sub-region containing sources Q, R, and S are also occasionally referred to by the collective moniker of W49\,Southwest, first named as such by \citet{1971AJ.....76..677W}. Figure \ref{fig:SOFIAdata} shows the \textit{SOFIA} observations at 20 and 37\,$\mu$m of W49A, with the main regions labeled (i.e. Central W49A, W49\,South, and W49\,Southwest), as well as all of the sources that are distributed outside of these main regions.  

While discussing the nature of the individual sources below, we will sometimes refer to the results of our spectral energy distribution (SED) model fitting (like derived luminosity, for instance) which are discussed in-depth in Section \ref{sec:data}. See that section for further information describing the SED model fitting algorithm used, as well as details regarding the assumptions, limitations, and results of the SED fitting.   

\subsubsection{W49\,South}\label{s:w49s}

W49~South is the brightest source in all of W49A in the infrared from $\sim$3\,$\mu$m out to $\sim$20\,$\mu$m. It is saturated in all \textit{Spitzer} IRAC and MIPS wavelengths, except in the IRAC channels 1 and 2 (3.6 and 4.5\,$\mu$m). In our 37\,$\mu$m data and at longer wavelengths seen by \textit{Herschel} it becomes the second brightest peak to source G. 

\citet{2000ApJ...540..316S} claim that there is a roughly circular halo of extended emission in their data at wavelengths from 8.0 to 20\,$\mu$m, and suggest the emission may be tracing a nearly complete 4$\arcsec$ radius dust shell. However our 37\,$\mu$m data show what appears to be a lane of lesser emission running NW-SE through the extended emission just east of the peak (Figure \ref{fig:W49South1}). \citet{2000ApJ...540..316S} discuss a protrusion to the northwest in their 20\,$\mu$m data, however, it is likely that this is an artifact of this darker lane creating a notch in the extended emission on the northern part of the source, and since the western side is brighter it looks like a protrusion. 

\begin{figure*}[htb!]
\epsscale{1.1}
\plotone{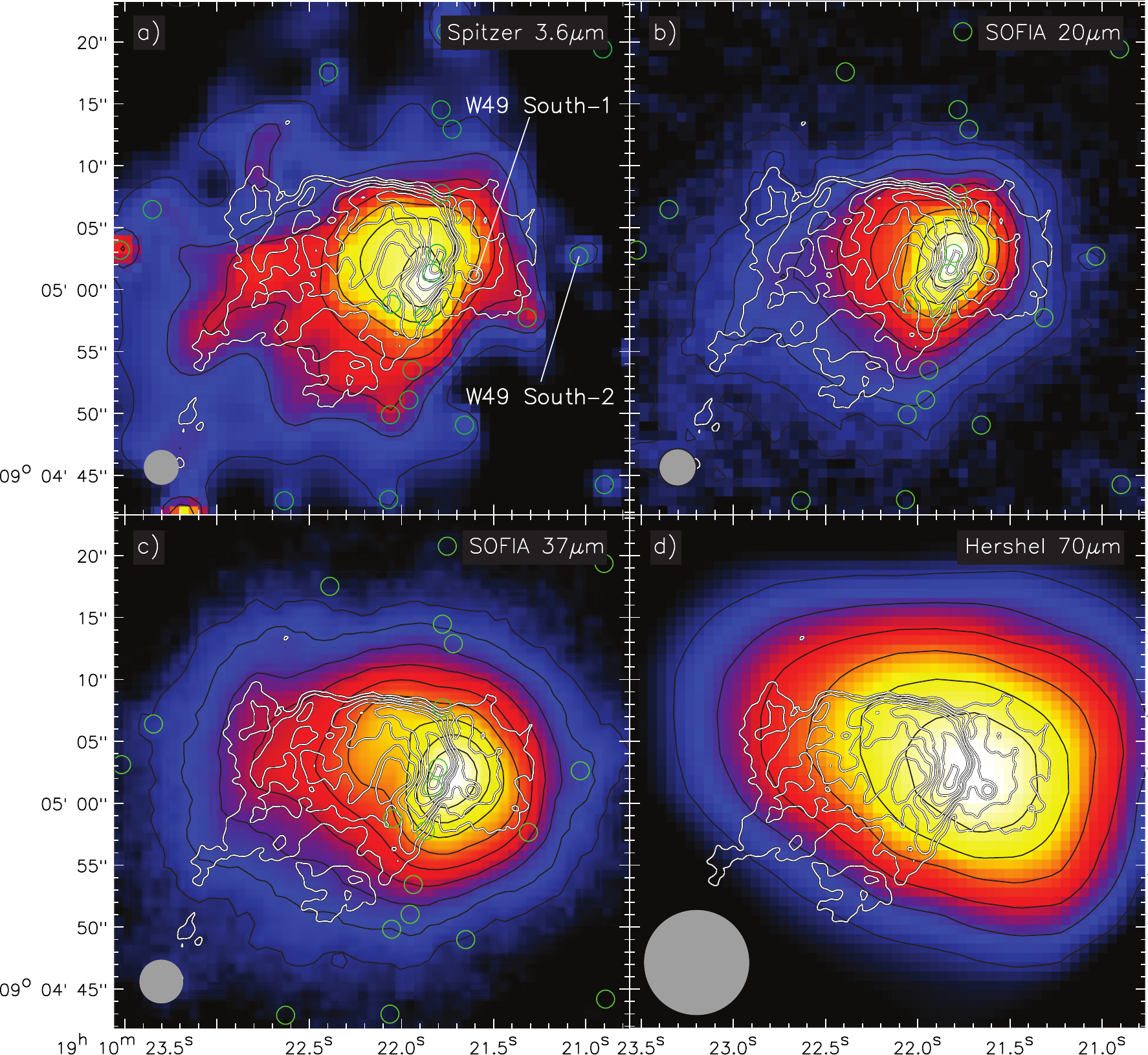}
\caption{W49\,South with false-color images and thin black contours showing emission at a) \textit{Spitzer} 3.6\,$\mu$m, b) \textit{SOFIA} 20\,$\mu$m, c) \textit{SOFIA} 37\,$\mu$m, and d) \textit{Herschel} 70\,$\mu$m. Overlaid on each panel are the 3.6\,cm radio continuum contours from \citet{1997ApJ...482..307D} as white contours, and the locations of near infrared point sources} identified by \citet{2015ApJ...813...25S} are indicated by the green circles. The resolution at each wavelength is shown by the gray circle in the lower left corner of each panel.  \label{fig:W49South2}
\end{figure*}

The radio continuum emission from this source is classified as a ``cometary'' shape \citep{1997ApJ...482..307D}, with a bright arc-shaped ridge of emission and a diffuse tail of emission toward the southeast. Overall, the infrared emission looks similar to the radio at wavelengths from 3.6\,$\mu$m through the mid-infrared \citep{2000ApJ...540..316S}, and out to the longest wavelengths we see with \textit{SOFIA} at 37\,$\mu$m (Figure \ref{fig:W49South2}), but with some distinctions discussed below.

The large-scale extended emission at 37\,$\mu$m (and at \textit{Herschel} 70\,$\mu$m) reaches almost an arcminute to the east of the peak and is pervaded by cm continuum emission (see Figure \ref{fig:W49South2}) as seen by \citet{1997ApJ...482..307D}. In our 20\,$\mu$m image and in the \textit{Spitzer} wavelengths, the extended emission of W49~South is elongated more toward the southeast (at a p.a. of $\sim$135$\arcdeg$), but the primary extension of emission at 37\,$\mu$m and 70\,$\mu$m is more toward the northeast (at a p.a. of $\sim$75$\arcdeg$). Since radio continuum emission is coming from the regions covered by both the 20 and 37\,$\mu$m extended emission, this implies variable extinction is at work, with higher values to the northeast. 

\begin{figure*}[htb!]
\epsscale{1.1}
\plotone{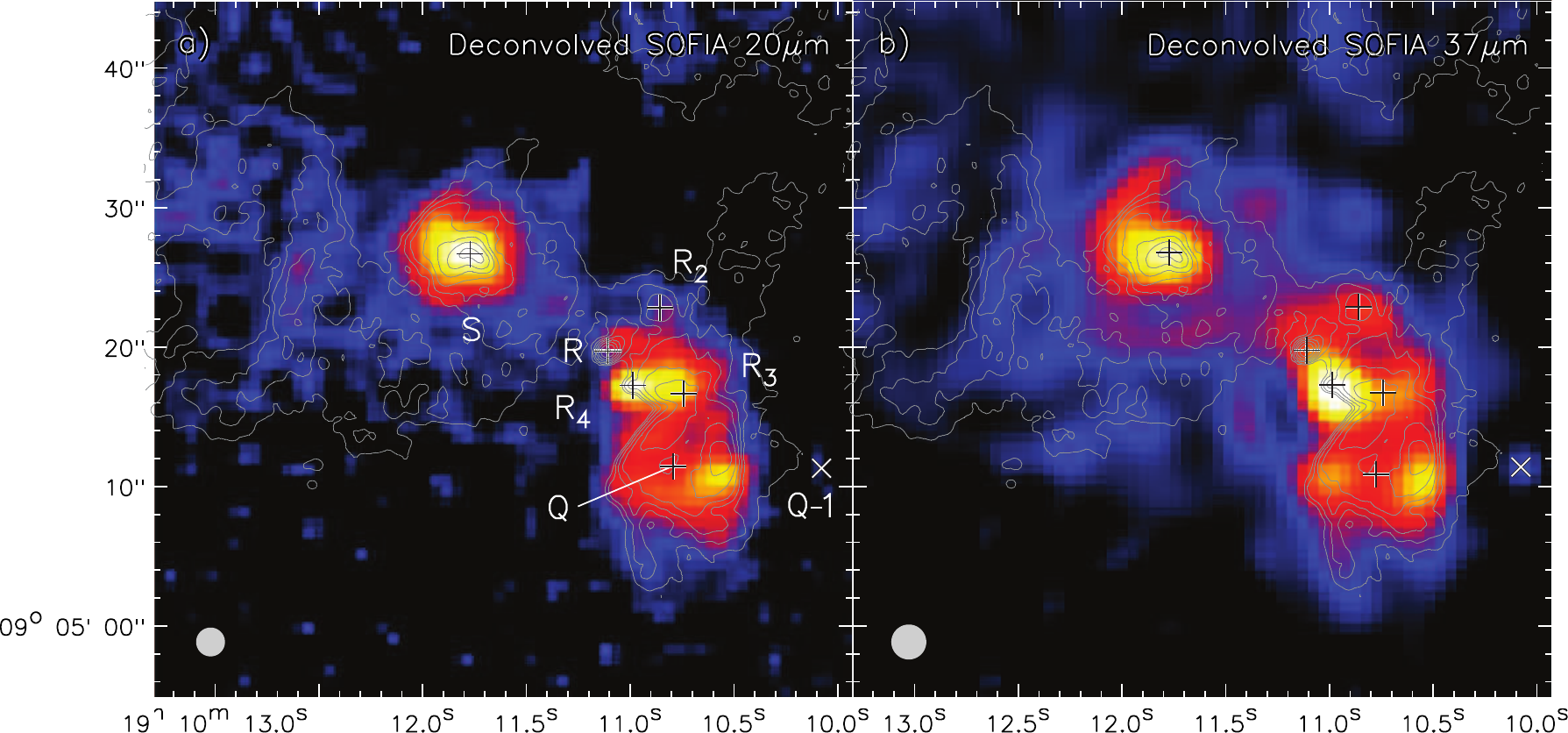}
\caption{W49\,Southwest with false-color images showing emission at a) \textit{SOFIA} 20\,$\mu$m and b) \textit{SOFIA} 37\,$\mu$m. The 3.6\,cm radio continuum contours from \citet{1997ApJ...482..307D} are overlaid on each panel with crosses showing the locations of the cm radio continuum peaks}, with the names of the peaks indicated. The location of the newly identified infrared peak Q-1 is shown by the x symbol. The resolution at each wavelength is shown by the gray circle in the lower left corner of each panel. \label{fig:W49Southwest1}
\end{figure*}

There exists a distinct radio continuum peak to the west of the radio arc named W49~South-1 \citep{1997ApJ...482..307D}. There is no noticeable peak in the infrared emission from this source above that from the cometary UC\ion{H}{2} emission in the \textit{SOFIA} 20\,$\mu$m data. It also seems not to be present in the \textit{Spitzer}-IRAC data either, or at the infrared wavelengths observed by \citet{2000ApJ...540..316S}. However the extended emission near the peak of W49\,South at 37\,$\mu$m definitely protrudes out in this direction (Figure \ref{fig:W49South1}). While the peak of infrared emission in the IRAC and \textit{SOFIA} bands seems co-located with the peak of the radio arc of the cometary UC\ion{H}{2} region, the \textit{Herschel} 70\,$\mu$m peak seems to be offset towards W49~South-1 (Figure \ref{fig:W49South2}). This may signal that this source is highly embedded and becomes a more important contributor to the bolometric luminosity of the region at longer wavelengths. Given the implied higher levels of extinction to the north and west of the radio arc, it is likely that the large-scale morphology of W49~South is due to a champagne-like flow as suggested by \citet{2018A&A...616A.107K}, rather than a bow shock from a moving source as was suggested by \citet{1997ApJ...482..307D}.  

\citet{2005A&A...430..481H} find 13 massive star candidates here in a revealed cluster they call Cluster No. 2 \citep{2003ApJ...589L..45A}, likely from an earlier epoch of star formation. \citet{2015ApJ...813...25S} claim to find no YSOs or MYSOs in the W49~South region, however given its large size ($d \sim 1.6$\,pc) and our derived luminosity ($1.6\times10^4\,L_{\sun}$) W49\,South must contain at least one MYSO. \citet{2015ApJ...813...25S} do identify several unclassified near-infrared sources in the area (Figure \ref{fig:W49South2}). Nine of these sources are located within or projected against the extended mid-infrared emitting region of W49~South. The only one of these sources that is present and resolved from the extended emission in our \textit{SOFIA} data is a source we call W49~South-2 (Figure \ref{fig:W49South1} and \ref{fig:W49South2}), which can be seen clearly in the \textit{Spitzer}-IRAC bands as well as our 20\,$\mu$m image. At 37\,$\mu$m there is an extension toward this region in the extended emission, but no clear source peak can be ascertained. Our SED modeling indicates that this source is also likely to be a MYSO, but since it is not emitting at cm radio continuum wavelengths it may be at an early evolutionary stage prior to the onset of a UC\ion{H}{2} region.  

\subsubsection{W49\,Southwest}\label{sec:sw}

\begin{figure*}[htb!]
\epsscale{1.1}
\plotone{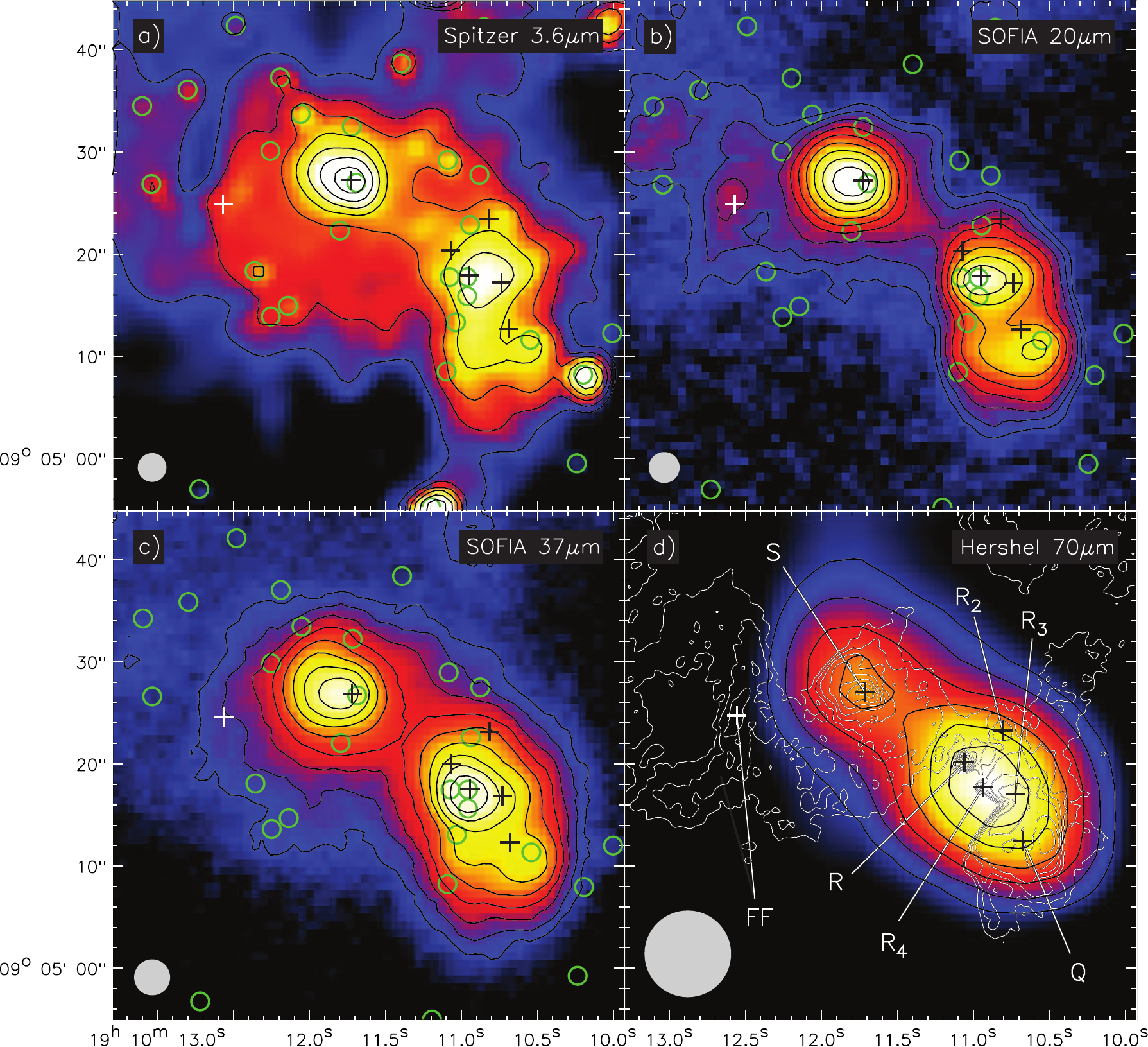}
\caption{W49\,Southwest with false-color images and thin black contours showing emission at a) \textit{Spitzer} 3.6\,$\mu$m, b) \textit{SOFIA} 20\,$\mu$m, c) \textit{SOFIA} 37\,$\mu$m, and d) \textit{Herschel} 70\,$\mu$m. The crosses are the peaks of the 3.6\,cm radio continuum sources from \citet{1997ApJ...482..307D}, and the locations of  near infrared point sources} identified by \citet{2015ApJ...813...25S} are indicated by the green circles. The 3.6\,cm radio continuum contours are shown along with the labels for the cm radio sources in panel d). The resolution at each wavelength is shown by the gray circle in the lower left corner of each panel. \label{fig:W49Southwest2}
\end{figure*}

The cm radio continuum regions R, S, and Q have been collectively referred to as W49\,Southwest. The radio source R was further divided into three subcomponents or peaks by \citet{1997ApJ...482..307D}, named R, R$_2$, and R$_3$. All of these radio peaks are seen as peaks or extensions in the deconvolved infrared emission in the \textit{SOFIA} 20 and 37\,$\mu$m data (Figure \ref{fig:W49Southwest1}). Images at all the infrared wavelengths seen by \textit{Spitzer}-IRAC and \textit{SOFIA} look fairly similar to the cm radio continuum emission \citep{1997ApJ...482..307D}, with the peaks and extended emission aligning fairly well.   
One of the largest distinctions between the cm radio continuum images and \textit{SOFIA} infrared images is that the bright and unresolved radio peak R is not the same location as the brightest infrared peak in the R region. There is a unresolved radio continuum source $\sim$3$\arcsec$ to the southwest of the R source (Figure \ref{fig:W49Southwest2}d) that was not identified by \citet{1997ApJ...482..307D} that we will refer to as R$_4$ (in keeping with the radio source nomenclature). This appears to be the radio peak that is associated with the brightest mid-infrared peak in the region. Additionally, we do see mid-infrared peaks in the deconvolved 20 and 37\,$\mu$m data that correspond to the approximate locations of radio peaks R, R$_2$, and R$_3$ (see Figure \ref{fig:W49Southwest1}). The R$_4$ source appears to dominate the region's emission at longer wavelengths, becoming brighter than source S at 37\,$\mu$m, and appearing to be the closest source location to the bright peak seen at 70\,$\mu$m (Figure \ref{fig:W49Southwest2}c \& d). 

While there is no definitive peak at the location of R in the \textit{Spitzer}-IRAC data, the deconvolved \textit{SOFIA} 20\,$\mu$m data show a definite peak at this location, and the deconvolved 37\,$\mu$m data shows a bright protrusion of emission in this direction (Figure \ref{fig:W49Southwest1}). R$_2$ shows the opposite behavior, with a protrusion towards this location at 20\,$\mu$m, and a definitive peak seen at this location at 37\,$\mu$m. Source R$_3$ is never really seen as a peak in the near- to mid-infrared and looks to be extended emission unresolved from and protruding to the west of source R$_4$.

\begin{figure*}[htb!]
\epsscale{1.1}
\plotone{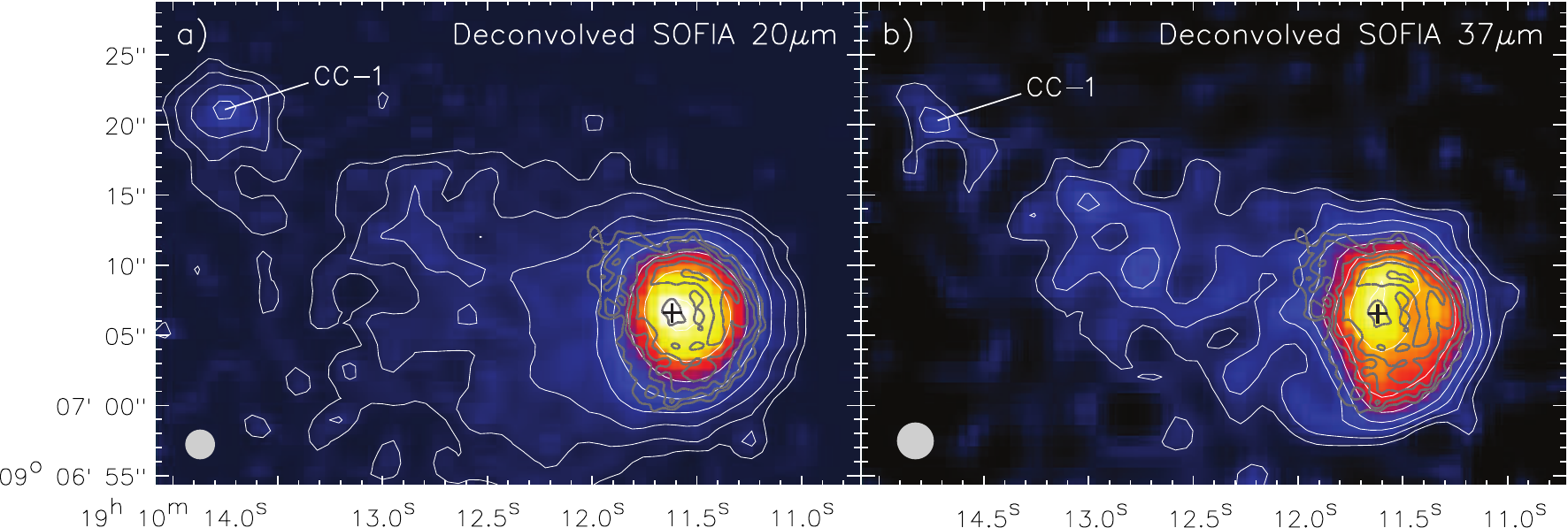}
\caption{W49A/CC with false-color images and thin white contours showing deconvolved emission at a) \textit{SOFIA} 20\,$\mu$m and b) \textit{SOFIA} 37\,$\mu$m. Overlaid in gray on each panel are the 3.6\,cm radio continuum contours from \citet{1997ApJ...482..307D}, however only the brightest 5 contours are shown to to isolate the bright core-halo radio structure. The black crosses show the radio peak position}. The resolution at each wavelength is shown by the gray circle in the lower left corner of each panel. \label{fig:W49A_CC}
\end{figure*}

Source S is the brightest infrared source in this sub-region of W49A at wavelengths less than 20\,$\mu$m and looks similar at all infrared wavelengths, and similar to the cm radio continuum emission see by \citet{1997ApJ...482..307D}. At all infrared wavelengths its core is elongated E-W, with a peak offset to the western part of the elongation. There is an arm of extended emission coming off of the eastern edge of the elongated core pointing towards the northwest, which is best seen in the 37\,$\mu$m deconvolved image (Figure \ref{fig:W49Southwest1}b). In radio cm continuum emission there is also an arm to the northwest coming off of source S (see radio contours in Figure \ref{fig:W49Southwest1}), but it does not seem to be co-located with the 37\,$\mu$m emission arm; this may be due to variable extinction or external heating of the arm.

Source Q in the radio appears to be two parallel lobe-like structures elongated southeast-to-northwest with a lane of lesser emission in between \citep[see][and Figure \ref{fig:W49Southwest1}]{1997ApJ...482..307D}. At infrared wavelengths we see extended emission throughout the same area covered by the radio continuum emission, and though the parallel lobe structures in the radio continuum are not as pronounced in the infrared, we do see relatively less infrared emission in the same location as the lane of decreased emission seen in the radio. This is contrary to what was reported by \citet{2000ApJ...540..316S} who state that the peak emission at mid-infrared wavelengths is coincident with a relative minimum at the center of the radio emission of Q. This was because their mid-infrared images were aligned to the radio continuum images by aligning the peak of infrared emission in the R region with the radio peak of source R, and not R$_4$; therefore their astrometry is offset from ours by about 3$\arcsec$. There is infrared emission within each of the parallel and elongated radio lobes, and while they do not peak in the same location as the radio lobes, they do seem to have the same position at all \textit{Spitzer}-IRAC and \textit{SOFIA} wavelengths (Figure \ref{fig:W49Southwest2}).  \citet{2000ApJ...540..316S} state that the morphology of the infrared emission from source Q looks like an edge-on torus at certain wavelengths, but it is unclear from our data if that is the case or not. Situated approximately 10$\arcsec$ to the west of the center of Q is a source resolved from the rest of the extended emission of the Q region at both \textit{SOFIA} wavelengths, but it is best seen at 37\,$\mu$m, which we call Q-1 (Figure \ref{fig:W49Southwest1}b). It is also faintly visible in the IRAC images and was listed as a potential YSO candidate by \citet{2015ApJ...813...25S}. Our SED fits to the photometry of Q-1 indicate that it is likely to be a MYSO ($M=8-12\,M_{\sun}$), though there is no radio continuum emission coming from the source. It may be a MYSO at a stage prior to the onset of a UC\ion{H}{2} region. 

\begin{figure*}[htb!]
\epsscale{1.1}
\plotone{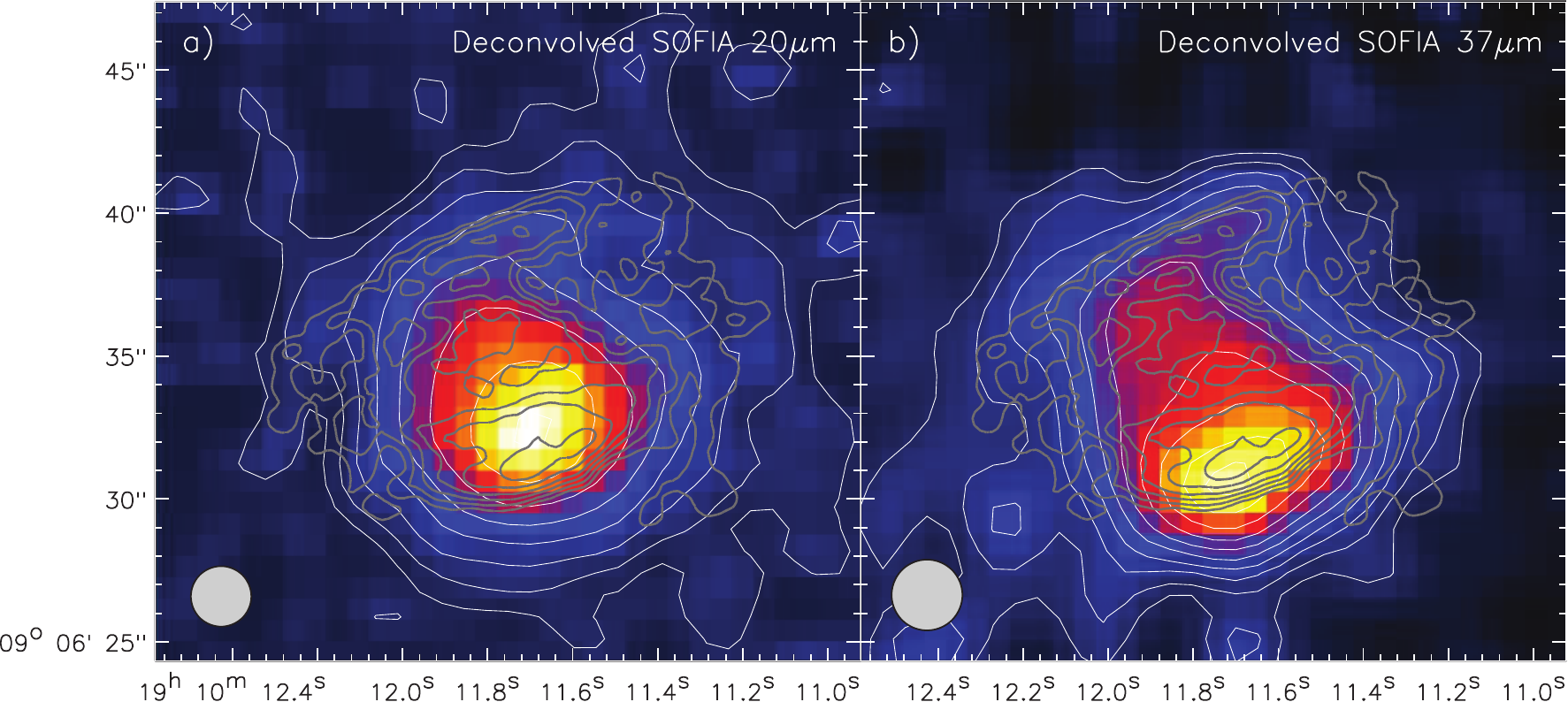}
\caption{W49A/DD with false-color images and thin white contours showing deconvolved emission at a) \textit{SOFIA} 20\,$\mu$m and b) \textit{SOFIA} 37\,$\mu$m. Overlaid on each panel are the 3.6\,cm radio continuum contours from \citet{1997ApJ...482..307D} as gray contours. The resolution at each wavelength is shown by the gray circle in the lower left corner of each panel. \label{fig:W49A_DD}}
\end{figure*}

The whole region surrounding W49\,Southwest is pervaded by multiple near-infrared sources \citep[][and see Figure \ref{fig:W49Southwest2}]{2005A&A...430..481H, 2015ApJ...813...25S}, which \citet{2005A&A...430..481H} suggest are from a previously formed cluster of stars associated with the UC\ion{H}{2} regions of S, R and Q Since 
the near-infrared sources have no significant mid-infrared emission (indicating no appreciable circumstellar dust) as seen in the SOFIA data, they are likely not massive YSOs and might not even be YSOs (i.e. 
they could be main sequence stars). This would added credence to the suggestion by \citet{2005A&A...430..481H} that the stars forming in the UC\ion{H}{2} regions now are likely not the first generation to form in the area.

\subsubsection{W49A/CC/DD/DD South}\label{sec:ccdd}
At cm radio continuum wavelengths, the brightest area of emission of CC has a core-halo morphology, with a more prominent arc on the western side, and a core with a peak offset to the east \citep{1997ApJ...482..307D}. The core of CC is claimed to be a MYSO candidate by \citet{2015ApJ...813...25S}. This bright core-halo structure is at the apex of a fan-shaped region of faint and diffuse emission that extends $\sim$40$\arcsec$ to the east and widens with distance from the CC peak. The bright core-halo structure is mimicked to some degree at all of the infrared wavelengths,though at 37\,$\mu$m the peak is very broad and the halo is not well resolved, even in the deconvolved data. Extended low-level infrared emission is seen throughout the fan-shaped extended emission of the CC region, but the knots change position with infrared wavelength and don't match up with the radio knots very well, indicating that these structures are likely to be externally heated and ionized knots of dust and gas. However, one peak within this extended region of emission, located $\sim$35$\arcsec$ east of the peak of CC and which we label as CC-1 in Figure \ref{fig:W49A_CC}, appears to be present in all \textit{Spitzer}-IRAC bands as well as the \textit{SOFIA} wavelengths. Our SED modeling shows that it is likely to be a MYSO.

For source DD, the 37\,$\mu$m emission of looks more similar to its radio continuum emission morphology than its 20\,$\mu$m image. DD has a bipolar appearance in the cm radio continuum images of \citet{1997ApJ...482..307D}, though both lobes appear flattened and the southern lobe is considerably brighter than the northern lobe (Figure \ref{fig:W49A_DD}). In the \textit{SOFIA} images, the northern lobe is more prominent at 37\,$\mu$m than at 20\,$\mu$m, however, like the radio continuum emission, the southern lobe is brighter than the northern lobe at both infrared wavelengths. In the \textit{Spitzer}-IRAC data, the southern lobe is also the most obvious. Given its large size ($d\sim0.7$\,pc), source DD is too large to be considered even a compact \ion{H}{2} region \citep[$\sim$0.1 $<d<$ 0.5\,pc;][]{1967ApJ...150..807M}. However, given the complex sub-structure, the large calculated luminosity of the source from our data ($\sim$2.0$\times10^5\,L_{\sun}$), and the bright cm radio continuum emission, this \ion{H}{2} region must house at least one MYSO.

Mid-infrared observations of the DD area by \citet{2000ApJ...540..316S} reveal another bright and extended source of emission about 10$\arcsec$ south of DD which they call DD South. They claim clear detections of this source at all four mid-infrared wavelengths they observed (12.3, 12.8, 13.2, and 20.6\,$\mu$m), with the integrated flux from the source at 20\,$\mu$m ($\sim$80\,Jy) being twice as bright as their measured 20\,$\mu$m flux from source G. They observe DD South as an irregular or core-halo morphology (similar to source S) with a diameter of about 12$\arcsec$. We do not detect this source at either \textit{SOFIA} wavelength (see red dots in Figure \ref{fig:SOFIAdata}) and it is not present in the \textit{Spitzer}-IRAC images. Given the large brightness claimed by \citet{2000ApJ...540..316S} for DD South at 20\,$\mu$m, our lack of a detection in the \textit{SOFIA} 20\,$\mu$m image is not due to our observations having a shallower detection limit. Since it has never been seen before in the infrared, DD South could be a flaring source, as has been seen in MYSOs like NGC~6334I \citep{2018ApJ...854..170H}. However, as we discuss in a later section (\ref{sec:other}),  it seems more plausible that this is an artifact in the data or produced in the data reduction and not a real source, since we also fail to detect the three other new infrared sources identified by \citet{2000ApJ...540..316S}. 

\subsubsection{Central W49A: Welch Ring Sources}

\begin{figure*}[htb!]
\epsscale{1.15}
\plotone{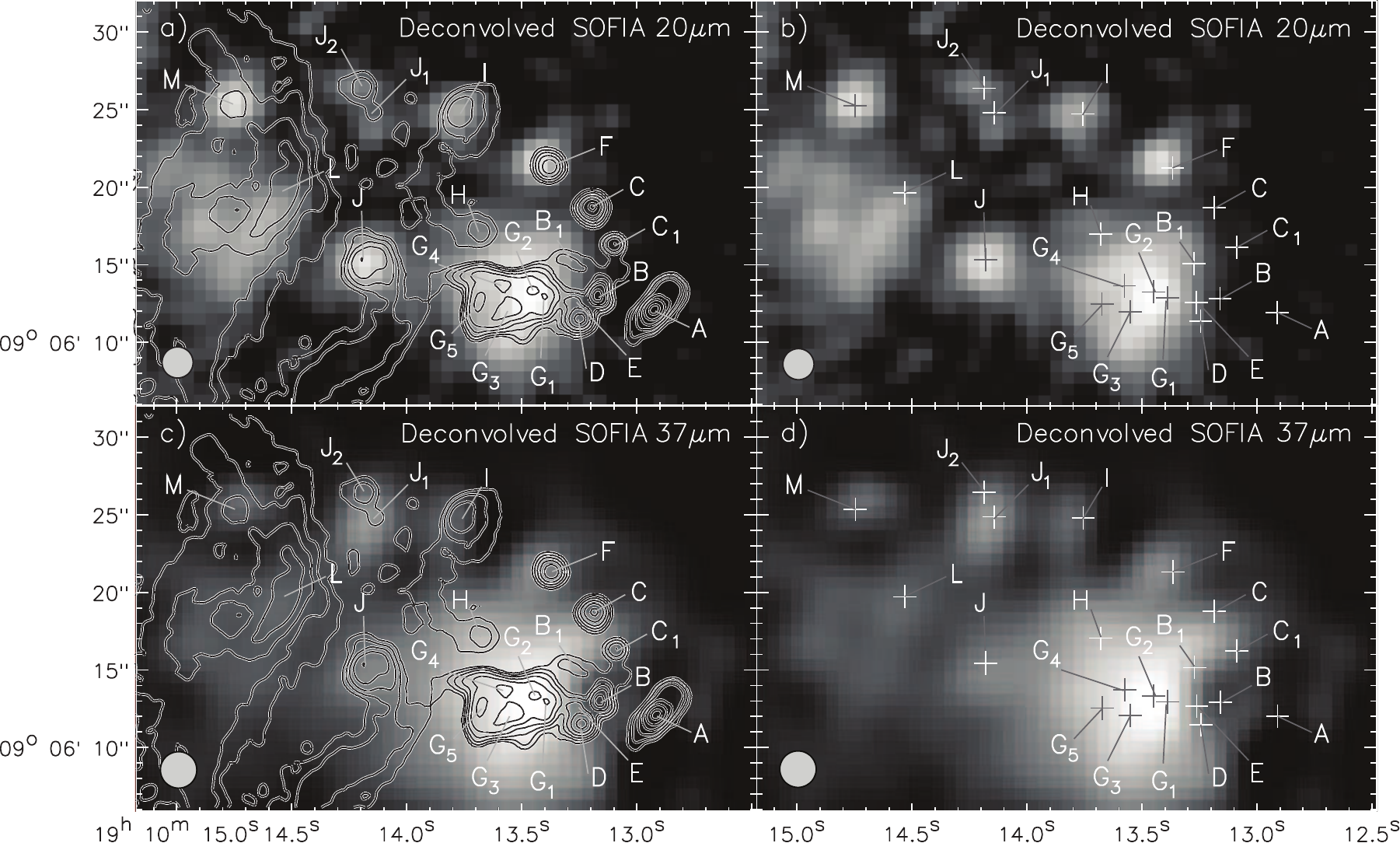}
\caption{Gray scale images of the ``Welch Ring'' area of W49A with deconvolved emission as seen by \textit{SOFIA} at 20\,$\mu$m (top row) and 37\,$\mu$m (bottom row). The left column of panels have overlays of the 3.6\,cm radio continuum emission from \citet{1997ApJ...482..307D} as gray contours with their peaks labeled. In the right column of panels the radio peak locations are indicated by the crosses. The resolution at each wavelength is shown by the gray circle in the lower left corner of each panel.  \label{fig:W49A_Ring}}
\end{figure*}

The central region of W49A contains more than two dozen identified cm radio continuum sources within one square arcminute. The most prominent sources are grouped into a feature known as the Welch Ring \citep{1987Sci...238.1550W}, and most of these sources are thought to be individual UC\ion{H}{2} regions (however as we will discuss later in this section and more in Section~\ref{sec:cps}, several of them may not be). 

\citet{2000ApJ...540..316S} pointed out that while many of these radio sources are detectable in the infrared, not all of them are. They posit that it is very likely that the extinction toward the western side of the ring of sources has such a high level of obscuration that the material is too optically thick even for their mid-infrared photons to be seen. They point to observations towards this region using molecular lines that show a higher concentration of molecular material on the western side of the ring \citep[e.g.,][]{1994ApJ...429L..37J,1993ApJ...413..571S}. Like \citet{2000ApJ...540..316S} and \citet{2009MNRAS.399..952S} we do not detect infrared emission coming from radio sources A or B at 20 or 37\,$\mu$m (Figure \ref{fig:W49A_Ring}), nor are they seen in any of the \textit{Spitzer}-IRAC bands. There is a weak infrared peak located 2$\arcsec$ to the west of A in all of the \textit{Spitzer}-IRAC bands; there is also emission here at 37\,$\mu$m. In fact, at 37\,$\mu$m we detect extended emission around sources A and B, with deficits in emission exactly at their locations. This likely means that, in addition to the high levels of environmental extinction (i.e from the large-scale molecular cloud structure), the self-extinction from circumstellar dust (i.e. disk and/or envelope material) for these sources may be a significant component in the overall level of extinction. This is most evident for source B, where there is a ring of mid-infrared emission at 37\,$\mu$m all around the source, but a dip in the emission right where the radio source peak is located (see Figure \ref{fig:W49A_Ring}d).

\citet{2000ApJ...540..316S} and \citet{2009MNRAS.399..952S} do not detect emission from radio sources B$_1$, D, or E. We also do not detect any point-like emission from the locations of these sources in our deconvolved 20$\mu$m image. In the deconvolved 37\,$\mu$m image there is extended emission towards these locations, and while it is possible that this is unresolved emission from the very bright G source, contributions to the emission in these areas due to sources B$_1$, D, and/or E cannot be ruled out. Interestingly, sources C and C$_1$, which also were not detected by \citet{2000ApJ...540..316S} and \citet{2009MNRAS.399..952S} at shorter mid-infrared wavelengths, appear to be associated with a protrusion of emission seen only at 37\,$\mu$m, though the brightest emission is located between the two sources, with the radio source peaks lying on the periphery of the infrared emission (see Figure \ref{fig:W49A_Ring}d). It could be that this infrared emission is not tracing emission coming from the exact locations of C or C$_1$ due to elevated levels of extinction direct along the line of sight to their radio peaks, but instead leaking out of an area of lesser extinction between them.

\begin{figure*}[htb!]
\epsscale{1.35}
\plotone{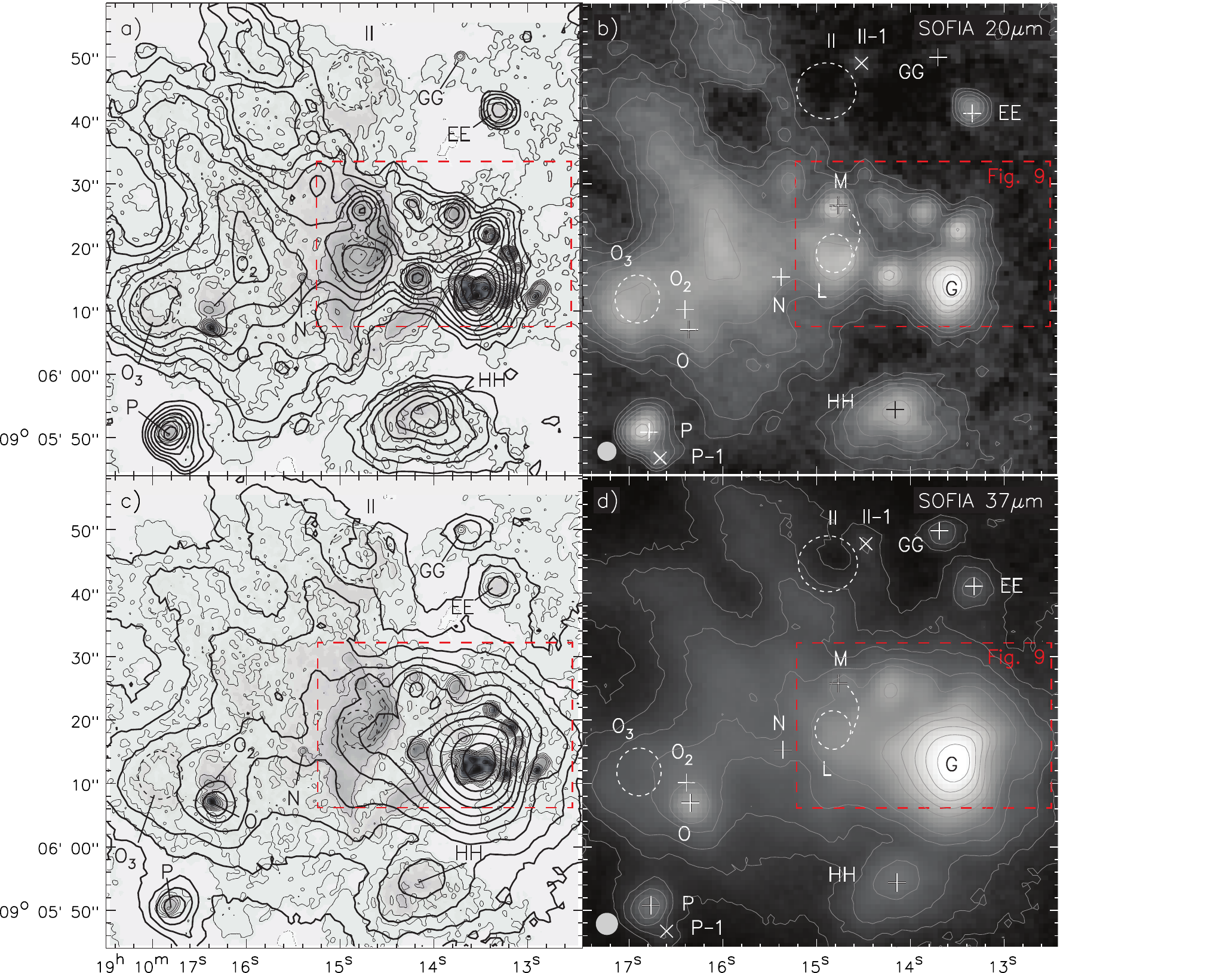}
\caption{Images showing the central region of W49A. a) The inverse gray scale image with light gray contours shows the 3.6\, cm radio continuum emission from \citet{1997ApJ...482..307D} overlaid with the 20\,$\mu$m emission from \textit{SOFIA} as thick black contours. b) A gray scale image with light gray contours shows the 20\,$\mu$m emission seen with \textit{SOFIA}. c) Same as a) except the overlaid thick black contours are the the 37\,$\mu$m emission from \textit{SOFIA}. d) A gray scale image with light gray contours shows the 37\,$\mu$m emission seen with \textit{SOFIA}. The location and sizes or the radio sources with ring-like appearances are shown by the dashed circles, and the location of the newly unidentified infrared sources are given by the x symbols. The sources within the red dashed box in all panels are not labeled, but this area is discussed in detail in Figure \ref{fig:W49A_Ring}. \label{fig:W49A_center}}
\end{figure*}

Source F is the brightest source in the area in the \textit{Spitzer}-IRAC images at 3.6 and 4.5\,$\mu$m, but G becomes more prominent at longer wavelengths, though F is still detected in our \textit{SOFIA} data at both 20 and 37\,$\mu$m. Our SED modeling of this source results in a rather low luminosity, and not a MYSO. As we discuss in detail in Section \ref{sec:cps}, the radio emission from source F is likely not due to free-free emission from a UC\ion{H}{2} region, but instead from non-thermal synchrotron emission (likely from an outflow).

Radio source G was resolved into five individual peaks by \citet{1997ApJ...482..307D} named G$_1$ to G$_5$ (Figure \ref{fig:W49A_Ring}). The higher spatial resolution follow-up observations at 3.6\,cm by \citet{2020AJ....160..234D} show that sources G$_1$ and G$_2$ break up even further into compact radio continuum objects (with source `G2a' being the brightest), and that G$_3$, G$_4$, and G$_5$ are not likely individual YSOs, but arcs of radio continuum emission tracing the edges of a more coherent cavity structure with a diameter of about 3$\arcsec$. Like \citet{2009MNRAS.399..952S} saw at 18.5\,$\mu$m and \citet{2000ApJ...540..316S} saw at 20.6\,$\mu$m, we do not resolve any of these sub-components at 20 and 37\,$\mu$m with SOFIA, and source G looks like a cometary \ion{H}{2} region whose peak is at the location of the G$_2$ radio peak. G$_2$ is also a compact infrared source in all \textit{Spitzer}-IRAC bands. We cannot resolve out mid-infrared sources associated with any other radio peak, even in the deconvolved data, and the region of brightest mid-infrared emission covers an area that encompasses all five radio peaks. However, the overall morphology of the extended infrared emission looks different than what is seen in the radio. The east-west extent is comparable in the two wavelength regimes, however, the cm radio continuum emission drops off precipitously both to the north and south, whereas in the mid-infrared (especially evident at 37\,$\mu$m) there is extended emission to the north and south (see Figures \ref{fig:W49A_Ring} and \ref{fig:W49A_center}). 

Source H can be seen as a broad, weak peak of emission in the \textit{Spitzer}-IRAC bands at 3.6 and 4.5\,$\mu$m. At the other two IRAC wavelengths, saturation effects from source F and G overwhelm the region. In the 20 and 37\,$\mu$m \textit{SOFIA} images, H appears as a protrusion of unresolved extended emission from G (Figure \ref{fig:W49A_Ring}). Radio sources I and J are seen at all infrared wavelengths from \textit{Spitzer} 3.6\,$\mu$m out to 37\,$\mu$m. At 37\,$\mu$m source J is partially resolved from the extended emission of source G (Figures \ref{fig:W49A_Ring}).

Radio sources J$_1$ and J$_2$ are only $\sim$2$\arcsec$ apart, with J$_2$ being the brighter of the two sources. In the \textit{Spitzer}-IRAC images, there is an infrared source that is elongated north-south with the northern half co-local with the peak of J$_2$. However, J$_1$ lies about 1$\farcs$5 southwest of J$_2$, so this southern half of the north-south elongated infrared emission seen with IRAC does not seem to be co-local with J$_1$. In the \textit{SOFIA} deconvolved 20\,$\mu$m image there are two separated sources, with the northern source at the location of J$_2$, and the other not quite at the location o f J$_1$ (Figure \ref{fig:W49A_Ring}). At 37\,$\mu$m there does not appear to be emission coming from J$_2$, but the peak is on the southern half of the elongated near-infrared source seen with \textit{Spitzer}-IRAC. This points to the possibility that J$_1$ is the more embedded or least evolved of the two sources.

In the radio, source L appears as a bright arc of emission which terminates to the south as the western side of a ring-like structure of emission, giving the overall appearance of a letter ``d'' (Figure \ref{fig:W49A_Ring}). The north half of the radio arc is not traced by infrared emission, however the ring-like structure is relatively bright at all wavelengths from 3.6 to 37\,$\mu$m. However, since the locations of the clumps that make up the ring structure do change with infrared wavelength, these structures are likely to be externally heated knots, rather than individual YSOs.    

Radio source M is seen at all \textit{Spitzer}-IRAC and \textit{SOFIA} wavelengths, and appears as a prominent compact point-like infrared source (Figure \ref{fig:W49A_Ring}). 

\subsubsection{Central W49A: Other sources}

Compact radio sources EE and GG lie to the northern of the Welch Ring (Figure \ref{fig:W49A_center}), and while EE is detected as a compact infrared source at all wavelengths from 3.6 to 37\,$\mu$m, the GG radio source is prominent only as a compact point source at 37\,$\mu$m, and is not detected at 20\,$\mu$m or shorter wavelengths. In the \textit{Herschel} 70\,$\mu$m image, source GG can be seen as a nearly resolved source north of the extended emission from the Welch Ring. This is consistent with GG being a very young, self-embedded MYSO. Our SED modeling does seem to confirm its massive nature (best fits yielding $M=16-24\,M_{\sun}$). 

Radio source II also lies just to the north of the Welch Ring (Figure \ref{fig:W49A_center}), and is seen as a weak partial ring of emission at cm radio wavelengths. This source only seems to have some weak emission in our 37\,$\mu$m image, even weaker emission at 20\,$\mu$m, and is not seen at shorter wavelengths. There does seem to be a point-like infrared source in the northwestern part of the ring, about 12$\arcsec$ east of GG, which we label II-1. This source is seen at all IRAC wavelengths, and our SED model fits show it to be a MYSO candidate. Given that this source has no detected cm radio continuum peak at this location would imply that II-1 is a very young MYSO in a phase prior to the onset of a UC\ion{H}{2} region. Our SED model fits imply it has a stellar mass of $M=8\,M_{\sun}$.

The compact radio sources N and O also appear to be highly embedded sources lying to the east of source G (Figure \ref{fig:W49A_center}). In the \textit{Spitzer}-IRAC data and the \textit{SOFIA} data, N is only seen at 37\,$\mu$m. In the \textit{Herschel} image at 70\,$\mu$m there is a protrusion of emission from the extended G source towards this location, and is thus likely associated with it. For source O, the shortest wavelength it is detected is at 5.8\,$\mu$m, though it is a relatively weak source at wavelengths 20\,$\mu$m and shorter. It becomes one of the brightest sources in the central W49A area at 37\,$\mu$m and is easily visible in the \textit{Herschel} 70\,$\mu$m image. Again, a lack of short wavelength emission and very bright far-infrared emission would indicate that these are both very young, self-embedded MYSOs (which our SED model fitting seems to confirm).

Also to the east of G are radio sources O$_2$ and O$_3$ (Figure \ref{fig:W49A_center}). The radio peak of O$_2$ is just north of O ($\sim$3$\arcsec$ away), surrounded by a fan-shaped region of extended radio continuum emission. There is no peak seen at the radio peak of O$_2$ at any infrared wavelength, and while there is some emission at each wavelength in the region, it looks different at each wavelength, signifying that we are likely seeing infrared emission from the diffuse dust in the overall cloud, and not emission coming directly from a YSO. Radio source O$_3$ is a ring of emission about 4$\arcsec$ in radius, and we see a similar ring structure at both 20 and 37\,$\mu$m.

P is a bright source elongated at a position angle of 135$\arcdeg$ in the \textit{SOFIA} images, with a second, much fainter infrared source located $\sim$5$\arcsec$ away to the southwest which we label P-1 (Figure \ref{fig:W49A_center}). Source P is associated with a compact radio continuum source, while P-1 is associated with a weak knot of radio continuum emission in the 3.6\,cm data. Both sources can be seen from 3.6\,$\mu$m out to 37\,$\mu$m, and there is even emission seen here at 70\,$\mu$m, though it is hard to tell, given the resolution of the \textit{Herschel} data, if emission is coming from both sources or not. We find from our SED model fitting that P is a MYSO and P-1 is an intermediate to high-mass YSO.

To the south of the G region lies radio source HH (Figure \ref{fig:W49A_center}). It is a large ($\sim$15$\arcsec\times$10$\arcsec$) triangular-shaped feature whose broad peak is the same from 3.6 to 37\,$\mu$m. HH might be a region containing multiple unresolved MYSOs given it's large size and the relatively high luminosity we derive for it from our SED model fitting (i.e., $\sim$2.6$\times10^5\, L_{\sun}$). 

\subsubsection{Other Sources in W49A}\label{sec:other}

Radio sources AA and BB are large ($d\sim20\arcsec$), weak, and extremely diffuse radio continuum regions \citep{1997ApJ...482..307D} with faint ring or bubble-like appearances at 3.6\,cm. The infrared emission extends over a comparable region to the radio at all infrared wavelengths from 3 to 37\,$\mu$m. In the \textit{SOFIA} images AA and BB appear as only faint patches of diffuse emission with highly broadened peaks at both 20 and 37\,$\mu$m within which we can see no clear substructure (Figure \ref{fig:SOFIAdata}). We derived luminosities for AA and BB in the same manner as we did for other extended sources (i.e., in Section~\ref{sec:alm}), and find that AA has a luminosity of only $\sim$6$\times10^4\,L_{\sun}$ (the equivalent of a B0 star) and BB has a luminosity of only $\sim$1$\times10^5\,L_{\sun}$ (the equivalent of a O9 star). Given their large, bubble-like radio morphologies, their relatively low infrared-derived luminosities, and their lack of compact and/or bright infrared or radio components or peaks, AA and BB may be more evolved \ion{H}{2} regions. Indeed, AA and BB have by far the lowest emission measures in W49A \citep[$EM\sim7-9\times10^5$\,cm$^{-6}$\,pc;][]{1997ApJ...482..307D}, and combined with their large sizes ($d > 1.0$\,pc), they do not have properties consistent with compact \ion{H}{2} regions \citep[$\sim0.1<d< 0.5$\,pc; $EM>10^8$\,cm$^{-6}$\,pc;][]{1967ApJ...150..807M}, and are more in line with the properties of evolved \ion{H}{2} regions \citep[e.g., Sh2-90;][]{2014A&A...566A.122S}. Consistent with this hypothesis, inspection of our multi-wavelength data of AA and BB reveal that both have an unresolved near-infrared source centrally located within their extended radio continuum regions. We cannot be completely certain if these sources are foreground stars, independent YSOs, or actually the central ionizing sources of the AA and BB \ion{H}{2} regions, however from the limited data available, they do appear to have characteristics consistent with being the central ionizing stars. For AA, the stellar source peak is located at $\alpha_{J2000}=$19:10:07.7, $\delta_{J2000}=$+9:05:45, and for BB the source peaks at $\alpha_{J2000}=$19:10:10.6, $\delta_{J2000}=$+9:05:45 and both are only detected in the {\it Spitzer} 3.6 and 4.5\,$\mu$m data. Neither source is detected at shorter wavelengths (i.e., in optical POSS2 images and {\it 2MASS} J/H/K images), which would be expected of stars at the distance of W49A due to the high line-of-sight extinction, and not of nearby foreground stars. Neither source is detected at {\it Spitzer} 5.8 or 8.0\,$\mu$m, nor in our {\it SOFIA} data, signifying sources with no circumstellar emission (i.e., consistent with revealed stars and not YSOs). 

Another newly-identified infrared source we name DD-1 is located $\sim$13$\arcsec$ southwest of the peak of DD (Figure \ref{fig:SOFIAdata}), with a faint peak surrounded by some modest extended infrared emission. This source is seen in all IRAC bands as well as with \textit{SOFIA}. The \textit{SOFIA} 20\,$\mu$m image shows a point source, whereas the 37\,$\mu$m image shows more of an extended source, but in both cases the source is faint. Again, the \textit{Herschel} 70\,$\mu$m image shows a tongue of emission towards this area as does the image at 3.6\,cm given in \citet{1997ApJ...482..307D}, although with very faint surface brightness. Our SED modeling reveal this source to be a MYSO. 

Radio source FF is a very faint partial ring of emission ($r$ $\sim$7$\arcsec$) whose western side is the most prominent arc, with a fainter arc on the eastern side (Figure \ref{fig:W49Southwest2}). Similarly, this is the case in the infrared, and the ring is better seen at 20\,$\mu$m than 37\,$\mu$m. It could be that is because this side of the ring is close to the very bright source S, which may be responsible, at least in part, for heating it (Figure \ref{fig:W49Southwest2}). Indeed, the western side of FF can be seen even down to 2\,$\mu$m \citep{2003ApJ...589L..45A}.

The large and bright radio source JJ at low resolution approximates a $20\arcsec\times30\arcsec$ rectangular shape (Figure \ref{fig:SOFIAdata}). The shape and size is grossly mimicked in the infrared, though the northeastern corner of the radio continuum region is depressed in infrared emission and the brightness distribution within the entire radio-emitting area fluctuates markedly with infrared wavelength. There are no well-defined peaks present that are co-local at near-infrared and mid-infrared wavelengths, though the radio peaks towards the northern corner of the triangle perhaps indicating the location of the MYSO(s) that are heating the entire clump.    

\begin{deluxetable*}{rrrrrrrrrrrl}
\tabletypesize{\scriptsize}
\tablecolumns{8}
\tablewidth{0pt}
\tablecaption{Observational Parameters of All Known Radio Continuum Sources in W49A}\label{tb:all}
\tablehead{\colhead{  }&
           \colhead{  }&
           \colhead{  }&
           \multicolumn{4}{c}{${\rm 20\mu{m}}$}&
           \multicolumn{4}{c}{${\rm 37\mu{m}}$}&
           \colhead{  }\\
           \cline{4-7} \cline{8-11} \\
           \colhead{ Source }&
           \colhead{ R.A. } &
           \colhead{ Dec. } &
		   \colhead{ FWHM } &
           \colhead{ $R_{\rm int}$ } &
           \colhead{ $F_{\rm int}$ } &
           \colhead{ $F_{\rm int-bg}$ } &
		   \colhead{ FWHM } &
           \colhead{ $R_{\rm int}$ } &
           \colhead{ $F_{\rm int}$ } &
           \colhead{ $F_{\rm int-bg}$ } \\
	   \colhead{  } &
	   \colhead{ J2000 } &
	   \colhead{ J2000 } &
	   \colhead{ ($\arcsec$) } &
	   \colhead{ ($\arcsec$) } &
	   \colhead{ (Jy) } &
	   \colhead{ (Jy) } &
	   \colhead{ ($\arcsec$) } &
	   \colhead{ ($\arcsec$) } &
	   \colhead{ (Jy) } &
	   \colhead{ (Jy) } \\
}
\startdata
A	&	19 10 12.88	&	9 06 12.0	&	   ND	&	3.1	&	$<$0.06	&\nodata		&	ND	&	3.1	&	$<$71.3	&	$<$11.0	&	\\
AA	&	19 10 07.77	&	9 05 44.2	&	diffuse	&	10.7	&	16.2	&	8.01	&	diffuse	&	10.7	&	58.8	&	42.0		\\
B	&	19 10 13.14	&	9 06 13.0	&	   U	&	1.5	&	$<$1.08	&	$<$0.73	&	U	&	1.5	&	$<$76.9	&	$<$74.9	\\
B$_1$	&	19 10 13.26	&	9 06 14.9	&	   U	&	1.5	&	$<$1.97	&	$<$0.90	&	U	&	1.5	&	$<$357	&	$<$355	&\\
BB	&	19 10 10.80	&	9 05 43.9	&	diffuse	&	7.7	&	9.32	&	3.09	&	diffuse	&	6.9	&	44.0	&	9.22	&\\
C	&	19 10 13.17	&	9 06 18.6	&	U	&	1.5	&	$<$0.93	&	$<$0.59	&	U	&	1.5	&	$<$172	&	$<$170		\\
C$_1$	&	19 10 13.08	&	9 06 16.2	&	U	&	1.5	&	$<$0.79	&	$<$0.45	&	U	&	1.5	&	$<$152	&	$<$150		\\
CC	&	19 10 11.53	&	9 07 05.8	&	diffuse	&	11.5	&	88.5	&	68.5	&	diffuse	&	11.5	&	284	&	241		\\
D	&	19 10 13.24	&	9 06 11.2	&	U	&	1.5	&	$<$2.08	&	$<$1.74	&	U	&	1.5	&	$<$347	&	$<$345		\\
DD	&	19 10 11.68	&	9 06 32.4	&	diffuse	&	10.7	&	47.0	&	27.3	&	diffuse	&	10.7	&	240	&	217		\\
E	&	19 10 13.26	&	9 06 12.4	&	U	&	1.5	&	$<$2.26	&	$<$1.92	&	U	&	1.5	&	$<$372	&	$<$370		\\
EE	&	19 10 13.25	&	9 06 41.0	&	4.48	&	5.4	&	6.46	&	2.45	&	5.97	&	4.6	&	40.9	&	27.4		\\
F	&	19 10 13.39	&	9 06 21.6	&	4.09	&	3.1	&	5.81	&	2.45	&	PR	&	3.1	&	$<$144	&	$<$139		\\
FF	&	19 10 13.19	&	9 05 25.6	&	7$\arcsec$ ring	&	13.8	&	45.4	&	14.2	&	7$\arcsec$ ring	&	13.8	&	$<$176	&	$<$143		\\
G\textsubscript{tot}\tablenotemark{\tiny{a}}	&	19 10 13.50	&	9 06 11.6	&	8.51	&	7.7	&	45.9	&	33.5	&	9.21	&	9.2	&	3700	&	3410		\\
GG	&	19 10 13.62	&	9 06 49.6	&	ND	&	3.8	&	$<$1.98	&	$<$0.18	&	4.45	&	3.8	&	22.8	&	11.1		\\
H	&	19 10 13.69	&	9 06 17.1	&	U	&	1.5	&	$<$1.75	&	$<$1.43	&	U	&	1.5	&	$<$321	&	$<$319		\\
HH	&	19 10 14.07	&	9 05 51.7	&	diffuse	&	10.7	&	39.9	&	16.9	&	diffuse	&	9.2	&	251	&	91.2		\\
I	&	19 10 13.76	&	9 06 24.4	&	4.46	&	3.1	&	4.18	&	1.49	&	4.6	&	2.3	&	53.2	&	38.1		\\
II	&	19 10 14.96	&	9 06 43.1	&	4$\arcsec$ ring	&	12.3	&	26.7	&	9.29	&	4$\arcsec$ ring 	&	12.3	&	268	&	136		\\
J	&	19 10 14.18	&	9 06 14.6	&	4.56	&	3.8	&	9.29	&	4.20	&	PR	&	3.8	&	$<$260	&	$<$104		\\
J$_1$+J$_2$\tablenotemark{\tiny{b}}	&	19 10 14.21	&	9 06 25.1	&	6.06	&	3.8	&	5.13	&	0.89	&6.40		&	4.6	&	238	&	57.9		\\
JJ	&	19 10 18.73	&	9 06 06.3	&	diffuse	&	21.5	&	203	&	120	&	diffuse	&	19.2	&	764	&	404		\\
KK	&	19 10 21.69	&	9 05 51.5	&	6.29	&	10.0	&	20.9	&	12.4	&	6.98	&	12.3	&	124	&	57.7		\\
L	&	19 10 14.76	&	9 06 17.0	&	4$\arcsec$ ring	&	6.1	&	24.0	&	9.43	&	diffuse	&	6.1	&	328	&	92.7		\\
LL	&	19 10 22.89	&	9 04 13.9	&	diffuse	&	11.5	&	33.9	&	16.4	&	diffuse	&	11.5	&	90.4	&	80.6		\\
M	&	19 10 14.73	&	9 06 25.1	&	4.06	&	3.8	&	8.33	&	3.24	&	5.4	&	3.1	&	65.1	&	10.8		\\
MM	&	19 10 23.56	&	9 06 02.4	&	diffuse	&	19.2	&	90.4	&	52.5	&	diffuse	&	21.5	&	416	&	273		\\
N	&	19 10 15.38	&	9 06 14.9	&	PR	&	3.1	&	$<$5.01	&	$<$2.99	&	4.4	&	3.1	&	44.2	&	4.64		\\
O	&	19 10 16.33	&	9 06 06.8	&	5.4	&	3.8	&	8.11	&	2.79	&	6.67	&	3.8	&	134	&	87.5		\\
O$_2$	&	19 10 16.30	&	9 06 11.6	&	PR	&	3.8	&	$<$9.57	&	$<$3.60	&	PR	&	3.8	&	$<$90.7	&	$<$66.8		\\
O$_3$	&	19 10 17.03	&	9 06 10.7	&	4$\arcsec$ ring	&	7.7	&	32.4	&	12.3	&	4$\arcsec$ ring 	&	7.7	&	$<$202	&	$<$123		\\
P	&	19 10 16.90	&	9 05 52.1	&	4.59	&	5.4	&	11.7	&	6.02	&	6.33	&	5.4	&	70.3	&	37.0		\\
Q	&	19 10 10.65	&	9 05 05.6	&	diffuse	&	7.7	&	41.4	&	35.8	&	diffuse	&	9.2	&	246	&	233		\\
R\textsubscript{tot}\tablenotemark{\tiny{c}}	&	19 10 10.80	&	9 05 18.5	&	6.92	&	6.9	&	58.9	&	47.9	&	11.5	&	6.9	&	485	&	408		\\
R	&	19 10 11.06	&	9 05 20.2	&	PR	&	1.5	&	$<$2.84	&	$<$2.34	&	PR	&	1.5	&	$<$38.5	&	$<$37.0		\\
R$_2$	&	19 10 10.80	&	9 05 23.2	&	PR	&	1.5	&	$<$1.23	&	$<$0.96	&	PR	&	1.5	&	$<$14.8	&	$<$13.3		\\
R$_3$	&	19 10 10.74	&	9 05 17.4	&	PR	&	1.5	&	$<$7.41	&	$<$7.22	&	PR	&	1.5	&	$<$39.9	&	$<$38.3		\\
S	&	19 10 11.76	&	9 05 26.6	&	5.22	&	6.9	&	77.4	&	65.1	&	7.68	&	8.4	&	402	&	302		\\
W49\,South	&	19 10 22.32	&	9 05 01.0	&	5.39	&	23.0	&	405	&	349	&	6.83	&	23.0	&	2290	&	2150		\\
W49\,South-1	&	19 10 21.60	&	9 05 01.1	&	U	&	1.5	&	$<$7.99	&	$<$7.88	&	U	&	1.5	&	$<$69.6	&	$<$68.2		\\
\enddata
\tablecomments{R.A. and Dec. are for the center of apertures used, not the source peaks. $F_{\rm int}$ indicates total flux inside the aperture. $F_{\rm int-bg}$ is for background subtracted flux. For the columns labeled `FWHM': `ND' means no detection (so the corresponding flux is the calculated as a 3$\sigma$  upper limit from the background), `U' means unresolved from other emission and `PR' means partially resolved from other emission (so flux values are upper limits due to contamination), `diffuse' means the source is large and cannot be fit by a Gaussian profile, `low S/N' means the detection is too faint to be fit by a Gaussian profile. Some sources are rings or partial rings or arcs and the radius of a ring that fits their shape is given.}
\tablenotetext{a}{G$_{tot}$ here refers to the collective flux of unresolved sources G$_1$, G$_2$, G$_3$, G$_4$, and G$_5$.}
\tablenotetext{b}{J$_{1}$+J$_{2}$ here refers to the collective flux of unresolved sources J$_1$ and J$_2$.}
\tablenotetext{c}{R$_{tot}$ here refers to the collective flux of unresolved sources R, R$_2$, and R$_3$ from the table above, as well as R$_4$, which can be found in Table\,\ref{tb:new}.}
\end{deluxetable*}

The north-central region of W49A has a long ($>$1$\arcmin$) ridge of emission snaking along at about the same declination east to west (Figure \ref{fig:SOFIAdata}). This ridge is easy to spot at all IRAC and \textit{SOFIA} wavelengths, and even at wavelengths as long as 70\,$\mu$m (with {\it Herschel}) and as short as 2\,$\mu$m \citep{2003ApJ...589L..45A}. There is also radio continuum emission coming from this ridge at cm wavelengths, and we labeled this the  ``Northern Ridge'' in Figure \ref{fig:SOFIAdata}. This could be a radiation-driven ridge of material snow-plowed by the combined emission of the O stars south of it. This would be consistent with the picture of W49A by \citet{2010A&A...520A..84P} where two expanding shells are responsible for some of the large-scale morphology of the region. 

Just south of the Northern Ridge on the eastern side is a infrared source we newly identify here as JJ-1 (Figure \ref{fig:SOFIAdata}). It is the only semi-compact infrared source that we can identify as a potential star-forming clump in or near the Northern Ridge. It is seen at all infrared wavelengths from IRAC to 70\,$\mu$m \textit{Herschel} data. Though it is hard to see in Figure \ref{fig:SOFIAdata}, at 37\,$\mu$m the source appears as an arc-shape with a broad peak offset to the southeast. At 20\,$\mu$m the source appears more point-like and peaking towards the center of the infrared emitting region. Interestingly, in the IRAC bands the source is seen as two structures, and arc of emission (coincident with the 37\,$\mu$m emission), and a point source (coincident with the 20\,$\mu$m peak). The arc of emission could be a partial dust shell around the point source. The source does display 3.6\,cm cm radio continuum emission comparable to the shape and extent of the 37\,$\mu$m emission, indicating it harbors at least one MYSO. 

To the east of JJ-1 lies an infrared source we label as KK-1 (Figure \ref{fig:SOFIAdata}). It is an unresolved point source at all infrared wavelengths from IRAC 3.6\,$\mu$m to \textit{SOFIA} 37\,$\mu$m, and there is even a tongue of emission in the extended \textit{Herschel} 70\,$\mu$m image towards this source. Though it displays no radio continuum emission, our SED modeling shows it to have the luminosity of a MYSO. Therefore, it may be a MYSO in a stage prior to the onset of a UC\ion{H}{2} region. 

Radio region KK has a radio peak coincident with peaks seen at all \textit{Spitzer}-IRAC and \textit{SOFIA} infrared wavelengths, as well as at \textit{Herschel} 70\,$\mu$m. KK is semi-compact and our SED models show it to be a MYSO. 

LL appears in the radio as broad source with an ill-defined peaked shaped like a kidney bean. This shape is mimicked in the infrared data. LL is situated at the end of a tail of infrared emission spurring off of W49\,South towards the south (Figure \ref{fig:SOFIAdata}). This tail is seen at all infrared wavelengths (though it is very faint in our 20\,$\mu$m image), but only at 70\,$\mu$m is the emission in the arm is brighter than the emission of LL, peaking half-way between W49\,South and LL. 

The radio region MM is large ($d\sim25\arcsec$) radio continuum region at low resolution \citep{1997ApJ...482..307D} with a diffuse extension towards the southeast (Figure \ref{fig:SOFIAdata}). This extension is also seen in the \textit{SOFIA} 37\,$\mu$m image of this source, whereas the morphology changes considerably at shorter wavelengths. The main peak in the \textit{SOFIA} data looks like a horizontal bar or unresolved binary peak. The \textit{Spitzer} 8\,$\mu$m image shows a barely resolved double source here, and at shorter IRAC wavelengths there is a third peak located just to the south ($\sim$4$\arcsec$) of the doublet which dominates a the shortest IRAC wavelengths.

\begin{deluxetable*}{rrrrrrrrrrrl}
\tabletypesize{\scriptsize}
\tablecolumns{11}
\tablewidth{0pt}
\tablecaption{Observational Parameters of All Newly Identified Sources in W49A}\label{tb:new}
\tablehead{\colhead{  }&
           \colhead{  }&
           \colhead{  }&
           \multicolumn{4}{c}{${\rm 20\mu{m}}$}&
           \multicolumn{4}{c}{${\rm 37\mu{m}}$}&
           \colhead{  }\\
           \cline{4-7} \cline{8-11} \\
           \colhead{ Source }&
           \colhead{ R.A. } &
           \colhead{ Dec. } &
		   \colhead{ FWHM } &
           \colhead{ $R_{\rm int}$ } &
           \colhead{ $F_{\rm int}$ } &
           \colhead{ $F_{\rm int-bg}$ } &
		   \colhead{ FWHM } &
           \colhead{ $R_{\rm int}$ } &
           \colhead{ $F_{\rm int}$ } &
           \colhead{ $F_{\rm int-bg}$ } \\
	   \colhead{  } &
	   \colhead{ J2000 } &
	   \colhead{ J2000 } &
	   \colhead{ ($\arcsec$) } &
	   \colhead{ ($\arcsec$) } &
	   \colhead{ (Jy) } &
	   \colhead{ (Jy) } &
	   \colhead{ ($\arcsec$) } &
	   \colhead{ ($\arcsec$) } &
	   \colhead{ (Jy) } &
	   \colhead{ (Jy) } \\
}
\startdata
CC-1	&	19 10 13.76	&	9 07 20.8	&	8.15	&	8.4	&	12.9	&	4.94	&	diffuse	&	8.4	&	46.8	&	22.5		\\
DD-1	&	19 10 11.34	&	9 06 20.9	&	2.42	&	3.1	&	1.78	&	0.20	&	4.3	&	3.8	&	14.5	&	2.82		\\
II-1	&	19 10 14.47	&	9 06 48.3	&	1.93	&	4.6	&	2.93	&	0.55	&	3.94	&	4.6	&	27.6	&	6.22		\\
JJ-1	&	19 10 19.02	&	9 06 41.7	&	4.86	&	9.2	&	21.4	&	6.09	&	10.2	&	10.0	&	141	&	34.3		\\
KK-1	&	19 10 21.29	&	9 06 27.6	&	5.46	&	3.8	&	1.45	&	0.26	&	4.93	&	4.6	&	16.6	&	7.76		\\
MM-1	&	19 10 25.43	&	9 05 43.2	&	5.61	&	10.0	&	19.6	&	6.02	&	7.72	&	10.0	&	50.9	&	25.3		\\
P-1	&	19 10 16.62	&	9 05 45.5	&	4.59	&	3.1	&	2.55	&	0.87	&	4.3	&	3.1	&	12.5	&	5.63		\\
Q-1	&	19 10 09.93	&	9 05 09.8	&	low S/N	&	3.1	&	1.03	&	0.21	&	low S/N	&	3.1	&	6.99	&	5.58		\\
R$_4$	&	19 10 10.92	&	9 05 16.8	&	PR	&	3.1	&	$<$25.2	&	$<$24.7	&	PR	&	3.1	&	$<$209	&	$<$207		\\
W49\,South-2	&	19 10 21.02	&	9 05 01.6	&	low S/N	&	3.8	&	2.57	&	0.49	&	PR	&	3.8	&	$<$24.7	&	$<$22.4		\\
\enddata
\tablecomments{\scriptsize Table note from Table 1 applies here.}
\end{deluxetable*}

Due to larger field of view than studied in \citet{1997ApJ...482..307D} and \citet{2000ApJ...540..316S}, we found a smaller ($d\sim4\arcsec$) but resolved infrared source at both \textit{SOFIA} wavelengths located about 58$\arcsec$ west of KK, which we label as MM-1 in Figure \ref{fig:SOFIAdata}. The source appears point-like at 37\,$\mu$m, but at shorter wavelengths there is diffuse but extended emission on both sides of the infrared peak at a p.a. of $\sim$135$\arcdeg$. At the shortest IRAC wavelengths it appears almost rectangular or bow-tie shaped. This source is also seen as an unresolved source in the \textit{Herschel} 70\,$\mu$m image. Since MM-1 is also off the field of the archival 3.6\,cm \textit{VLA} data as well, it is unknown if this source is a cm radio continuum emitter. Given the large derived luminosity from our SED modeling, we conclude that this source is powered by at least one MYSO and the extended emission could be due to outflow cavities which are often seen in the mid-infrared \citep[e.g.,][]{2006ApJ...642L..57D,2017ApJ...843...33D}.

\citet{2000ApJ...540..316S} found four new mid-infrared sources named for their proximity to known sources; BB East, DD South, EE East, and HH West (see red dots in Figure \ref{fig:SOFIAdata}). DD South is large ($d\sim10\arcsec$) and estimated to be 80\,Jy at 20\,$\mu$m, and thus should be very obvious in or \textit{SOFIA} images, yet it does not appear in either our 20 or 37\,$\mu$m data, nor is there a source there in any of the IRAC bands. Source BB East is about the same extent with a peak surface brightness of $\sim$1\,Jy$\cdot$arcsec$^{-2}$, which should also be easily detectable in our images and yet nothing is present in either our 20 or 37\,$\mu$m data or in any of the IRAC bands. EE east and HH west are much smaller and fainter sources, however both are estimated to have a peak surface brightness of 0.25\,Jy$\cdot$arcsec$^{-2}$ at 20\,$\mu$m, and so we should even be able to detect that level of emission with a S/N of $\sim$30 in our 20\,$\mu$m \textit{SOFIA} data. However, we see no hint of either source in either \textit{SOFIA} band, nor are there IRAC sources at these locations. Given that all of these sources should be easily detected in our images and are not, and given that none of the four sources have components in any of the IRAC bands either, it could be that none of these sources are real. 

\section{Results and Data Analysis}\label{sec:data}

Tabulating the detections and non-detections from sources discussed in Section \ref{sec:W49Asources}, we have produced Tables 1 and 2. Table 1 is a list of all previously known radio continuum sources, as given by \citet{1997ApJ...482..307D}. In this table we identify in right ascension and declination the aperture centers (not the source peaks or centers) used for the photometry of each source as well and the aperture radii used at each wavelength ($R$\textsubscript{int}). We give the integrated flux densities at both wavelengths within those apertures ($F$\textsubscript{int}), as well as background subtracted estimates of the flux densities of sources. We apply the same aperture photometry practices as we did in our previous studies to ascertain the aperture sizes to use for flux extraction. To quickly summarize, we choose an aperture radius where the flux from the azimuthally-averaged radial profile of a source just begins to level out. If the source is surrounded by extended emission, this background is only a local minimum. The background flux estimate is taken from the statistics of the data within an annulus just outside that aperture, the thickness of which is determined by the range of radii where the background remains at a constant level. These background subtracted flux estimates are given in Table 1 in the columns labeled $F$\textsubscript{int-bg}. 

If the radial profile can be fit by a Gaussian, we report that value in the FWHM column of Table 1. If we cannot fit a Gaussian profile, a reason is given in the FWHM column. `ND' means it was a non-detection, and a 3-$\sigma$ upper-limit value is given for $F_{\rm int}$ (and no value is given for $F_{\rm int-bg}$). This only applies to sources A (at 20 and 37\,$\mu$m) and GG (at 20\,$\mu$m only). We used the background statistics within an aperture that would fit the radio size of source A (3.1$\arcsec$) centered on the location of the radio source peak to estimate the upper-limit fluxes. `PR' and `U' mean `partially resolved' and `unresolved', respectively. This means there is too much contamination (or variability in the contamination) from nearby sources to estimate a flux density from the source alone. We therefore employed an aperture size for these sources that was either just large enough to contain the radio source extent, or was the point-source radius (1$\farcs$5) derived from the average FWHM ($\sim$3$\farcs$1) at 20 and 37\,$\mu$m for \textit{SOFIA}, whichever was larger. The background subtracted fluxes for unresolved and partially resolved sources is taken from the closest area showing a global minimum (rather than local minimum used for resolved sources) because we cannot determine with any accuracy a good local minimum to use. This means that the $F$\textsubscript{int-bg} values in Table 1 are true upper limits to the source fluxes for unresolved and partially resolved sources. If the source is large and flocculent, it is labeled `diffuse'. Several sources appear to be wind-blown bubbles or otherwise appear as a ring-shape, a broken ring, or a large arc/partial ring. These sources are fit by a circle whose radius best fit the structure, and they are reported in Table 1 with the words `ring' along with the radius value of that fit.

All sources newly identified for the first time in this work, summing 10 new sources in all, are listed in Table 2. The observational parameters listed in that table were obtained in the same way as described for Table 1. Four of these newly identified sources (CC-1, JJ-1, P-1, and R$_4$) are coincident with radio continuum emission peaks at 3.6\,cm, as seen in the maps of \citet{1997ApJ...482..307D} but were not labeled or discussed in that work, and DD-1 is contained in a region of extended 3.6\,cm radio continuum emission with no definitive peak. MM-1 was off-field in the available radio data. Four newly identified mid-infrared sources, II-1, KK-1, Q-1, and W49\,South-2 have no detected 3.6\,cm emission at their locations. 
In addition to radio source A, some other radio sources identified by \citet{1997ApJ...482..307D} were not detected in the \textit{SOFIA} data. Sources B and O$_2$ were not detected at either \textit{SOFIA} wavelength as a peak, though there is some diffuse and/or extended mid-infrared emission in their areas. Radio sources C, C$_1$, D, E, and H may have associated emission at \textit{SOFIA} wavelengths that is not well-resolved from the bright and extended emission of the G complex. Furthermore, compact radio sources GG and N are only clearly detected at 37\,$\mu$m.  

Finally, though we had the sensitivity to detect them, we do not detect mid-infrared emission from previously identified mid-infrared sources BB East, DD South, EE East, and HH West \citep{2000ApJ...540..316S}. 

\subsection{Physical Properties of Sub-Components and Point Sources: SED Model Fitting and Determining MYSO Candidates}\label{sec:cps}

Defining the embedded MYSO population in W49A is one of the primary goals of this study. Thermal infrared observations are able to penetrate through overlying extinction from the larger star-forming cloud complex and are sensitive to the emission from the dust localized near to and enshrouding young and forming stars (or bound star systems). In fact, our previous papers from this G\ion{H}{2} region survey (\citetalias{2019ApJ...873...51L} and \citetalias{2020ApJ...888...98L}) have shown that, at the \textit{SOFIA}-FORCAST 20 and 37\,$\micron$ wavelengths (and $\sim3\arcsec$ of angular resolution), we can detect not only the known MYSOs and potential MYSOs within these G\ion{H}{2} regions (as found via their free-free radio continuum emission), but we also uniquely find MYSOs in their earliest embedded stages of evolution, prior to the onset of detectable UC\ion{H}{2} regions\footnote{We also can detect lower-mass YSOs that are heavily embedded and possessing stars that are non-ionizing (and thus not detectable in the cm radio continuum).}. 

In order to identify the MYSO candidates in W49A, we will use SED model fitting. Therefore, the first step is to determine which features in the data are likely to be internally heated sources and measure their flux densities at as many wavelengths as we can to fill out their SEDs. We have already described in the previous section how we performed aperture photometry on our sources. This was performed on the natural resolution images because (as discussed in Section \ref{sec:obs}) the deconvolved images can be subject to larger flux errors. As a consequence, some groups of sources identified in the 3 times higher resolution (0.8$\arcsec$) radio continuum images or partially resolved in our deconvolved \textit{SOFIA} images are lumped together into a single source if they are not resolved from other nearby sources in the natural resolution \textit{SOFIA} images (specifically J$_1$+J$_2$, G\textsubscript{tot}, W49\,South, and R\textsubscript{tot}). Source candidates were found as resolved sources or peaks in the 20 and 37\,$\mu$m natural resolution images, and then cross referenced with the \textit{Spitzer}-IRAC, \textit{Herschel}-PACS, and cm radio data for spatial coincidences. Only source candidates that were spatially coincident with sources or emission peaks at longer and/or shorter wavelengths made it to the final source list. In the end we identified 24 compact infrared sources from the sources in Tables \ref{tb:all} and \ref{tb:new}, and they are listed in Table \ref{tb:sedp}. 
To be considered a ``compact'' source the emission had to be compact enough and/or resolved enough from nearby sources to have a measurable FWHM in Tables \ref{tb:all} and \ref{tb:new}, though we will additionally include the resolved (yet low S/N) sources Q-1 and W49\,South-2. One needs to note that the far distance of W49A (11.1\,kpc) makes it difficult to resolve close binary and multiple systems since one FORCAST pixel ($\sim$0$\farcs768$) corresponds to a spatial size $\sim$0.04\,pc. We also know that several of our ``compact sources'' contain multiple sources, because of resolution issues. For instance, we do not resolve out radio sources G$_1$, G$_2$, G$_3$, G$_4$, and G$_5$ even in the deconvolved \textit{SOFIA} data, and these sources are treated as a single source G\textsubscript{tot}. We will discuss more the impacts of multiplicity on our analyses later in this section when we discuss the SED model fitting. 

We performed additional aperture photometry for these 24 compact sources with \textit{Spitzer}-IRAC 3.6, 4.5, 5.8, and 8.0\,$\mu$m as well as {\it Herschel}-PACS 70 and 160\,$\mu$m archival data. We applied the same photometry methods to the IRAC data as we did on the \textit{SOFIA} data. {\it Herschel}-PACS fluxes for the sources were obtained from the fixed size apertures, i.e., $R_{\rm int}=5$\,pixels for both 70 and 160\,$\mu$m data, without background subtraction. Following \citetalias{2019ApJ...873...51L} and \citetalias{2020ApJ...888...98L}, we consider the {\it Herschel}-PACS fluxes as upper limits due to the large and uncertain levels of contamination from the environmental extended emission that makes it difficult to identify the accurate 70 and 160\,$\mu$m PSFs and flux densities. For some sources either the both the 70\,$\mu$m and seven 160\,$\mu$m photometry apertures (F, G\textsubscript{tot}, I, J, KK, and N) or just the 160\,$\mu$m apertures (EE and M) enclosed saturated pixels. For these sources we do not include these saturated data points in the SED fitting. We tabulate the {\it Spitzer} and {\it Herschel} photometry data in the Appendix (Table\,\ref{tb:cpsb} and \ref{tb:cpsc}, respectively). 

\begin{figure}
\center
\begin{tabular}[b]{c@{\hspace{-0.1in}}c}
\includegraphics[width=3.2in]{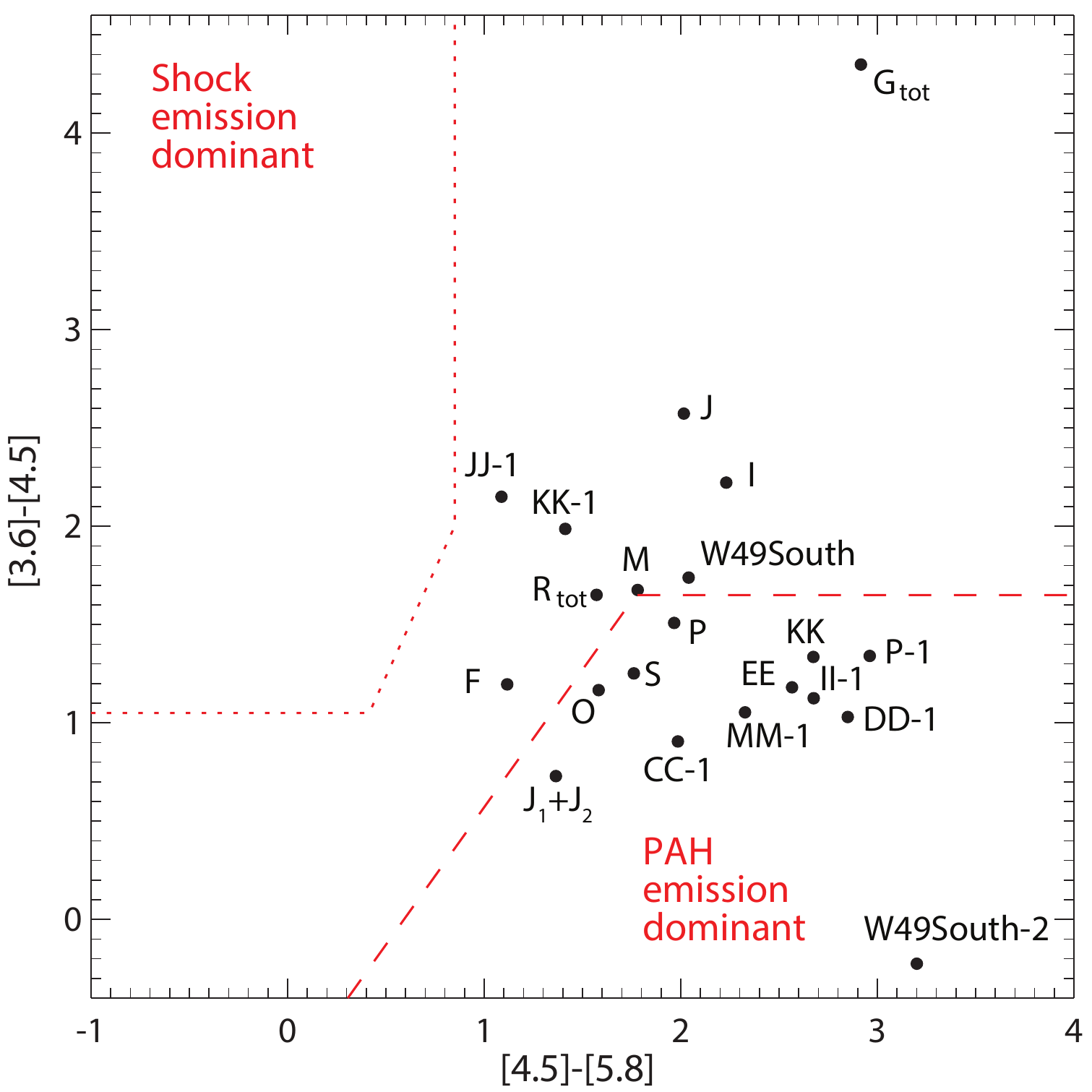}\\
\end{tabular}
\caption{\footnotesize A color-color diagram utilizing our background-subtracted \textit{Spitzer}-IRAC 3.6, 4.5, 5.8 and 8\,$\mu$m source photometry to distinguish ``shocked emission dominant'' and ``PAH emission dominant'' YSO candidates from our list of sub-components and point sources. Above the dotted line (upper-left) indicates shock emission dominant regime. Below the dashed line (bottom-right) indicates PAH dominant regime. This method was adopted from \citet{2009ApJS..184...18G}. Sources GG, N, and Q-1 are not included in this diagram due to weak or non-detection of {\it Spitzer}-IRAC bands.}
\label{fig:ccd}
\end{figure}

\begin{deluxetable*}{rccccrclrclcl}
\tabletypesize{\small}
\tablecolumns{12}
\tablewidth{0pt}
\tablecaption{SED Fitting Parameters of Selected Compact Infrared Sources in W49A}\label{tb:sedp}
\tablehead{\colhead{   Source   }                                              &
           \colhead{  $L_{\rm obs}$   } &
           \colhead{  $L_{\rm tot}$   } &
           \colhead{ $A_v$ } &
           \colhead{  $M_{\rm star}$  } &
           \multicolumn{3}{c}{$A_v$ Range}&
           \multicolumn{3}{c}{$M_{\rm star}$ Range}&
           \colhead{ Best }&
           \colhead{Notes}\\
	   \colhead{        } &
	   \colhead{ ($\times 10^3 L_{\sun}$) } &
	   \colhead{ ($\times 10^3 L_{\sun}$) } &
	   \colhead{ (mag.) } &
	   \colhead{ ($M_{\sun}$) } &
       \multicolumn{3}{c}{(mag.)}&
       \multicolumn{3}{c}{($M_{\sun}$)}&
       \colhead{  Models   } &
       \colhead{   }
}
\startdata
                  CC-1 &     11.6 &    108 &     26.5 &     16 &  26.5 & - &  79.5 &  12 & - &  32 &  7 & MYSO$\dagger$ \\
                  DD-1 &      1.77 &     11.7 &     68.9 &      8 &  33.5 & - &  76.8 &   8 & - &   8 &  8 & MYSO  \\
                    EE &     18.0 &     36.5 &     72.9 &     16 &  10.6 & - & 132 &  12 & - &  64 &  7 & MYSO  \\
                     F &      3.53 &      2.59 &     26.5 &      2 &  12.6 & - &  33.5 &   1 & - &   8 & 10 &       \\
  G\textsubscript{tot} &    275 &    841 &     16.8 &     64 &   2.70 & - &  92.2 &  24 & - &  96 & 20 & MYSO$\dagger$  \\
                    GG &     39.9 &     99.2 &    252 &     16 & 252 & - & 424 &  16 & - &  24 & 11  & MYSO \\
                     I &     27.7 &     41.5 &     55.6 &     12 &  47.7 & - & 117 &  12 & - &  24 &  7 & MYSO\\
                  II-1 &      3.59 &     12.1 &     23.8 &      8 &  23.8 & - &  60.4 &   8 & - &   8 &  6 & MYSO\\
                     J &     37.5 &    457   &    101   &     48 &  24.3 & - & 101 &  16 & - &  96 & 12 & MYSO\\
       J$_{1}$+J$_{2}$ &     29.3 &   2040   &    233   &    128 & 176 & - & 233 &  16 & - & 160 &  7 & MYSO\\
                    JJ-1 &     13.4 &    50.9   &     53.0 &     12 &  13.2 & - &  79.5 &  12 & - &  12 &  10 & MYSO\\       
                    KK &     39.8 &    113   &     53.0 &     16 &  26.5 & - &  79.5 &  12 & - &  16 &  5 & MYSO\\
                  KK-1 &      1.17 &     13.6 &     36.9 &     12 &  23.5 & - &  69.6 &   8 & - &  24 &  6 & MYSO$\dagger$\\
                     M &      5.75 &     15.3 &     53.0 &      8 &   1.70 & - &  53.0 &   8 & - &  32 & 15 & MYSO\\
                  MM-1 &     13.9 &     20.2 &      1.70 &     12 &   0.80 & - &  26.5 &  12 & - &  24 & 16 & MYSO\\
                     N &      9.54 &     46.7 &    218   &     12 &  13.2 & - & 579 &   8 & - & 128 &  7 & MYSO$\ddagger$\\
                     O &     66.0 &    119    &     47.7 &     24 &  47.7 & - &  53.0 &  24 & - &  24 & 11 & MYSO\\
                     P &     20.7 &    549    &     26.5 &     48 &   1.70 & - &  28.5 &  12 & - &  48 &  8 & MYSO\\
                   P-1 &      2.53 &     13.3 &     53.0 &      8 &  26.5 & - &  72.9 &   2 & - &  12 & 25 & pMYSO\\
                   Q-1 &      4.06 &     26.6 &    127   &     12 &   8.40 & - & 127 &   8 & - &  12 & 14 & MYSO\\
  R\textsubscript{tot} &    186 &   2380      &     33.5 &    128 &  17.6 & - &  58.7 &  64 & - & 128 &  7 & MYSO\\
                     S &    175 &    749      &     26.5 &     48 &   2.70 & - &  53.0 &  24 & - &  96 & 27 & MYSO\\
              W49\,South &    359 &   1560    &     16.8 &     96 &   2.70 & - & 134 &  32 & - & 128 & 10 & MYSO$\dagger$\\
            W49\,South-2 &     20.8 &     39.94 &     71.5 &     12 &  47.0 & - &  79.5 &   8 & - &  16 & 12 & MYSO\\
\enddata
\tablecomments{\footnotesize A ``MYSO'' in the right column denotes a MYSO candidate. A ``pMYSO'' indicates that their is greater uncertainty in the derived physical parameters and that these sources are possible MYSO candidates.}
\tablenotetext{\dagger}{The SED fits are poor for these sources, likely due to the presence of multiple unresolved sources. However the fits are all under-fitting the data (and therefore underestimating the luminosities), and yet still have luminosities indicative of MYSOs.}
\tablenotetext{\ddagger}{Source N has only one nominal data point for the SED fits. However, the fits are somewhat constrained by the Spitzer and Herschel upper limits.}
\end{deluxetable*}

\begin{figure*}
\begin{tabular}[b]{c@{\hspace{-0.1in}}c}
\includegraphics[width=6.5in]{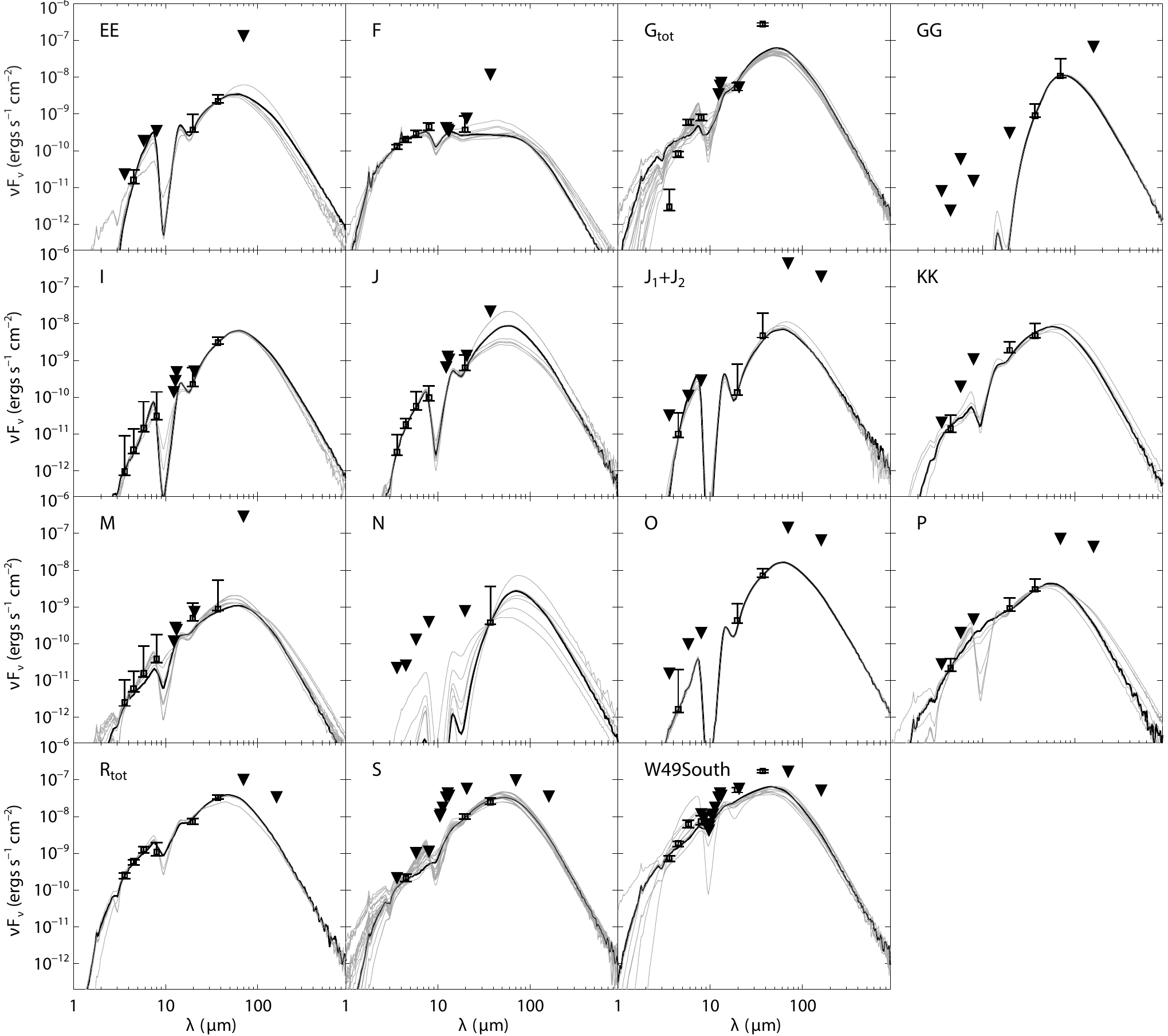}\\
\end{tabular}
\caption{\footnotesize SED fitting with ZT MYSO models of compact sources in W49A. Black lines are the best fit model to the SEDs, and the system of gray lines are the remaining fits in the group of best fits (from Table\,\ref{tb:sedp}). Upside-down triangles are data that are used as upper limits in the SED fits.}
\label{fig:fd1}
\end{figure*}

\begin{figure*}
 \begin{center}
 \includegraphics[width=4.7in]{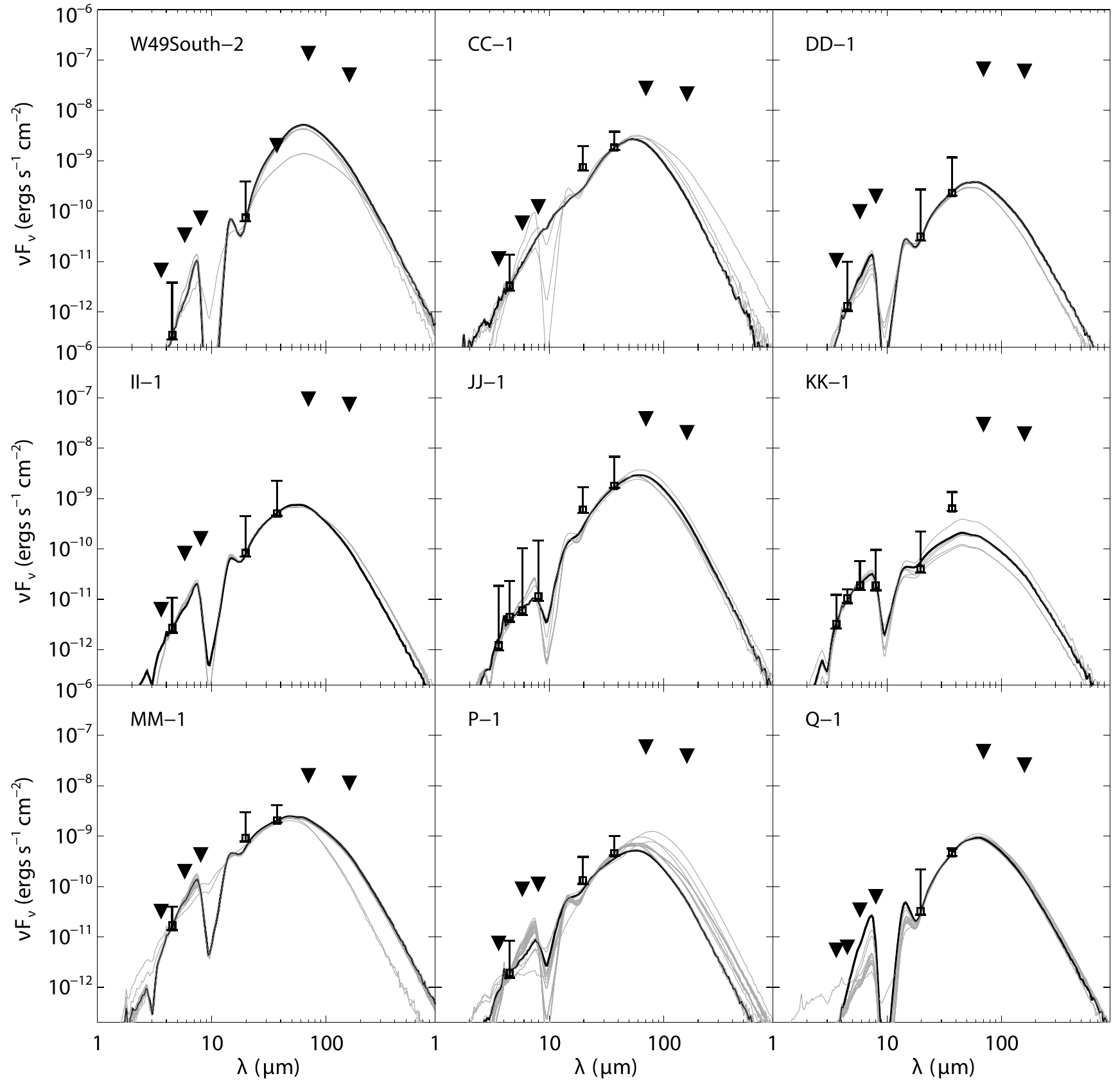}\\
 \end{center}
\caption{\footnotesize Same as Fig.~\ref{fig:fd1} but for newly identified sources in this study.}
\label{fig:fd2}
\end{figure*}

To determine how to handle the {\it Spitzer}-IRAC photometry data that we would use in our SED model fitting, we first tested the possibility of flux contamination in the 3.6, 5.8 and 8\,$\mu$m bands due to the presence of emission from polycyclic aromatic hydrocarbons (PAHs) and for contamination in the 4.5\,$\mu$m band due to shock-excited molecular hydrogen emission by adopting the color-color diagram analysis of \cite{2009ApJS..184...18G} as shown in Figure~\ref{fig:ccd}. This analysis finds no ``shock emission dominant'' sources in W49A. For the twelve ``PAH emission dominant'' sources seen in Figure~\ref{fig:ccd} we defined their IRAC 3.6, 5.8, and 8.0\,$\mu$m flux densities as upper limits, while the 4.5\,$\mu$m flux densities were considered as nominal data points. 
 
FORCAST 20 and 37\,$\mu$m fluxes are assumed to be nominal data points for all sources except GG and N (for which 20\,$\mu$m measurements are upper limits) and F, J, and W49\,South-2 (for which 37\,$\mu$m measurements are upper limits) due to weak emission and/or strong background emissions. We consider the subtracted background flux of each source and each nominal band flux as the upper-limit error since the total photometric error does not exceed the background flux and the choice of the level of background to subtract off is the highest source of uncertainty. The lower-limit errors of all derived IRAC and FORCAST flux densities are set to the calculated total photometric error, which are estimated to be 20$\%$ for all IRAC bands, and 15 and 10$\%$ for the FORCAST 20 and 37\,$\mu$m data, respectively. These flux density uncertainty estimates are consistent with those found in \citetalias{2019ApJ...873...51L} and \citetalias{2020ApJ...888...98L}.
 
We also incorporated the infrared photometry data of \citet{2000ApJ...540..316S} in our SED models for several sources. We only included data from that work for sources with definitive measurements (i.e., no upper limits) and only for the sources that appear in our Table \ref{tb:sedp}. Explicitly the data included in the fits were for W49\,South (at 8.0. 8.6, 9.1, 9.7, 9.8, 10.5, 10.7, 11.3, 12.3, 12.8, 13.2, and 20.6\,$\mu$m), S (at 10.5, 10.7, 11.3, 12.3, 12.8, 13.2, and 20.6\,$\mu$m), F, G, I, J, and M (these latter five sources at 12.3, 12.8, 13.2, and 20.6\,$\mu$m). When comparing the \citet{2000ApJ...540..316S} photometry at 20.6\,$\mu$m to our SOFIA data at 20\,$\mu$m we noticed that the values from \citet{2000ApJ...540..316S} were systematically in between our non-background-subtracted fluxes and our background-subtracted fluxes. For the brighter sources like W49\,South, G, and S, the differences between our non-background-subtracted fluxes and our background-subtracted fluxes is small and thus so is the difference of either of these values to those of \citet{2000ApJ...540..316S}, i.e., the same to within the combined photometric errors. For the fainter sources, our background subtracted flux values can be up to half those quoted by \citet{2000ApJ...540..316S}. Given that \citet{2000ApJ...540..316S} detected sources that are likely not real (i.e., DD South, EE East, HH West, and BB East), and if this were due to improper background subtraction, the photometry for the rest of the sources could be off as well. It may also be that our method of background subtraction might be over-subtracting flux from the actual sources we are performing photometry on (as we discuss in Section \ref{sec:data}). Given that the photometry from \citet{2000ApJ...540..316S} was consistently higher than our photometry at 20\,$\mu$m, we chose to utilize the data as upper limits only (including their quoted 10\% photometric error). 

Based on the {\it Spitzer}, \textit{SOFIA}, {\it Herschel}, and \citet{2000ApJ...540..316S} photometry and the uncertainties, we constructed near-infared to far-infrared SEDs of the 24 selected compact sources intending to fit theoretical SED models of MYSOs \citep[][hereafter ZT models]{2013ApJ...775...79Z}. Each model fit provides normalized minimum $\chi^2$ values (so called $\chi_{\rm nonlimit}^2$). We selected the group of the best fit models as we did in previous studies (see \citetalias{2019ApJ...873...51L} and \citetalias{2020ApJ...888...98L} for details). The number of the best fit models and the ranges of the derived parameters based on the models are listed in Table~\ref{tb:sedp}. Figures~\ref{fig:fd1} and \ref{fig:fd2} show the observed SEDs, and the fits of the best models of the 24 sources are summarized in Table~\ref{tb:sedp}. 

One obvious issue to address is that the ZT models assume a single central stellar source, and given the extreme distance to W49A, it is highly likely that at least some of our compact sources house multiple unresolved stellar components. It is for this reason that we do not tabulate or discuss the model parameters like disk size or accretion rate, and instead concentrate on the values for internal mass and source luminosity. In \citetalias{2019ApJ...873...51L}, we showed that the derived stellar masses of a compact source could be trusted even if a \textit{SOFIA} defined aperture contained multiple YSOs due to the limited angular resolution of the data. IRS2E in W51A was the representative case. There were four NIR-defined sources \citep{2016ApJ...825...54B} contained within the photometric aperture we used for the 20 and 37\,$\micron$ data. The stellar mass we derived for IRS2E in \citetalias{2019ApJ...873...51L} (64\,$M_{\sun}$ for best fit model, 64--128\,$M_{\sun}$ for the range) agreed well with the total stellar masses of all four NIR YSOs combined (80\,$M_{\sun}$) as derived by \citet{2016ApJ...825...54B}. We have a similar case in W49A for source $R$\textsubscript{tot}. As seen in $\S$\ref{sec:sw}, $R$\textsubscript{tot} contains three different cm radio sources identified in \citet{1997ApJ...482..307D} with a total stellar mass, of 85\,$M_{\sun}$. However, adding the additional radio source that was not identified in that work, R$_4$ (which we measure its 3.6\,cm flux density to be 9\,mJy, which translates into a B0 ZAMS star) this adds another $\sim$18\,$M_{\sun}$, for a total of 103\,$M_{\sun}$. The derived stellar mass of the ZT model's best fit is 128\,$M_{\sun}$ with a mass range of 64--128\,$M_{\sun}$ for $R$\textsubscript{tot} which is in fairly good agreement with the estimation from the radio data. 

On the other hand, we see in Figures \ref{fig:fd1} and \ref{fig:fd2} that there are a few compact sources that are not fit very well with the MYSO models: G\textsubscript{tot}, W49\,South, CC-1 and KK-1. It is likely that these sources are not being fit because they have multiple unresolved MYSOs or stellar sources responsible for their heating, perhaps even at different evolutionary stages or embedded in variable extinction, which would lead to multiple temperature components that could skew the SED data. Source CC-1, in particular, is at the larger end of our ``compact'' sources size limit and given its size and relatively bright and extended radio emission, it is very likely excited by more than one massive star. So it could be that the ZT models only do a good job of estimating the overall mass of a binary or proto-cluster if the sources are coeval or subject to the same extinction (e.g., R\textsubscript{tot}). 

In our previous papers we used the the derived data like those compiled in Table \ref{tb:sedp} to determine which sources are MYSO or potential MYSO candidates. The conditions for a source to be considered a MYSO candidate in our previous papers has been that it must 1) have an SED reasonably fit by the MYSO models, 2) have a M$_{\rm star}\ge8\,$M$_{\sun}$ for the best model fit model, and 3) yield a M$_{\rm star}\ge8\,$M$_{\sun}$ in all of the model fits in the group of best fit models. In those previous papers we categorized ``potential MYSOs'' (pMYSOs) as sources that fulfilled only the first two of the MYSO criteria. However, as we have been discussing, W49A is so distant compared to those previously observed G\ion{H}{2} regions, that we are likely seeing some compact sources where multiple stellar components may be affecting the SED shape so that they cannot be properly fit by our single-star models. Specifically, G\textsubscript{tot}, W49\,South, CC-1, and KK-1 are not well-fit by the MYSO SED models. However, since the best fits to the data for all four sources satisfy criteria 2) and 3) above, and in all four cases the SED fits under-fit the data (thus indicating even larger luminosities for these source than those given in the table), all four sources are likely to contain at least one MYSO.

Because of its great distance and large measured extinction along the line of sight, one would expect that the only YSOs we would be capable of detecting in the mid-infrared would have to be massive. In fact all sources we identify in Table \ref{tb:sedp} have associated 3.6\,cm continuum emission (a potential indicator of a possible ionizing MYSO) except for II-1, KK-1, Q-1, and W49\,South-2. However, SED model fitting for all of these sources yield derived luminosities of MYSOs. Therefore, given their lack of radio continuum emission and high luminosities, II-1, KK-1, Q-1, and W49\,South-2 are the only MYSOs in Table \ref{tb:sedp} that could potentially be in an extremely young evolutionary phase prior to the onset of an \ion{H}{2} region.   

P-1 is the only source that fulfills the pMYSO criteria. It is well-fit by the MYSO SED models, and the best fit is for a massive star ($8\,M_{\sun}$), but the range of best fits includes stellar masses less than $8\,M_{\sun}$. 
Of all of the sources in Table \ref{tb:sedp}, the derived values for source N should be viewed with healthy skepticism since the fits are to data with only one nominal data point (i.e. the SOFIA 37\,$\mu$m data point). The SED fits are still somewhat constrained by the large number of upper limits at shorter and longer wavelengths, however, the derived values from the group of best fits given in Table \ref{tb:sedp} vary the most for source N than all others (especially the A$_V$ range). Even with these loose constraints, however, all fits are indicating that the source contains a MYSO (as would be expected for sources that are only seen at 37\,$\mu$m and longer wavelengths). 

Source F is a special case where it is well fit by the ZT models and the best fit mass is only 2\,M$_{\sun}$, and it has models that fit only up to 8\,M$_{\sun}$. Therefore, source F would not qualify as a MYSO or pMYSO given our above criteria. Interestingly, source F is the least luminous source in our list, yet based upon its cm continuum flux it is estimated to be ionized by a O7.5 star \citep{1997ApJ...482..307D}. However, given the 1.3 and 3.6\,cm radio fluxes from \citet{1997ApJ...482..307D}, we calculate a radio spectral slope of $\alpha_{radio} = -0.22$, which is atypical for a UC\ion{H}{2} region, and more indicative of synchotron emission arising in shocks resulting from the interaction of a collimated stellar wind with a surrounding magnetized medium \citep[e.g., see IRAS 16547–4247 from][]{2003ApJ...587..739G}. Therefore, F may be a less-massive YSO with non-thermal radio cm continuum emission coming from a jet/outflow. Interestingly, all sources in the eastern side of the Welch Ring (sources F, H, I, J, J$_2$, and M) have very negative radio spectral slopes ($-0.22 > \alpha_{radio} > -0.76$) indicating their emissions are dominated by synchotron emission\footnote{Sources dominated by thermal Bremsstrahlung emission have radio spectral indices of $\alpha_{radio} \ge -0.1$, while non-thermal sources dominated by sychrotron emission have $\alpha_{radio} < -0.1$ \citep{2016MNRAS.460.1039P}.}.   
Taking all of the above into account, we show in the last column of Table \ref{tb:sedp} our estimates for which sources we believe are MYSO and pMYSO candidates. The only source with not enough evidence to be categorized as either a MYSO or pMYSO is source F.   

In this study, therefore, we classify 22 sources satisfying our criteria of housing a MYSO out of the 24 \textit{SOFIA}-FORCAST defined compact sources. Overall, we determined 23 sources to be either a MYSO or pMSYO ($\sim96\%$). In \citetalias{2019ApJ...873...51L} and \citetalias{2020ApJ...888...98L}, $87\%$ of the point sources were found to likely be MYSOs or pMYSOs in W51A region ($d\sim5.4$\,kpc), while M17 ($d\sim2.0$\,kpc) showed only $44\%$ sources as MYSOs and pMYSOs. The main reason of the difference was suggested as the results of the distance of W51A and M17 so that more of low to intermediate mass YSOs could be detected in M17 due to its relatively closer distance. W49A is about two times farther than W51A from Sun, and we are measuring an even higher rate of MYSO detection from the \textit{SOFIA}-FORCAST mid-infrared imaging. More interesting however, is the drop it in the rate of detection of pre-UC\ion{H}{2} region MYSOs between the W51A study and this one. Almost half of the sources determined to be MYSOs for W51A have no detectable radio continuum emission, whereas that is the case for only four MYSOs (II-1, KK-1, W49\,South-2, and Q-1) in this W49A study. Again, this is likely to be due to the differences in distance. Pre-UC\ion{H}{2} region sources will be compact, and are usually found close to other radio-emitting and/or mid-infrared-emitting MYSOs, and thus we are less likely to resolve them from their neighbors at the distance of W49A. Also, not only is there a larger amount of interstellar extinction in the general case of comparing a region 5\,kpc away to one 11\,kpc away, but in the particular case of W49A we have the obscuring dust of the Milky Way's Sagittarius spiral arm, which crosses the line of sight to W49A twice. This interstellar and spiral arm extinction heavily affects the transmission of infrared emission from sources within W49A, but does not really affect the cm radio continuum emission from those same sources. Therefore, this is the likely reason we are mainly detecting the larger, radio-continuum emitting sources of W49A with our \textit{SOFIA} data.  
 
\begin{figure}
\begin{center}
\includegraphics[width=3.3in]{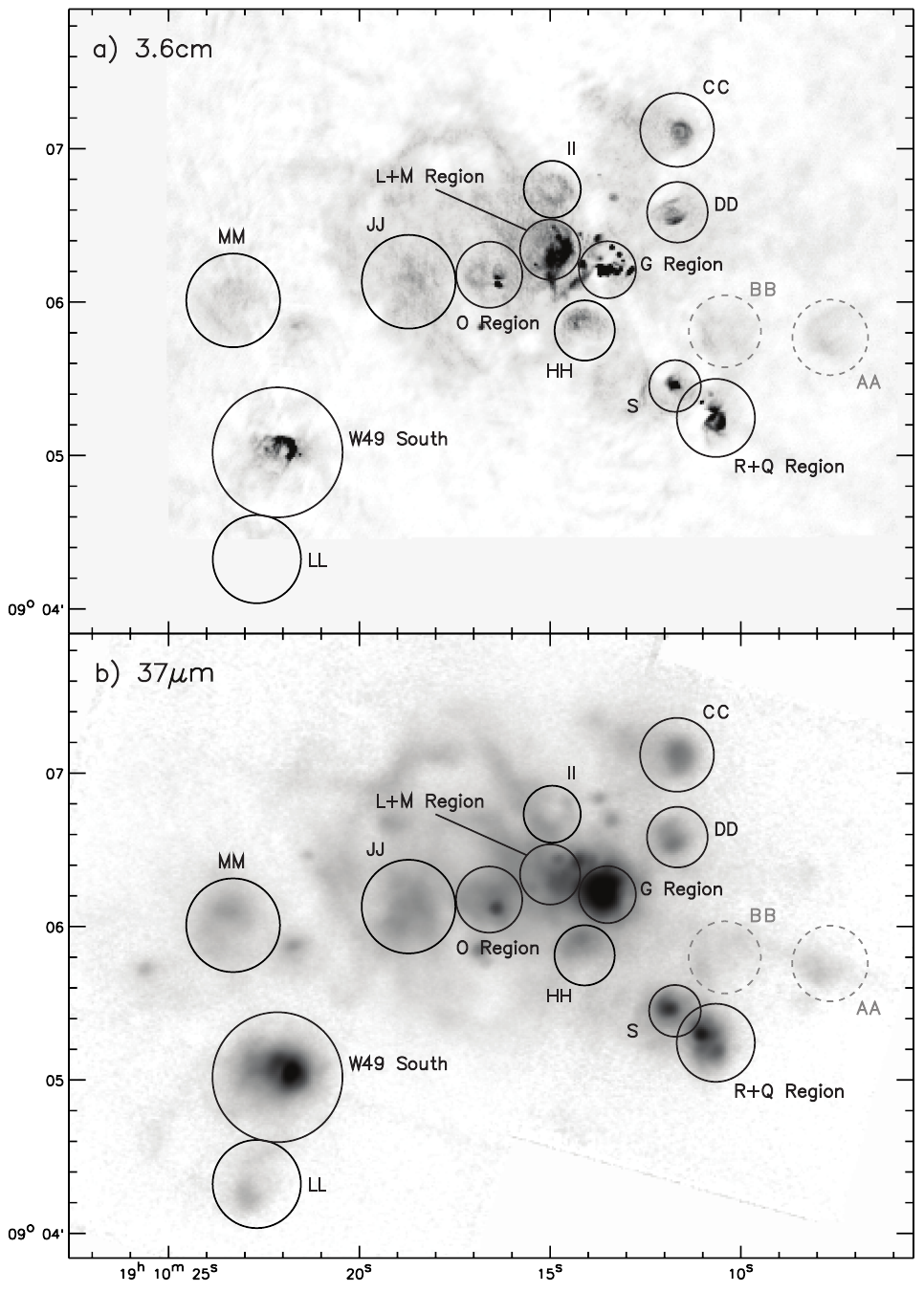}\\
\end{center}
\caption{\footnotesize The major radio-defined sub-regions within W49A. Labeled in black are the sources chosen for the extended region analyses. The circles are demonstrating the size and positions of the apertures used for photometry of the extended regions, as given in Table~\ref{tb:major}. Sources AA and BB are shown here in gray for completeness, but they are likely not star-forming clumps.} a) 3.6\,cm radio map from \citet{1997ApJ...482..307D}. b) The 37\,$\mu$m {\it SOFIA} map.
\label{fig:radio}
\end{figure} 
 
\begin{deluxetable*}{rrrrccccc}
\tabletypesize{\scriptsize}
\tablecolumns{9}
\tablewidth{0pt}
\tablecaption{Observational Parameters of Major Sub-Regions Within W49A}\label{tb:major}
\tablehead{\colhead{  }&
           \colhead{  }&
           \colhead{  }&
           \colhead{  }&
           \multicolumn{2}{c}{${\rm 20\mu{m}}$}&
           \multicolumn{2}{c}{${\rm 37\mu{m}}$}\\
           \cmidrule(lr){5-6} \cmidrule(lr){7-8}
           \colhead{ Source }&
           \colhead{ R.A. } &
           \colhead{ Dec. } &
           \colhead{ $R_{\rm int}$ } &
           \colhead{ $F_{\rm int}$ } &
           \colhead{ $F_{\rm int-bg}$ } &
           \colhead{ $F_{\rm int}$ } &
           \colhead{ $F_{\rm int-bg}$ } \\
	   \colhead{  } &
	   \colhead{ J2000 } &
	   \colhead{ J2000 } &
	   \colhead{ ($\arcsec$) } &
	   \colhead{ (Jy) } &
	   \colhead{ (Jy) } &
	   \colhead{ (Jy) } &
	   \colhead{ (Jy) }
}
\startdata
         AA & 19 10 07.6 & 9 05 45.6 & 15.0 &  33.5 &   9.57 & 106 &  58.4  \\
         BB & 19 10 10.4 & 9 05 47.7 & 14.2 &  36.2 &   8.93 & 165 &  50.3  \\
         CC & 19 10 11.6 & 9 07 07.8 & 14.5 &   112 &   74.8 & 385 &   246  \\
         DD & 19 10 11.6 & 9 06 35.1 & 12.0 &  58.5 &   29.3 & 280 &   147  \\
    G Region& 19 10 13.4 & 9 06 12.6 & 11.2 &  72.7 &   44.7 &4110 &  3720  \\
         HH & 19 10 14.0 & 9 05 48.6 & 12.0 &  52.5 &   17.8 & 369 &   168  \\
         II & 19 10 14.9 & 9 06 44.7 & 11.1 &  32.1 &   1.56 & 286 &  41.9  \\
         JJ & 19 10 18.7 & 9 06 07.8 & 18.6 &   228 &    123 & 992 &   417  \\
  L+M Region& 19 10 15.0 & 9 06 20.7 & 11.9 &  75.4 &   23.6 & 968 &   338  \\
         LL & 19 10 22.7 & 9 04 19.2 & 17.5 &  70.6 &   21.1 & 189 &  98.7  \\
         MM & 19 10 23.3 & 9 06 00.6 & 18.5 &   119 &   51.0 & 505 &   283  \\
    O Region& 19 10 16.6 & 9 06 10.8 & 13.1 &  96.7 &   33.7 & 724 &   390  \\
  R+Q Region& 19 10 10.6 & 9 05 14.7 & 15.4 &   127 &   96.2 & 903 &   776  \\
          S & 19 10 11.7 & 9 05 27.0 & 10.2 &  95.9 &   70.6 & 493 &   305  \\
   W49\,South & 19 10 22.1 & 9 05 01.1 & 25.5 &  445 &    350 &2440 &  2180  \\
W49\,Southwest$\dagger$	&	19 10 11.28	&	9 05 15.6		&	21.5   &	223	&	180	&	1420	&	1280		\\
W49\,North$\dagger$	&	19 10 15.94	&	9 06 30.8		&	69.0	&	1190	&	812	&	11500	&	10100		\\
\enddata
\tablecomments{\scriptsize R.A. and Dec. are for the center of apertures used, not the source peaks. $R_{\rm int}$ gives the size of the radius used for aperture photometry. $F_{\rm int}$ indicates total flux inside the aperture. $F_{\rm int-bg}$ is for background subtracted flux.}
\tablenotetext{\dagger}{W49\,North and W49\,Southwest are not used in the extended region analyses, but are included here for completeness. W49\,Southwest is the combined S, R and Q regions. W49\,North is basically the entire central region of W49A, not including W49\,Southwest or W49\,South.}
\end{deluxetable*} 
 
\begin{deluxetable*}{rccccccc}
\tabletypesize{\small}
\tablecolumns{8}
\tablewidth{0pt}
\tablecaption{Derived Parameters of Major Sub-Regions in W49A}\label{tb:sede}
\tablehead{\colhead{   Source   }                                              &
           \colhead{  $M_{\rm vir}$   } &
           \colhead{  $M$   } &
           \colhead{ $L$ } &
           \colhead{  $T_{\rm cold}$  } &
           \colhead{  $T_{\rm warm}$  } &
           \colhead{  $L/M$  } &
           \colhead{ $\alpha_{\rm vir}$ }\\
	   \colhead{        } &
	   \colhead{ ($M_{\sun}$) } &
	   \colhead{ ($M_{\sun}$) } &
	   \colhead{ ($\times 10^4 L_{\sun}$) } &
	   \colhead{ (K) } &
	   \colhead{ (K) } &
       \colhead{  $L_{\sun}/M_{\sun}$  } &
       \colhead{     }
}
\startdata
               AA &   2312.70 &    679.20 &      6.1 &     55.5 & 272.6 &  \textit{44.8}$\dagger$ &   3.4\\
               BB &   3061.80 &   1245.70 &     16.1 &     43.3 & 297.1 &  \textit{64.7}$\dagger$ &   2.5\\
               CC &   2323.70 &    763.80 &     26.5 &     61.4 & 269.8 & 173.6 &   3.0\\
               DD &   2252.90 &    875.40 &     19.5 &     54.7 & 266.1 & 111.5 &   2.6\\
         G Region &   4477.10 &   5760.50 &    622.0 &     71.4 & 245.3 & 539.5 &   0.8\\
               HH &   3092.80 &    807.90 &     32.8 &     47.3 & 260.0 & 202.9 &   3.8\\
               II &   3674.70 &   1621.60 &     26.9 &     38.4 & 365.7 &  83.0 &   2.3\\
               JJ &   2374.30 &   2510.30 &     68.8 &     53.3 & 252.4 & 137.0 &   0.9\\
       L+M Region &   3151.00 &   2266.10 &     82.8 &     48.0 & 282.2 & 182.8 &   1.4\\
               LL &   1056.30 &    813.30 &     19.4 &     47.5 & 262.6 & 119.0 &   1.3\\
               MM &   2107.30 &    826.50 &     28.8 &     58.9 & 261.8 & 174.6 &   2.5\\
         O Region &   3341.40 &   1459.50 &     51.5 &     54.8 & 282.7 & 176.3 &   2.3\\
       R+Q Region &   2796.20 &   1053.60 &     59.3 &     74.8 & 296.0 & 281.6 &   2.7\\
                S &   2174.70 &    622.30 &     26.3 &     82.3 & 261.6 & 211.5 &   3.5\\
         W49South &   5777.70 &   3241.60 &    164.0 &     78.9 & 271.9 & 253.6 &   1.8\\       
\enddata
\tablenotetext{\dagger}{\scriptsize As discussed in Section\,\ref{sec:other}, AA and BB are likely not star-forming clumps. Since the the $L/M$ analysis only holds for star-forming clumps, these values do not accurately represent the evolutionary state of these sources.}
\end{deluxetable*} 
 
\subsection{Physical Properties of Extended Sources: Kinematic Status and Global History}\label{sec:es}

\subsubsection{The Relative Evolutionary States of the Sub-Regions of W49A}\label{sec:alm}

In \citetalias{2019ApJ...873...51L} and \citetalias{2020ApJ...888...98L}, we compared two independent tracers of molecular clump evolution, the virial parameter ($\alpha_{\rm vir}$) and the luminosity-to-mass ratio ($L/M$), of the larger extended sub-regions in W51A and M17. We assume these large and extended radio continuum sub-regions are candidates for being star-forming clumps (rather than individual cores) housing embedded (proto)clusters of massive stars that are ionizing the extended \ion{H}{2} regions seen in radio continuum. The higher $\alpha_{\rm vir}$ as well as $L/M$ values are assumed to demonstrate relatively older proto/young stellar clusters, and plotting the $\alpha_{\rm vir}$ vs. $L/M$ parameters for these sources in W51A and M17 yielded a relatively linear correlation. We repeated this same analysis here, this time toward the radio-defined extended sources of W49A. 

We identified 15 extended radio continuum regions throughout W49A by identifying separate regions in the 37 and 70\,$\mu$m maps that correlate with the major radio continuum regions identified by \citet{1997ApJ...482..307D} in their 3.6\,cm map. As seen in Figure~\ref{fig:radio}, separate regions identified in the radio continuum maps do not always appear as separate sources in the infrared. We thus had to group some closely positioned radio continuum sources/regions into one region for study. In most cases, these radio regions do share a common diffuse radio continuum envelope as seen in the 3.6\,cm map (i.e., Figure~\ref{fig:SOFIAdata}a). Specifically, we group most of the western Welch Ring sources into one source called the ``G Region'',  we group together the bright L and M sources into the ``L+M Region'', all of the O sources (i.e., O, O$_2$, and O$_3$) into the ``O Region'', and the R and Q complexes into the ``R+Q Region''. We tabulate the observational parameters from the \textit{SOFIA} data for the major sub-regions within W49A in Table\,\ref{tb:major}. For all extended sources, the value of the aperture to used for the photometry was determined by looking at the cm radio maps and finding an aperture that encompasses all of the cm emission from each source as well as the extended dust emission as seen in the SOFIA 20 and 37\,$\mu$m maps and determining the smallest aperture radius that would encompass each source at all of these wavelengths. These radii were used for all photometry performed on data at SOFIA and Spitzer wavelengths (i.e. $R_{int}$ in Tables\,\ref{tb:major} and \ref{tb:cpsx}). Descriptions of how we determine the apertures for each source in the {\it Spitzer} and {\it Herschel} data are detailed in Appendix\,\ref{sec:appendixflux}. 

\begin{figure}
\begin{center}
\includegraphics[width=3.3in]{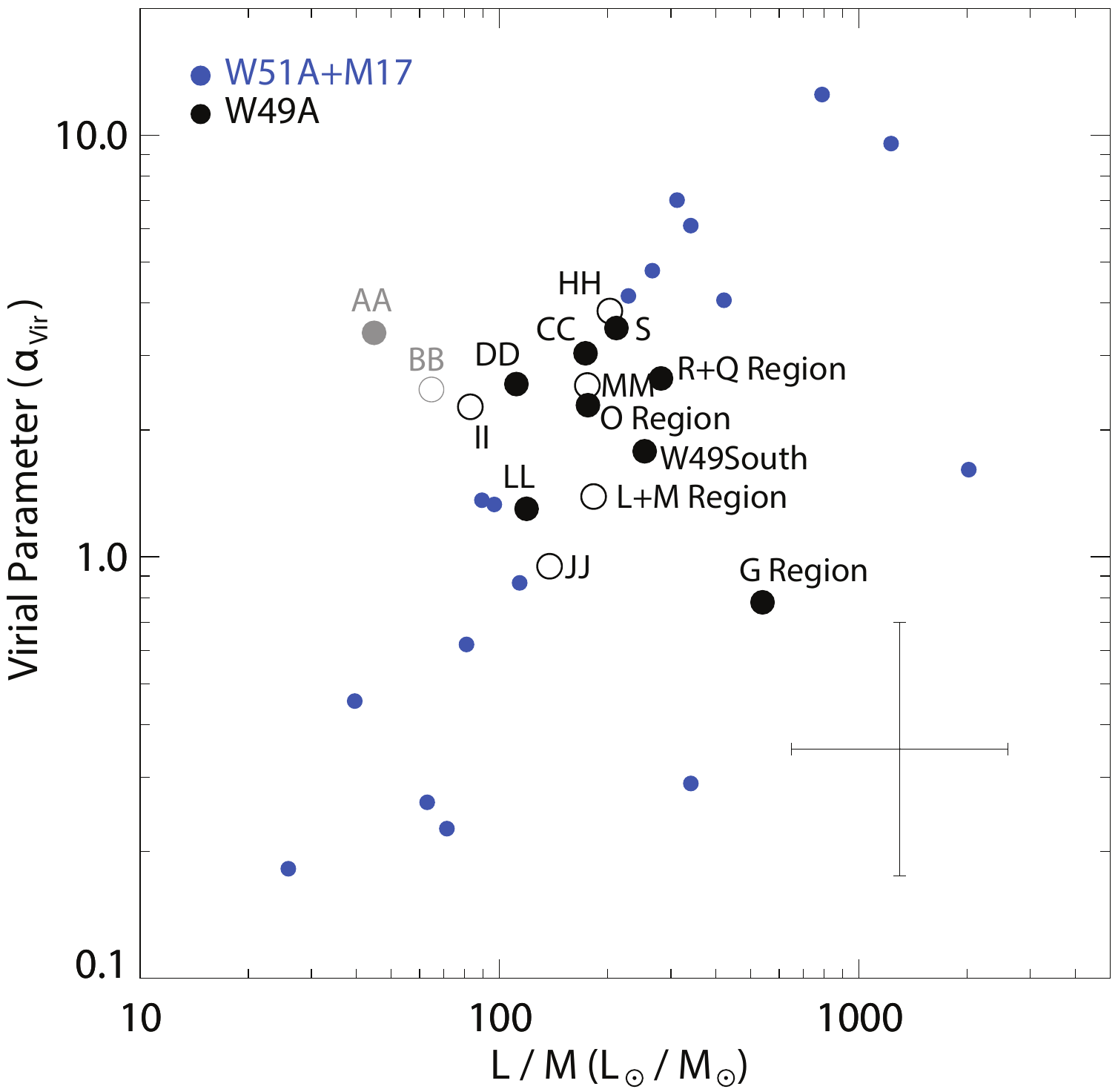}\\
\end{center}
\caption{\footnotesize $\alpha_{\rm vir}$ vs. $L/M$ of extended sources in W49A. $L/M$ is calculated based on the infrared data where the background levels are determined from the outer annuli of individual sources for the filled circles. Backgrounds of open circles are estimated from the central region of W49A. Sources AA and BB are shown here in gray for completeness, but they are likely not star-forming clumps.}
\label{fig:fd3c}
\end{figure}

To perform our evolutionary analyses, first the masses of each major extended region were derived based on the pixel-by-pixel graybody fitting method of \citet{2016ApJ...829L..19L} where the far-infrared data from {\it Herschel} and JCMT were used to determine the cold dust components. Then the bolometric luminosities, $L$, for these extended sources were calculated based on a two temperature graybody fit with the integrated total fluxes of each source and each band (see Tables~\ref{tb:major}, as well as Appendix~\ref{sec:appendixflux}). For both graybody fitting methods, the background fluxes of well-resolved extended sources (filled circles of Figure~\ref{fig:fd3c}) were estimated from the immediate outer annuli of each source. For sources that were not well-resolved or displayed strong nearby environmental emission, we calculated the background fluxes based on an average from representative areas in the center of W49A that were relatively free of bright source emission. Since the background subtraction and, thus, parameters derived from their fluxes are less certain for these sources, we plot them with a different symbol (open circles) in Figure~\ref{fig:fd3c}.  Also, following the techniques of our previous papers, we utilized the $^{13}$CO(1-0) data of FOREST Unbiased Galactic plane Imaging survey with the Nobeyama 45 m telescope \citep[FUGIN,][]{2017PASJ...69...78U} in order to derive the kinematic property, $\alpha_{\rm vir}$, for each of these extended sources. We summarize in Table~\ref{tb:sede} the physical parameters we derived for each extended source under the assumption that they are each a star-forming clump, i.e., the virial mass ($M_{\rm vir}$), clump mass ($M$), bolometric luminosity ($L$), the derived warm and cold temperature components ($T_{\rm cold}$ and $T_{\rm warm}$), the luminosity-to-mass ratio ($L/M$), and the viral parameter ($\alpha_{\rm vir}$).

As shown in Table~\ref{tb:sede}, the extended sources in W49A have masses spanning from $622\,M_{\sun}$ to $5760\,M_{\sun}$ while the mean mass, $\overline{M}$, is $\sim1636\,M_{\sun}$. For comparison, the $\overline{M}$ of both M17 and W51A are larger at $\sim2100\,M_{\sun}$ and $\sim3500\,M_{\sun}$, respectively. Furthermore, the mass range of the extended sources in M17 ($\sim20\,M_{\sun}<M<4340\,M_{\sun}$) skews lower than W49A, while the W51A mass range skews higher ( $\sim107\,M_{\sun}<M<9930\,M_{\sun}$). Additionally, the mass range of extended sources in W49A also has less scatter than M17 and W51A, with a standard deviation of the clump masses of only $\sim1378\,M_{\sun}$ (compared to $\sim2090\,M_{\sun}$ for M17 and $\sim3510\,M_{\sun}$ for W51A).

The level of the kinematic stability within G\ion{H}{2} regions can be inspected via virial analysis of the individual extended sources they contain. For instance, in \citetalias{2019ApJ...873...51L}, we derived the $\alpha_{\rm vir}$ values for 13 radio-defined extended sources in W51A. The calculated values of $\alpha_{\rm vir}$ derived for W51A spanned a range between 0.18 and 12.5. We found 8 of them are gravitationally bound ($\alpha_{\rm vir}<2$) while 6 of them were self-collapsing ($\alpha_{\rm vir}<1$). The other 5 sources were extremely unbound ($\alpha_{\rm vir}\gg2$). The large variation of these $\alpha_{\rm vir}$ values for W51A indicate that multi-generational star formation is occurring within the G\ion{H}{2} region, consistent with results from previous studies. Similarly, the extended sources in M17 also showed a large spread of $\alpha_{\rm vir}$ spanning from 0.29 to 9.56. However, as we can see from Table~\ref{tb:sede}, W49A has a $\alpha_{\rm vir}$ range between 0.8 to 3.8, where only two sub-regions (the G Region and JJ) have a value of $\alpha_{\rm vir} < 1$ consistent with presently undergoing self gravitational collapse. Three sources (LL, W49\,South, and the L+M Region) are expanding but still gravitationally bound ($ 2> \alpha_{\rm vir} > 1$) and all of the other the sources appear to be gravitationally unbound ($\alpha_{\rm vir} > 2$). However, we caution against over-interpretation of the data in the case of W49A, since even the sub-region with the highest virial parameter (i.e., $\alpha_{\rm vir} = 3.8$) is within the estimated factor of two error (see the error bar in Figure~\ref{fig:fd3c}) of being gravitationally bound (i.e., $\alpha_{\rm vir} < 2$). This again, is in contrast to our results of W51A and M17 which showed extended sub-regions with $\alpha_{\rm vir}$ values firmly categorized as self collapsing even when taking into account the factor of two error (i.e., having $\alpha_{\rm vir}<0.5$) and firmly categorized as gravitationally unbound (i.e., having $\alpha_{\rm vir}>4$).

Although the virial analysis tentatively shows the evolution and the kinematic state of each molecular clump simultaneously, the $L/M$ would show the evolutionary state of the clump only \citep[e.g.,][]{2007ApJ...654..304K}. Excluding extended radio regions AA and BB (because our $L/M$ analysis only holds for star-forming clumps; see Section~\ref{sec:other}), the minimum and maximum $L/M$ values of the extended sources in W49A are $83\,L_{\sun}/M_{\sun}$ and $540\,L_{\sun}/M_{\sun}$, respectively. For the comparison, W51A and M17 showed the ranges $26 \lesssim{L/M} \lesssim 800\,L_{\sun}/M_{\sun}$ and $300 \lesssim {L/M} \lesssim 2000\,L_{\sun}/M_{\sun}$, respectively. Note that \citet{2013ApJ...779...79M} inspected the $L/M$ of the 303 mid- to high-mass star forming clumps of Milky Way from their unbiased CO surveys where the clumps showed $0.1 \lesssim {L/M} \lesssim 1000\,L_{\sun}/M_{\sun}$ across entire evolutionary states. 

As we did in our previous studies, we plot the $\alpha_{\rm vir}$ vs. $L/M$ for the star-forming clumps in W49A in Figure~\ref{fig:fd3c} where we place the clumps of W51A and M17 as well. The plot clearly shows that the relative age spread of the clumps in W49A is smaller than W51A and M17 cases. It also shows that the clumps in W49A tend to be absent of an extremely young (i.e., sources in the lower left of the plot) population compared to W51A and M17. The small overall spread in the data points on this plot appears to indicate the star-forming clumps in W49A tend be be more coeval, unlike the cases of W51A and M17. 

However, there is one source in Figure~\ref{fig:fd3c} that stands out. The source with the lowest virial parameter, the G region, contradictorily has the highest $L/M$ value. Consistent with its low virial parameter value, the G region is very likely a youthful region, given that it is in the area of highest extinction in W49A and houses the highest concentration of MYSOs in the entire G\ion{H}{2} Region. The other two blue points on the graph in the lower right that accompany the G region are the northern and southern bar sources of M17. Our conjecture for M17 was these regions are further subject to intense external heating and ionization that could skew the $L/M$ values higher than expected for a self luminous source alone. This could be the case with the G region of W49A, since it is located in an area with the highest level of environmental radio continuum emission. 

\subsubsection{The History of Stellar Cluster Formation in W49A}\label{sec:hist}

There have been many studies looking into how star formation proceeded within W49A, with contradictory conclusions. \citet{2003ApJ...589L..45A} and \citet{2005A&A...430..481H} used a near-infrared color-color diagram analysis to deduce the relative age distribution of the stellar clusters and UC\ion{H}{2} regions and concluded that W49A experienced independent multi-seeded star formation activities. Based on the expansion ages of the UC\ion{H}{2} regions within W49A, \citet{2002ApJ...564..827C} suggested that the star formation of W49A started from the periphery then moved to the central region (i.e. the Welch Ring). \citet{2010A&A...520A..84P}, however, suggested expanding shells centered within the Welch ring triggered the star formation in the other regions of W49A so that the direction of the sequential star formation would be opposite to what \citet{2002ApJ...564..827C} suggested. \citet{2019A&A...622A..48R} studied the kinematic properties of the radio recombination lines \citep[as a part of THOR survey;][]{2018PASP..130k4301W} in W49A and concluded that it has undergone expanding and re-collapsing events where these event formed multiple generations of young stellar clusters. It is worthy to note that they claimed the relative ages among the different generations of the stellar clusters were very small. 

In \citetalias{2019ApJ...873...51L}, we suggested that the sequential triggering star formation would be easily recognized in the $\alpha_{\rm vir}$ vs. $L/M$ plot with a simple linear correlation so that the younger clumps/clusters are located at lower-left (low $\alpha_{\rm vir}$ and $L/M$) and the older ones would be opposite. Then by inspecting how these young and old regions are distributed within the GHII region one could speculate as to whether the data are consistent with triggering scenarios or not. As shown in Figure~\ref{fig:fd3c}, the data points for the sub-regions in W49A do not extend to the very young or very old corners of the plot, but are instead grouped fairly tightly together near the center of the plot. This seems to indicate that, globally, the star-forming sub-regions within W49A are relatively coeval given the small spread in both evolutionary indicators, $L/M$ and $\alpha_{\rm vir}$. This implies non-sequentially triggered, multi-seeded star formation activity throughout W49A, as suggested by \citet{2003ApJ...589L..45A} and \citet{2005A&A...430..481H}. The coeval nature of all extended sub-regions in W49A would be inconsistent with sequentially triggered (i.e. longer timescale) formation scenarios like both the outside-in formation scenario \citep[i.e.,][]{2002ApJ...564..827C} and the inside-out formation scenario \citep[i.e.,][]{2010A&A...520A..84P}.  

There is an additional origin scenario that states the global star formation throughout W49A might have been initiated due to a cloud-cloud collision occurring along our line of sight to the region. This scenario was originally suggested by \citet{1977ApJ...212..664M} based on molecular and recombination line observations, with follow-up studies supporting the idea \citep[e.g.,][]{1996MNRAS.281..294B, 1993ApJ...413..571S}. Unfortunately, our data cannot prove such a scenario, and more direct evidence of the cloud-cloud collision scenario will need targeted observations, such as comparing velocity profiles of the 158\,$\mu$m [\ion{C}{2}] line emission and CO isotopologue bands \citep{2018MNRAS.478L..54B, 2021PASJ...73S.239L}. However, such a global-triggering scenario could explain why so many sub-regions spread throughout such a large volume could have begun star-formation activities at the same time.

\section{Summary}\label{sec:sum}

In this, our third paper from our mid-infrared imaging survey of Milky Way Giant \ion{H}{2} regions, we obtained \textit{SOFIA}-FORCAST 20 and 37\,$\micron$ maps toward W49A, covering the entire the infrared-emitting area of the region at $\sim$3$\arcsec$ spatial resolution. The 37\,$\mu$m images are the highest spatial resolution infrared observations of W49A yet obtained at wavelengths beyond 25\,$\mu$m. We compared these \textit{SOFIA}-FORCAST images with previous multi-wavelength observations from the near-infrared to radio wavelengths from various ground- and space-based telescopes in order to inspect the morphological and physical properties of the compact and extended sources within in W49A.  We itemize below our main conclusions from this study.  

1) The infrared observations from \textit{SOFIA} show the region to have structured but extended dust emission spread over a $\sim$5.0$\arcmin\times 3.5\arcmin$ area, which corresponds generally to the extent of the cm radio continuum emission seen by \citet{1997ApJ...482..307D}. While most of the infrared features are resolved at both \textit{SOFIA} wavelengths, the dust is more pronounced and extended at 37\,$\mu$m compared to 20\,$\mu$m, indicative of widespread cool dust.

2) The most well-known feature in W49A is the ring of radio continuum sources referred to as the Welch Ring. We do not detect 20\,$\mu$m emission coming from the locations of radio sources on the western side of this ring, namely sources A, B, B$_1$, D, or E. However, in the deconvolved 37\,$\mu$m image there is extended emission towards the locations of B$_1$, D, and E, and while it is possible that this is simply unresolved emission from the very bright G source, contributions to the emission in these areas due to sources B$_1$, D, and/or E cannot be ruled out. It is presumed that the reason for the non-detections in the infrared is because the extinction is higher on the western side of the ring of radio sources. We determined the radio spectral slope of the Welch Ring radio sources and discovered several of the sources in the eastern side have radio emission indicative of non-thermal synchrotron emission (likely from ionized jets/outflows), rather than free-free Bremsstrahlung emission from UC\ion{H}{2} regions. 

3)  The two sources with the largest derived luminosities are W49~South and source G. W49~South is the brightest peak in all of W49A in the infrared from $\sim$3\,$\mu$m out to $\sim$20\,$\mu$m. However, our data show that at 37\,$\mu$m and longer wavelengths (as seen by \textit{Herschel}) it becomes the second brightest peak to source G. This suggests that, while both contain young MYSOs, G is likely at a relatively earlier and more embedded phase of evolution. 

4) We have identified 10 new sources from the \textit{SOFIA} data: CC-1, DD-1, II-1, JJ-1, KK-1, MM-1, P-1, Q-1, R$_4$, and W49\,South-2. 

5) Though we had the sensitivity to detect them, we do not detect mid-infrared emission from previously identified mid-infrared sources BB East, DD South, EE East, and HH West \citep{2000ApJ...540..316S}. These sources are also not seen in the \textit{Spitzer}-IRAC data  and therefore might not be real sources.

6) Radio sources AA and BB are large ($d>1$\,pc), extremely diffuse radio continuum regions with faint ring or bubble-like appearances, relatively low infrared-derived luminosities, and possess the lowest radio emission measures of any sub-region within W49A. They both have a stellar source centrally located within their extended radio continuum regions, which may be singularly responsible for their ionization. We suggest that, based upon these data, AA and BB are evolved \ion{H}{2} regions and not star-forming sub-regions.

7) We  performed SED modeling on 24 identified compact infrared sources using photometry of {\it SOFIA}, {\it Spitzer}, and {\it Herschel} data. We found 22 sources satisfying our criteria of housing a MYSO, and we determined 23 sources to be either a MYSO or pMSYO ($\sim96\%$). In our two previously studied G\ion{H}{2} regions, $87\%$ of the point sources were found to likely be MYSOs or pMYSOs in W51A region ($d\sim5.4$\,kpc), while M17 ($d\sim1.98$\,kpc) showed only $44\%$ sources as MYSOs and pMYSOs. We suggest that the main reasons for detecting such a high fraction of MYSOs with \textit{SOFIA} is due to the combination of the much larger extinction towards W49A and its extreme distance (11.1\,kpc), which would make less massive (and therefore less luminous) YSOs difficult to detect.

8) While almost half of the sources determined to be MYSOs for W51A have no detectable radio continuum emission, that is the case for only four MYSOs (II-1, KK-1, Q-1, and W49\,South-2) in this W49A study. This is likely a consequence of the much larger distance and extinction towards W49A. 

9) It has been speculated that W49A may be a relatively young G\ion{H}{2} region overall \citep{1987Sci...238.1550W}. Our evolutionary analyses show that W49A appears to be absent of an extremely young population compared to W51A and M17. This appears to indicate that W49A is neither an extremely young or old G\ion{H}{2} region globally.  Our evolutionary analyses also show that the relative age spread of the star-forming clumps in W49A is smaller than the W51A/M17 cases, and thus are much more coeval. This coeval nature of the extended sub-regions in W49A is inconsistent with internally-triggered, sequential formation scenarios, and may be more consistent with a single global triggering event.

\acknowledgments
This research is based on observations made with the NASA/DLR Stratospheric Observatory for Infrared Astronomy (\textit{SOFIA}). \textit{SOFIA} is jointly operated by the Universities Space Research Association, Inc. (USRA), under NASA contract NAS2-97001, and the Deutsches \textit{SOFIA} Institut (DSI) under DLR contract 50 OK 0901 to the University of Stuttgart. This work is also based in part on archival data obtained with the \textit{Spitzer Space Telescope}, which is operated by the Jet Propulsion Laboratory, California Institute of Technology under a contract with NASA. This work is also based in part on archival data obtained with \textit{Herschel}, an European Space Agency (ESA) space observatory with science instruments provided by European-led Principal Investigator consortia and with important participation from NASA. Financial support for this work was provided by NASA through \textit{SOFIA} awards 05\_0008 and 06\_0011 issued by USRA. 

\vspace{5mm}
\facility{\textit{SOFIA}(FORCAST)}

\clearpage

\appendix
\section{Data release}

The FITS images used in this study are publicly available at: {\it https://dataverse.harvard.edu/dataverse/SOFIA-GHII}. 

The data include the \textit{SOFIA} FORCAST 20 and 37\,$\mu$m final image mosaics of W49A and their exposure maps.

\section{Additional Photometry of Compact and Extended Sources in W49A}\label{sec:appendixflux}

In addition to the fluxes derived from the \textit{SOFIA}-FORCAST data, we used some additional photometry data in our SED analyses from \citet{2000ApJ...540..316S}, as well as measured fluxes for our sources from both \textit{Spitzer}-IRAC and \textit{Herschel}-PACS.

\subsection{Compact Sources}

As mentioned in \S\ref{sec:data}, we performed optimal extraction photometry for the FORCAST 20 and 37\,$\mu$m images to define the location of all compact sources, and to determine the aperture radii to be used for photometry. Using these source locations, we employed the optimal extraction technique on the \textit{Spitzer}-IRAC 8\,$\mu$m data for all sources to find the optimal aperture to be used for all IRAC bands (since the source sizes are typically similar or smaller at the shorter IRAC bands). As we have done for the FORCAST images, we estimated the background emission from the annuli that showed the least contamination from nearby sources, i.e. showing relatively flat radial intensity profile (\S\ref{sec:data}). Table\,\ref{tb:cpsb} shows the photometry values we derive for all compact sources from the \textit{Spitzer}-IRAC bands.

Table\,\ref{tb:cpsc} shows the photometry result for the \textit{Herschel}-PACS bands for the compact sources. In general, this aperture size cannot be determined accurately using the optimal extraction technique due to the ubiquity of extended emission from nearby sources that are overlapping the source being measured. Indeed, only one compact source (GG) could be resolved in PACS band images, so for all other sources we use fixed aperture radii for both PACS bands ($R_{\rm int}$=16$\farcs$0 for 70\,$\mu$m and $R_{\rm int}$=22$\farcs$5 for 160\,$\mu$m). We compared our aperture sizes to those typically used in the Hi-GAL Compact Source Catalogue \citep{2017MNRAS.471..100E,2016A&A...591A.149M}. That catalogue employs aperture sizes comparable to the ones we used in this study. We therefore believe that the fixed aperture sizes we employ here are reasonable, especially since the data are only being used to provide upper limits to our SED model fits in most cases. 

\subsection{Extended Sources}

Table \,\ref{tb:cpsx} shows the \textit{Spitzer}-IRAC photometry values for all extended sources. We performed a color-color analysis similar to that in Figure\,\ref{fig:ccd} to determine that all sources are PAH contaminated. Thus all flux values at 3.6, 5.8, and 8.0\,$\mu$m are considered upper limits. The uncontaminated 4.5\,$\mu$m flux values were used as nominal data points and background subtraction was performed only for these data (see Table\,\ref{tb:cpsx}). 

The \textit{Herschel}-PACS photometry of extended sources can be found in Table~\ref{tb:cpsy} and Table~\ref{tb:cpsZ}. As we already note from compact source photometry, there are certain areas covered by saturated pixels in some bands. For the extended sources where the area of saturated pixels was less than 10\% of the photometric aperture area, we performed simple 2D Gaussian fitting of the sources to estimate flux values for each saturated pixel. If more than 10\% of area of an extended source is covered by saturated pixels, we did not derive a flux value for that band and therefore exclude those data from the SED fitting (Table~\ref{tb:cpsy}). For regions that were resolved well enough at 70 and or 160\,$\mu$m in the {\it Herschel} data, we found the determined the best aperture for each source given its radial profile and derived flux measurements with background subtraction (Table\,\ref{tb:cpsZ}) and used these values in the evolutionary analyses in Section \,\ref{sec:alm}. For unresolved sources, we used fixed apertures of 16$\arcsec$ for 70\,$\mu$m, 22.5$\arcsec$ at 160\,$\mu$m, 30$\arcsec$ at 250\,$\mu$m, 40$\arcsec$ at 350\,$\mu$m, and 57.5$\arcsec$ at 500\,$\mu$m.

\begin{deluxetable*}{rrrrrrrrrr}
\tabletypesize{\small}
\tablecolumns{8}
\tablewidth{0pt}
\tablecaption{{\it Spitzer}-IRAC Observational Parameters of Compact Sources in W49A}\label{tb:cpsb}
\tablehead{\colhead{  }&
           \colhead{  }&
           \multicolumn{2}{c}{${\rm 3.6\mu{m}}$}&
           \multicolumn{2}{c}{${\rm 4.5\mu{m}}$}&
           \multicolumn{2}{c}{${\rm 5.8\mu{m}}$}&
           \multicolumn{2}{c}{${\rm 8.0\mu{m}}$}\\
           \cmidrule(lr){3-4} \cmidrule(lr){5-6} \cmidrule(lr){7-8} \cmidrule(lr){9-10}\\
           \colhead{ Source }&
           \colhead{ $R_{\rm int}$ } &
           \colhead{ $F_{\rm int}$ } &
           \colhead{ $F_{\rm int-bg}$ } &
           \colhead{ $F_{\rm int}$ } &
           \colhead{ $F_{\rm int-bg}$ } &
           \colhead{ $F_{\rm int}$ } &
           \colhead{ $F_{\rm int-bg}$ } &
           \colhead{ $F_{\rm int}$ } &
           \colhead{ $F_{\rm int-bg}$ } \\
	   \colhead{  } &
	   \colhead{ ($\arcsec$) } &
	   \colhead{ (Jy) } &
	   \colhead{ (Jy) } &
	   \colhead{ (Jy) } &
	   \colhead{ (Jy) } &
	   \colhead{ (Jy) } &
	   \colhead{ (Jy) } &
	   \colhead{ (Jy) } &
	   \colhead{ (Jy) } 
}
\startdata
      CC-1 &  2.4 &   0.0134 &   0.0034 &   0.0206 &   0.0050 &   0.1117 &   0.0199 &   0.3242 &   0.0678  \\
      DD-1 &  3.0 &   0.0125 &   0.0012 &   0.0150 &   0.0020 &   0.1883 &   0.0175 &   0.5248 &   0.0581  \\
      EE &  4.8 &   0.0271 &   0.0127 &   0.0447 &   0.0241 &   0.3525 &   0.1637 &   0.9083 &   0.3928  \\
         F &  3.0 &   0.1731 &   0.1626 &   0.3301 &   0.3130 &   0.6908 &   0.5603 &   1.5504 &   1.1992  \\
G\textsubscript{tot} &  2.4 &   0.0107 &   0.0036 &   0.1424 &   0.1259 &   1.3734 &   1.1827 &   2.5986 &   2.1640  \\
        GG &  3.8 &   $<$0.0095 &      \nodata &   $<$0.0035 &      \nodata &   $<$0.1130 &      \nodata &   $<$0.0400 &      \nodata  \\
         I &  2.4 &   0.0110 &   0.0011 &   0.0205 &   0.0056 &   0.1467 &   0.0281 &   0.3718 &   0.0816  \\
      II-1 &  3.6 &   0.0074 &   0.0023 &   0.0160 &   0.0041 &   0.1559 &   0.0307 &   0.4208 &   0.0756  \\
         J &  2.4 &   0.0115 &   0.0039 &   0.0397 &   0.0269 &   0.2782 &   0.1100 &   0.5491 &   0.2624  \\
J$_{1}$+J$_{2}$ &  4.2 &   0.0380 &   0.0120 &   0.0570 &   0.0150 &   0.2083 &   0.0339 &   0.7562 &   0.0988  \\
      JJ-1 &  3.0 &   0.0222 &   0.0014 &   0.0342 &   0.0067 &   0.2004 &   0.0117 &   0.3876 &   0.0306  \\
        KK &  3.6 &   0.0243 &   0.0096 &   0.0493 &   0.0210 &   0.3705 &   0.1581 &   2.8366 &   0.7992  \\
      KK-1 &  2.4 &   0.0147 &   0.0039 &   0.0235 &   0.0156 &   0.1106 &   0.0367 &   0.2560 &   0.0500  \\
      MM-1 &  6.0 &   0.0381 &   0.0151 &   0.0592 &   0.0255 &   0.3799 &   0.1392 &   1.1234 &   0.4616  \\
         M &  2.4 &   0.0126 &   0.0031 &   0.0265 &   0.0091 &   0.1657 &   0.0302 &   0.4667 &   0.1035  \\
         N &  3.0 &   $<$0.0260 &      \nodata &   0.0379 &   0.0025 &   0.2467 &   0.0058 &   $<$1.0296 &      \nodata  \\
         O &  2.4 &   0.0184 &   0.0013 &   0.0300 &   0.0025 &   0.1854 &   0.0069 &   0.5129 &   0.0091  \\
         P &  3.6 &   0.0330 &   0.0128 &   0.0584 &   0.0328 &   0.3758 &   0.1286 &   1.2071 &   0.5455  \\
       P-1 &  2.4 &   0.0089 &   0.0013 &   0.0126 &   0.0029 &   0.1701 &   0.0286 &   0.2934 &   0.0367  \\
       Q-1 &  2.4 &   $<$0.0065 &      \nodata &   $<$0.0093 &      \nodata &   0.0654 &   0.0044 &   0.1663 &   0.0085  \\
R\textsubscript{tot} &  3.6 &   0.3491 &   0.3108 &   1.0205 &   0.9094 &   2.9816 &   2.4747 &   5.1631 &   2.9199  \\
         S &  4.8 &   0.2527 &   0.1573 &   0.4104 &   0.3186 &   1.9483 &   1.0326 &   2.9391 &   1.2742  \\
  W49\,South & 16.2 &   1.0395 &   0.8972 &   3.2082 &   2.8473 &  15.4830 &  11.9295 &  28.5123 &  19.7979  \\
W49\,South-2 &  1.8 &   0.0080 &   0.0010 &   0.0058 &   0.0005 &   0.0645 &   0.0065 &   0.1929 &   0.0442  \\
\enddata
\end{deluxetable*}
 
\begin{deluxetable*}{rrrrr}
\tabletypesize{\small}
\tablecolumns{8}
\tablewidth{0pt}
\tablecaption{{\it Herschel}-PACS Observational Parameters of Compact Sources in W49A}\label{tb:cpsc}
\tablehead{\colhead{  }&
           \multicolumn{2}{c}{${\rm 70\mu{m}}$}&
           \multicolumn{2}{c}{${\rm 160\mu{m}}$}\\
           \cmidrule(lr){2-3} \cmidrule(lr){4-5} \\
           \colhead{ Source }&
           \colhead{ $R_{\rm int}$ } &
           \colhead{ $F_{\rm int}$ } &
           \colhead{ $R_{\rm int}$ } &
           \colhead{ $F_{\rm int}$ }  \\
	   \colhead{  } &
	   \colhead{ ($\arcsec$) } &
	   \colhead{ (Jy) } &
	   \colhead{ ($\arcsec$) } &
	   \colhead{ (Jy) } 
}
\startdata
                  CC-1 & 16.0 &   633 & 22.5 &  1120  \\
                  DD-1 & 16.0 &  1528 & 22.5 &  3166  \\
                    EE & 16.0 &  3016 & 22.5 &      \nodata  \\
                     F & 16.0 &  \nodata & 22.5 &      \nodata  \\
  G\textsubscript{tot} & 16.0 &  \nodata & 22.5 &      \nodata  \\
                    GG &  9.6  &   252\tablenotemark{\small{a}} & 22.5 &  3524  \\
                     I & 16.0 &      \nodata & 22.5 &      \nodata  \\
                  II-1 & 16.0 &  2180 & 22.5 &  3890  \\
                     J & 16.0 &      \nodata & 22.5 &      \nodata  \\
       J$_{1}$+J$_{2}$ & 16.0 &      \nodata & 22.5 &      \nodata  \\
                  JJ-1 & 16.0 &      879 & 22.5 &      1076  \\
                    KK & 16.0 &      \nodata & 22.5 &      \nodata  \\
                  KK-1 & 16.0 &   679 & 22.5 &  1011  \\
                     M & 16.0 &  6587 & 22.5 &      \nodata  \\
                  MM-1 & 16.0 &   367 & 22.5 &   595  \\
                     N & 16.0 &      \nodata & 22.5 &      \nodata  \\
                     O & 16.0 &  3261 & 22.5 &  3413  \\
                     P & 16.0 &  1633 & 22.5 &  2263  \\
                   P-1 & 16.0 &  1356 & 22.5 &  2058  \\
                   Q-1 & 16.0 &  1103 & 22.5 &  1357  \\
  R\textsubscript{tot} & 16.0 &  2302 & 22.5 &  1776  \\
                     S & 16.0 &  2239 & 22.5 &  1861  \\
              W49\,South & 16.0 &  3815 & 22.5 &  2688  \\
            W49\,South-2 & 16.0 &  3110 & 22.5 &  2674  \\
\enddata
\tablenotetext{a}{$F_{\rm int-bg}$ (nominal data point) since source GG could be resolved in PACS 70$\micron$ map. The $F_{\rm int}$ of 70$\micron$ is 731.86~Jy.}
\end{deluxetable*}

 \begin{deluxetable*}{rrrrrrrr}
\tabletypesize{\small}
\tablecolumns{8}
\tablewidth{0pt}
\tablecaption{{\it Spitzer}-IRAC Observational Parameters of Major Extended Sub-Regions in W49A}\label{tb:cpsx}
\tablehead{\colhead{  }&
           \colhead{  }&
           \colhead{${\rm 3.6\mu{m}}$}&
           \multicolumn{2}{c}{${\rm 4.5\mu{m}}^*$}&
           \colhead{${\rm 5.8\mu{m}}$}&
           \colhead{${\rm 8.0\mu{m}}$}\\
           \cmidrule(lr){4-5} \\
           \colhead{ Source }&
           \colhead{ $R_{\rm int}$ } &
           \colhead{ $F_{\rm int}$ } &
           \colhead{ $F_{\rm int}$ } &
           \colhead{ $F_{\rm int-bg}$ } &
           \colhead{ $F_{\rm int}$ } &
           \colhead{ $F_{\rm int}$ } \\
	   \colhead{  } &
	   \colhead{ ($\arcsec$) } &
	   \colhead{ (Jy) } &
	   \colhead{ (Jy) } &
	   \colhead{ (Jy) } &
	   \colhead{ (Jy) } &
	   \colhead{ (Jy) } 
}
\startdata
         AA & 15.0 &   0.151 &   0.196 &   0.093 &   1.41 &   4.34  \\
         BB & 14.2 &   0.280 &   0.344 &   0.164 &   2.62 &   7.22  \\
         CC & 14.5 &   0.773 &   0.893 &   0.644 &   3.79 &  10.7   \\
         DD & 12.0 &   0.258 &   0.392 &   0.213 &   2.64 &   7.39  \\
          G & 11.2 &   0.327 &   0.714 &   0.565 &   3.92 &   8.72  \\
         HH & 12.0 &   0.233 &   0.355 &   0.111 &   2.59 &   7.19  \\
         II & 12.3 &   0.150 &   0.225 &   0.107 &   2.19 &   6.00  \\
         JJ & 21.5 &   1.019 &   1.51  &   0.570 &   9.62 &  31.2   \\
        L+M & 11.9 &   0.317 &   0.540 &   0.237 &   3.88 &  10.9   \\
         LL & 17.5 &   0.229 &   0.318 &   0.150 &   2.43 &   6.90  \\
         MM & 21.5 &   0.522 &   0.779 &   0.359 &   5.25 &  15.4   \\
          O & 13.1 &   0.584 &   0.842 &   0.393 &   5.11 &  14.6   \\
        R+Q & 15.4 &   0.901 &   1.881 &   1.70  &   9.10 &  19.5   \\
          S & 10.2 &   0.442 &   0.745 &   0.455 &   3.70 &  10.2   \\
   W49South & 25.5 &   1.59  &   3.67  &   3.17  &  17.7  &  37.2   \\        
\enddata
\tablenotetext{*}{For {\it Spitzer}-IRAC data, only fluxes at 4.5\,$\mu$m are uncontaminated by PAH emission. Fluxes at all other wavelengths are used as an upper limits and therefore do not have any background subtraction performed.}
\end{deluxetable*}

\begin{deluxetable*}{rrrrrr}
\tabletypesize{\small}
\tablecolumns{6}
\tablewidth{0pt}
\tablecaption{{\it Herschel}-PACS Observational Parameters of Major Extended Sub-Regions in W49A (no background subtraction)}\label{tb:cpsy}
\tablehead{\colhead{  }&
           \colhead{ ${\rm 70\mu{m}}$ } &
           \colhead{ ${\rm 160\mu{m}}$ } &
           \colhead{ ${\rm 250\mu{m}}$ } &
           \colhead{ ${\rm 350\mu{m}}$ } &
           \colhead{ ${\rm 500\mu{m}}$ }  \\
           \colhead{ Source  } &
           \colhead{ $F_{\rm int}$ } &
           \colhead{ $F_{\rm int}$ } &
           \colhead{ $F_{\rm int}$ } &
           \colhead{ $F_{\rm int}$ } &
           \colhead{ $F_{\rm int}$ } \\
	   \colhead{  } &
	   \colhead{ (Jy) } &
	   \colhead{ (Jy) } &
	   \colhead{ (Jy) } &
	   \colhead{ (Jy) } &
	   \colhead{ (Jy) } 
}
\startdata
         CC &        1018 &        1532 &         292* &         229* &         247*  \\
         DD &        1453 &        1012 &        \nodata &       \nodata &         676*  \\
   G Region &       16676* &       10136* &      \nodata &       \nodata &         812*  \\
         HH &         834 &        1227 &        \nodata &       \nodata &         588*  \\
         II &        1365 &        1471 &        \nodata &       \nodata &         545*  \\
         JJ &        2103 &        1453 &         619 &         334 &         219  \\
 L+M Region &        3968 &        3041* &       \nodata &       \nodata &         762*  \\
         LL &        1129 &        1037 &         244 &         142 &          97.5  \\
         MM &        1029 &         698 &         275 &         131 &          98.4  \\
   O Region &        3321 &        1236 &        \nodata &         532 &         355*  \\
  R+Q Region&        1634 &         738 &         491 &         249 &         143*  \\
          S &        1263 &         795 &         467 &         262* &         200*  \\
 W49\,South &        5150 &        2836 &        1090* &         336 &         188  \\
\enddata
\tablecomments{Photometry values at 250 and 350\,$\mu$m are not given for sources DD, G Region, HH, II, and L+M Region since more than 10\% of the area within the source aperture are saturated pixels. The 250\,$\mu$m value for the O Region is also not given for the same reason.}
\tablenotetext{*}{These sources contain saturated pixels inside the photometry aperture covering less than 10\% of the area. Simple 2D Gaussian fitting have been performed to estimate the flux values of the pixels.}
\end{deluxetable*}
 
\begin{deluxetable*}{rrrrrr}
\tabletypesize{\small}
\tablecolumns{6}
\tablewidth{0pt}
\tablecaption{{\it Herschel}-PACS Observational Parameters of Major Extended Sub-Regions in W49A (background subtracted)}\label{tb:cpsZ}
\tablehead{\colhead{  }&
           \multicolumn{2}{c}{${\rm 70\mu{m}}$}&
           \multicolumn{2}{c}{${\rm 160\mu{m}}$}\\
           \cmidrule(lr){2-3} \cmidrule(lr){4-5} \\
           \colhead{ Source } &
           \colhead{ $R_{\rm int}$ } &
           \colhead{ $F_{\rm int-bg}$ } &
           \colhead{ $R_{\rm int}$ } &
           \colhead{ $F_{\rm int-bg}$ } \\
	   \colhead{  } &
	   \colhead{ ($\arcsec$) } &
	   \colhead{ (Jy) } &
	   \colhead{ ($\arcsec$) } &
	   \colhead{ (Jy) } 
}
\startdata
         CC &          19.2 &         501 &          27.0 &         581  \\
         DD &          16.0 &         508 &          22.5 &        \nodata  \\
   G Region &          25.6 &       13596 &          27.0 &        7852  \\
         LL &          16.0 &         768 &          22.5 &         757  \\
   O Region &          16.0 &        1725 &          22.5 &        \nodata  \\
 R+Q Region &          16.0 &        1343 &          18.0 &         723  \\
          S &          12.8 &         979 &          13.5 &         191  \\
 W49\,South &          22.4 &        4452 &          22.5 &        1689  \\
\enddata
\end{deluxetable*}


\clearpage

\begin{thebibliography}{}

\bibitem[Alves \& Homeier(2003)]{2003ApJ...589L..45A} Alves, J. \& Homeier, N.\ 2003, \apjl, 589, L45
\bibitem[Barbosa et al.(2016)]{2016ApJ...825...54B} Barbosa, C.~L., Blum, R.~D., Damineli, A., Conti, P.~S., \& Gusm{\~a}o, D.~M.\ 2016, \apj, 825, 54 
\bibitem[Becklin et al.(1973)]{1973ApL....13..147B} Becklin, E.~E., Neugebauer, G., \& Wynn-Williams, C.~G.\ 1973, \aplett, 13, 147
\bibitem[Bisbas et al.(2018)]{2018MNRAS.478L..54B} Bisbas, T.~G., Tan, J.~C., Csengeri,~T., et al. 2018, \mnras, 478, 54
\bibitem[Buckley \& Ward-Thompson (1996)]{1996MNRAS.281..294B} Buckley, H.~D., \& Ward-Thompson, D.\ 1996, \mnras, 281, 294
\bibitem[Conti \& Blum(2002)]{2002ApJ...564..827C} Conti, P.~S. \& Blum, R.~D.\ 2002, \apj, 564, 827
\bibitem[Conti \& Crowther(2004)]{2004MNRAS.355..899C} Conti, P.~S. \& Crowther, P.~A.\ 2004, \mnras, 355, 899. doi:10.1111/j.1365-2966.2004.08367.x
\bibitem[Clarke et al.(2015)]{2015ASPC..495..355C} Clarke, M., Vacca, W.~D., \& Shuping, R.~Y.\ 2015, Astronomical Data Analysis Software an Systems XXIV (ADASS XXIV), 495, 355
\bibitem[De Buizer(2006)]{2006ApJ...642L..57D} De Buizer, J.~M.\ 2006, \apjl, 642, L57
\bibitem[De Buizer et al.(2017)]{2017ApJ...843...33D} De Buizer, J.~M., Liu, M., Tan, J.~C., et al.\ 2017, \apj, 843, 33
\bibitem[De Pree et al.(1997)]{1997ApJ...482..307D} De Pree, C.~G., Mehringer, D.~M., \& Goss, W.~M.\ 1997, \apj, 482, 307
\bibitem[De Pree et al.(2000)]{2000ApJ...540..308D} De Pree, C.~G., Wilner, D.~J., Goss, W.~M., et al.\ 2000, \apj, 540, 308
\bibitem[De Pree et al.(2020)]{2020AJ....160..234D} De Pree, C.~G., Wilner, D.~J., Kristensen, L.~E., et al.\ 2020, \aj, 160, 234
\bibitem[Dickel \& Goss(1990)]{1990ApJ...351..189D} Dickel, H.~R. \& Goss, W.~M.\ 1990, \apj, 351, 189
\bibitem[Dreher et al.(1984)]{1984ApJ...283..632D} Dreher, J.~W., Johnston, K.~J., Welch, W.~J., et al.\ 1984, \apj, 283, 632
\bibitem[Elia et al.(2017)]{2017MNRAS.471..100E} Elia, D., Molinari, S., Schisano, E. et al.\ 2017, \mnras, 471, 100
\bibitem[Garay et al.(2003)]{2003ApJ...587..739G} Garay, G., Brooks, K.~J., Mardones, D., et al.\ 2003, \apj, 587, 739
\bibitem[Gutermuth et al.(2009)]{2009ApJS..184...18G} Gutermuth, R. A., Megeath, S. T., Myers, P. C. et al., 2009, ApJS, 184, 18
\bibitem[Herter et al.(2013)]{2013PASP..125.1393H} Herter, T.~L., Vacca, W.~D., Adams, J.~D., et al.\ 2013, \pasp, 125, 1393
\bibitem[Homeier \& Alves(2005)]{2005A&A...430..481H} Homeier, N.~L. \& Alves, J.\ 2005, \aap, 430, 481
\bibitem[Hunter et al.(2018)]{2018ApJ...854..170H} Hunter, T.~R., Brogan, C.~L., MacLeod, G.~C., et al.\ 2018, \apj, 854, 170
\bibitem[Jackson \& Kraemer(1994)]{1994ApJ...429L..37J} Jackson, J.~M. \& Kraemer, K.~E.\ 1994, \apjl, 429, L37
\bibitem[Kim et al.(2018)]{2018A&A...616A.107K} Kim, W.-J., Urquhart, J.~S., Wyrowski, F., et al.\ 2018, \aap, 616, A107
\bibitem[Krumholz \& Tan(2007)]{2007ApJ...654..304K} Krumholz, M.~R., \& Tan, J.~C.\ 2007, \apj, 654, 304
\bibitem[Lada \& Lada(2003)]{2003ARA&A..41...57L} Lada, C.~J. \& Lada, E.~A.\ 2003, \araa, 41, 57. doi:10.1146/annurev.astro.41.011802.094844
\bibitem[Lim et al.(2016)]{2016ApJ...829L..19L} Lim, W., Tan, J.~C., Kainulainen, J., Ma, B., \& Butler, M.~J.\ 2016, \apjl, 829, L19 
\bibitem[Lim \& De Buizer(2019)]{2019ApJ...873...51L} Lim, W. \& De Buizer, J. M.\ 2019, \apj, 873, 51 [Paper 1]
\bibitem[Lim et al.(2020)]{2020ApJ...888...98L} Lim, W., De Buizer, J.~M., \& Radomski, J.~T.\ 2020, \apj, 888, 98 [Paper 2]
\bibitem[Lim et al.(2021)]{2021PASJ...73S.239L} Lim, W., Nakamura, F., Wu, B., et al.\ 2021, PASJ, 73, 239
\bibitem[Ma et al.(2013)]{2013ApJ...779...79M} Ma, B., Tan, J.~C., \& Barnes, P.~J.\ 2013, \apj, 779, 79
\bibitem[Mezger et al.(1967)]{1967ApJ...150..807M} Mezger, P.~G., Schraml, J., \& Terzian, Y.\ 1967, \apj, 150, 807
\bibitem[Mois{\'e}s et al.(2011)]{2011MNRAS.411..705M} Mois{\'e}s, A.~P., Damineli, A., Figuer{\^e}do, E., et al.\ 2011, \mnras, 411, 705. doi:10.1111/j.1365-2966.2010.17713.x
\bibitem[Molinari et al.(2016)]{2016A&A...591A.149M} Molinari, S., Schisano, E., Elia, D. et al.\ 2016, \aap, 591, 149
\bibitem[Mufson \& Liszt(1977)]{1977ApJ...212..664M} Mufson, S.~L. \& Liszt, H.~S.\ 1977, \apj, 212, 664
\bibitem[Peng et al.(2010)]{2010A&A...520A..84P} Peng, T.-C., Wyrowski, F., van der Tak, F.~F.~S., et al.\ 2010, \aap, 520, A84
\bibitem[Purser et al.(2016)]{2016MNRAS.460.1039P} Purser, S.~J.~D., Lumsden, S.~L., Hoare, M.~G., et al.\ 2016, \mnras, 460, 1039
\bibitem[Plume et al.(2004)]{2004ApJ...605..247P} Plume, R., Kaufman, M.~J., Neufeld, D.~A., et al.\ 2004, \apj, 605, 247
\bibitem[Rugel et al.(2019)]{2019A&A...622A..48R} Rugel, M. R.; Rahner, D.; Beuther, H., et al.\ 2019, \aap, 622, 48
\bibitem[Samal et al.(2014)]{2014A&A...566A.122S} Samal, M.~R., Zavagno, A., Deharveng, L., et al.\ 2014, \aap, 566, A122. doi:10.1051/0004-6361/201321794
\bibitem[Saral et al.(2015)]{2015ApJ...813...25S} Saral, G., Hora, J.~L., Willis, S.~E., et al.\ 2015, \apj, 813, 25
\bibitem[Serabyn et al.(1993)]{1993ApJ...413..571S} Serabyn, E., Guesten, R., \& Schulz, A.\ 1993, \apj, 413, 571
\bibitem[Smith et al.(1978)]{1978A&A....66...65S} Smith, L.~F., Biermann, P., \& Mezger, P.~G.\ 1978, \aap, 66, 65
\bibitem[Smith et al.(2000)]{2000ApJ...540..316S} Smith, N., Jackson, J.~M., Kraemer, K.~E., et al.\ 2000, \apj, 540, 316
\bibitem[Smith et al.(2009)]{2009MNRAS.399..952S} Smith, N., Whitney, B.~A., Conti, P.~S., de Pree, C.~G., \& Jackson, J. M.\ 2009, \mnras, 399, 952
\bibitem[Umemoto et al.(2017)]{2017PASJ...69...78U} Umemoto, T., Minamidani, T.; Kuno, N., et al.\ 2017, PASJ, 69, 78
\bibitem[Urquhart et al.(2013)]{2013MNRAS.435..400U} Urquhart, J.~S., Thompson, M.~A., Moore, T.~J.~T., et al.\ 2013, \mnras, 435, 400
\bibitem[Wang et al.(2018)]{2018PASP..130k4301W} Wang, L.-L., Luo, A.-L., Hou, ., et al.\ 2018, \pasp, 130, 430
\bibitem[Ward-Thompson \& Robson(1990)]{1990MNRAS.244..458W} Ward-Thompson, D. \& Robson, E.~I.\ 1990, \mnras, 244, 458
\bibitem[Webster et al.(1971)]{1971AJ.....76..677W} Webster, W.~J., Altenhoff, W.~J., \& Wink, J.~E.\ 1971, \aj, 76, 677
\bibitem[Welch et al.(1987)]{1987Sci...238.1550W} Welch, W.~J., Dreher, J.~W., Jackson, J.~M., et al.\ 1987, Science, 238, 1550
\bibitem[Westbrook et al.(1976)]{1976ApJ...209...94W} Westbrook, W.~E., Werner, M.~W., Elias, J.~H., et al.\ 1976, \apj, 209, 94
\bibitem[Westerhout(1958)]{1958BAN....14..215W} Westerhout, G.\ 1958, \bain, 14, 215
\bibitem[Young et al.(2012)]{2012ApJ...749L..17Y} Young, E.~T., Becklin, E.~E., Marcum, P.~M., et al.\ 2012, \apjl, 749, L17
\bibitem[Zhang et al.(2013)]{2013ApJ...775...79Z} Zhang, B., Reid, M.~J., Menten, K.~M., et al.\ 2013, \apj, 775, 79

\end{thebibliography}
\end{document}